\pdfoutput=1 
\documentclass[aps,superscriptaddress,preprintnumbers,prd,twocolumn]{revtex4-2}
\usepackage{amsmath}
\usepackage{amssymb}
\usepackage{braket}
\usepackage{enumerate}
\usepackage{hyperref}
\hypersetup{colorlinks=true}
\usepackage{tikz}
\usepackage{quantikz}
\usetikzlibrary{arrows}
\usetikzlibrary{calc}
\usetikzlibrary{decorations.markings}
\usetikzlibrary{math}

\begin{document}

\preprint{UMD-PP-021-03}

\title{Shearing approach to gauge-invariant Trotterization}

\author{Jesse R. Stryker}
\email[]{jstryker@lbl.gov}

\affiliation{
{Physics Division, Lawrence Berkeley National Laboratory, Berkeley, California 94720, USA.}}

\affiliation{Maryland Center for Fundamental Physics, University of Maryland, College Park, Maryland 20742, USA.}

\date{\today}

\begin{abstract}
  Universal quantum simulations of gauge field theories are exposed to the risk of gauge symmetry violations when it is not known how to compile the desired operations exactly using the available gate set.
  In this article, we show how time evolution can be compiled in an Abelian gauge theory---if only approximately---without compromising gauge invariance, by graphically motivating a block-diagonalization procedure.
  When gauge-invariant interactions are associated with a ``spatial network'' in the space of discrete quantum numbers, it is seen that cyclically shearing the spatial network converts simultaneous updates to many quantum numbers into conditional updates of a single quantum number;
  ultimately, this eliminates any need to pass through (and acquire overlap onto) unphysical intermediate configurations.
  Shearing is explicitly applied to gauge-matter and magnetic interactions of lattice quantum electrodynamics.
  The features that make shearing successful at preserving Abelian gauge symmetry may also be found in non-Abelian theories, bringing one closer to gauge-invariant simulations of quantum chromodynamics.
\end{abstract}

\maketitle

\section{Introduction}
The real-time dynamics of quantum field theories (QFTs) holds the details of rich nonperturbative physics, such as the fragmentation of jets at the Large Hadron Collider or early universe bubble nucleation in extensions of the Standard Model.
Analytic methods that can quantitatively predict nonperturbative phenomena from the underlying quantum field theories are scarce, and this deficit underlies the existence of what is today a sophisticated, numerical lattice QFT program.
For decades, the lattice QFT program has pushed the envelope of high-performance computing and fundamental physics, but its scope has generally been restricted to properties for which a QFT's Euclidean or imaginary-time path integral is not hampered by a sign problem.
The calculation of real-time dynamics has largely been precluded by sign problems, except for special cases \cite{Luscher:1986pf,Luscher:1990ux}.
The fact is that real-time dynamics explicitly involves events in Minkowskian, rather than Euclidean, spacetime.
Any general computational solution---if one exists---will require the use of nontraditional lattice methods.

In recent years, a new computational avenue for dynamics has opened up
\cite{
Martinez:2016yna,
Klco:2018kyo,
Kokail:2018eiw,
Schweizer:2019lwx,
Klco:2019evd,
Yang:2020yer,
Kreshchuk:2020kcz,
Klco:2019xro,
Gustafson:2021imb,
Bauer:2021gup,
Atas:2021ext,
ARahman:2021ktn,
Ciavarella:2021nmj,
Huffman:2021gsi,
Xu:2021tey,
Ciavarella:2021lel,
Nguyen:2021hyk,
Illa:2022jqb,
Mildenberger:2022jqr,
Pardo:2022hrp,
Farrell:2022wyt,
Farrell:2022vyh,
Gupta:2024gnw,
Davoudi:2024wyv,
Ciavarella:2024fzw,
Crippa:2024hso,
Zhang:2024fgv,
Than:2024zaj,
Schuhmacher:2025ehh,
Davoudi:2025rdv%
} based on simulating field theoretic degrees of freedom
\cite{%
Jordan:2012xnu,
Jordan:2011ci,
Bhattacharyya:2018bbv,
Klco:2018zqz,
Klco:2019yrb,
Barata:2020jtq,
IgnacioCirac:2010us,
Casanova:2011wh,
Jordan:2014tma,
HamedMoosavian:2017koz,
Lamm:2019uyc,
Mueller:2019qqj,
Harmalkar:2020mpd,
Kharzeev:2020kgc,
Bauer:2019qxa,
Davoudi:2021ney,
Banuls:2019bmf,
Byrnes:2005qx,
Stryker:2018efp,
Lamm:2019bik,
Alexandru:2019nsa,
Zohar:2019ygc,
Shaw:2020udc,
Chakraborty:2020uhf,
Liu:2020eoa,
Paulson:2020zjd,
Ji:2020kjk,
Davoudi:2020yln,
Bender:2020ztu,
Haase:2020kaj,
Kan:2021nyu,
Zohar:2011cw,
Zohar:2013zla,
Zohar:2014qma,
Zohar:2016wmo,
Bender:2018rdp,
Davoudi:2019bhy,
Mil:2019pbt,
Banerjee:2012pg,
Tagliacozzo:2012vg,
Mezzacapo:2015bra,
Zache:2018jbt,
Zohar:2012xf,
Mathur:2016cko,
Raychowdhury:2019iki,
Raychowdhury:2018osk,
Dasgupta:2020itb,
Kreshchuk:2020dla,
Buser:2020cvn,
Kan:2022esj,
Illa:2022jqb,
Farrell:2022wyt,
Farrell:2022vyh,
DAndrea:2023qnr,
Sukeno:2023uhx,
Bauer:2023qgm,
Than:2024zaj%
}
and their interactions with laboratory-controllable quantum systems, i.e., quantum simulators and computers \cite{Feynman:1981tf,Lloyd:1996aai}.
Quantum computers are naturally suited for simulating Hamiltonian mechanics, as opposed to path integrals, and thereby appear to be sign problem-free.
This paper is specifically concerned with simulation by universal digital quantum computers (as opposed to analog ones, as well as quantum annealers), which are the model for architectures like those being engineered by Google \cite{Arute:2019zxq}, IBM \cite{Chow:2012zps}, IonQ \cite{Debnath:2016xdi}, Rigetti \cite{Reagor:2018csq}, and others.

Digital quantum simulation requires a quantum algorithm or set of instructions to be executed by the quantum computer.
For QFTs, this may entail truncating and mapping the field degrees of freedom onto quantum bits (qubits) and mimicking Schr\"{o}dinger-picture time evolution through an appropriate sequence of unitary gates.
At the end of the evolution, measurements would be made to extract the observables of interest.
How exactly QFTs and their dynamics would best be ``digitized'' is a subject of active research
\cite{%
Wiese:2014rla,
Zohar:2016iic,
Kaplan:2018vnj,
Klco:2018zqz,
Lamm:2019bik,
Alexandru:2019nsa,
Raychowdhury:2019iki,
Klco:2019evd,
Davoudi:2020yln,
Bender:2020ztu,
Meurice:2020pxc,
Kreshchuk:2020dla,
Barata:2020jtq,
Bauer:2019qxa,
Bauer:2021gup,
Ciavarella:2021nmj,
Meurice:2021ujn,
Ciavarella:2022zhe,
Pardo:2022hrp,
Irmejs:2022gwv,
Kadam:2022ipf,
DAndrea:2023qnr,
Davoudi:2024wyv,
Calajo:2024qrc,
Ciavarella:2024fzw,
Mariani:2024osg,
Carena:2024dzu,
Mathew:2024bed,
Li:2024ide,
Kadam:2024zkj,
Grabowska:2024emw,
Ma:2025ysk%
}.

Among field theories that will be studied by quantum computers, gauge theories such as quantum chromodynamics are uniquely complex due to their hallmark local constraints---namely gauge invariance, which is identified with Gauss's law and charge conservation.
Unlike certain analog devices, in which there can be opportunities to map gauge symmetry to a physical symmetry of the simulator \cite{Zohar:2012xf}, qubits and the operations done on them generally do not discriminate between gauge-violating interactions and gauge-conserving ones.
Gauge symmetry, therefore, is a concern for both the wave functions and the effectively simulated interactions.
We do note, however, that there is evidence of dynamical robustness against gauge-violating interactions in certain contexts 
\cite{%
Halimeh:2020kyu,
Jensen:2022hyu,
Gustafson:2023swx%
}.
One may also consider deliberately engineering the simulation to suppress possible gauge violations,
for example by
exploiting the quantum Zeno effect
\cite{%
Stannigel:2013zka,
raimond2010phase,
raimond2012quantum,
signoles2014confined,
Halimeh:2020ecg,
Halimeh:2021lnv,
VanDamme:2021njp,
Halimeh:2022mct,
Ball:2024xmw,
Wauters:2024shc,
Ball:2024uqu%
},
designing (effective) energy penalties 
\cite{%
Surace:2023ycp,
Cheng:2024pdu,
DePaciani:2025uzj%
},
or averaging over gauge transformations 
\cite{%
Lamm:2020jwv,
Kasper:2020owz,
Tran:2020azk,
Nguyen:2021hyk%
}
-- among a variety of other creative ideas
\cite{%
Zhao:2022dma,
Surace:2023qwo,
Domanti:2023qht,
Okuda:2024jzh,
Schmale:2024ebh,
Ballini:2024qmr,
DePaciani:2025uzj%
}.
Until the costs and benefits of approaches that permit some gauge violations are fully understood, exactly gauge-invariant protocols, such as those presented below, remain highly desirable.

Symmetry preservation aside, a crucial consideration in selecting any QFT digitization is the implementation of its time evolution.
Gauge theory Hamiltonians are often given in the form $H=\sum_j H_j$, where each $H_j$ is a manifestly gauge-invariant operator involving fields from a site, link, or unit square (plaquette) of a spatial lattice.
The time evolution operator $ e^{ -i t H} $ generally will not be a native operation and must be constructed from the available gate set.
The simplest and most well-studied approaches take advantage of Lie-Trotter-Suzuki product formulas \cite{trotterProductSemigroups59,Suzuki:1976be,Childs:2019hts}:
at first order, $e^{-i \, t H} \simeq \left[\prod_j e^{-i \, t H_j / s } \right]^s $ for a sufficiently large number of Trotter steps, $s$.
This \emph{Trotterization} of the time evolution operator is inexact for finite $s$, but it is gauge-invariant because each $H_j$ is.

In practice, the complexity of the $H_j$ may call for a second level of approximation:
$e^{-i \, \delta t H_j} \rightarrow \prod_k V_{jk}(\delta t) $ in the limit of small $\delta t$, where each factor $ V_{jk} (\delta t) $ is an implementable sequence of elementary gates.
There is no reason to expect \emph{a priori} that the individual $ V_{jk} $ will be gauge-invariant, nor is there any guarantee $ \prod_k V_{jk}(\delta t) $ conserves gauge symmetry either.
Protecting gauge invariance through this second layer of approximation (which is also referred to as Trotterization) is a nontrivial aspect of designing the quantum circuit.

In this paper, a solution is introduced for Trotterizing Abelian gauge theories while preserving gauge symmetry.
This is achieved by finding collections $V_{jk} (\delta t)$ that \emph{are} individually gauge-invariant, implying gauge-invariant time evolution.
The key insight is that the transitions induced by gauge-invariant operators can be thought of as a geometric graph or ``spatial network'' of parallel edges.
Through appropriate changes of basis, the edges can be aligned parallel to the axis of a single quantum number, effectively turning the problem of tightly correlated transitions of multiple quantum numbers into transitions of a single quantum number conditioned by Boolean logic.
This method, based on `cyclic shears,' is illustrated by applying it to hopping terms (minimally coupled fermion-photon interactions) and plaquette operators encountered in compact U(1) gauge theories, such as lattice QED.

\section{Example: U(1) hopping terms}
In Ref.~\cite{Shaw:2020udc} (and later, Ref.~\cite{Kan:2022esj}), complete and scalable algorithms with bounded errors were given for time evolution of the lattice Schwinger model (QED in 1D space), truncated in the basis of electric fields and fermion occupation numbers.
The $H_j$ were link-localized electric energies and hopping terms, and site-localized mass terms.
The propagators $e^{-i t H_j / s}$ associated to the electric and mass terms could be decomposed straightforwardly and exactly using typical universal gate sets because they are diagonal in the computational basis.
The off-diagonal hopping terms do not enjoy the same benefit, making it less obvious how to exactly decompose their associated propagators as circuits.

Concretely, let $\hat{\psi}$ and $\hat{\chi}$ be the fermionic modes at two neighboring lattice sites, $\hat{U}$ the gauge-link variable joining them, $x>0$ the interaction strength, and $ \hat{T}_{\text{hop}} = x \, \hat{\psi}^\dagger \hat{\chi} \hat{U} $ the associated hopping term.
Conjugate to a link operator is the integer electric field $\hat{E}$ along that link;
we work in the eigenbasis of $\hat{E}$, with $ \hat{U} = \sum_E \ket{E+1} \bra{E} $.
We also assume a hard cutoff for which $ E_{\text{min}} \leq E \leq E_{\text{max}} $ and $ \hat{U} \ket{ E_{\text{max}} } = \hat{U}^\dagger \ket{ E_{\text{min}} } = 0 $.

A Jordan-Wigner transformation can be used to turn the fermionic operators into qubit (spin $\tfrac{1}{2}$) operators, e.g.,
\begin{align}
  \hat{T}_{\text{hop}} = x \, \sigma^-_\psi \sigma^+_\chi \hat{U} ;
\end{align}
we are using a computational basis in which $ \sigma^+ = \ket{0} \bra{1} $ and 0 and 1 denote eigenvalues of $ \hat{n}_{\psi} \equiv \hat{\psi}^\dagger \hat{\psi}$ or $ \hat{n}_{\chi} \equiv \hat{\chi}^\dagger \hat{\chi}$.
Electric eigenstates, truncated to $\eta$ qubits per link, are interchangeably labeled by nonnegative integers $ \mathcal{E} = E - E_{\text{min}} $.
Arithmetic involving $ \hat{\mathcal{E}} $ is always modulo $ 2^{\eta} $.

In multiqubit quantum computation, it is common for \emph{Pauli operator} to refer to any tensor product of Pauli matrices $I$, $X$, $Y$, or $Z$ acting on individual qubit spaces (e.g., $X \otimes X$ or $I \otimes Z \otimes Z$).
Pauli operators form a Hermitian basis for all multiqubit operators.
The expression of an operator with respect to this basis is its \emph{Pauli decomposition}.
The exponential of a Pauli operator is considered easy to simulate, but directly Trotterizing the Pauli decomposition can be prohibitively inefficient.
Generally, one will divide up and manipulate different parts of the Hamiltonian before arriving at Pauli operators intended for hardware implementation.

Diagonalized operators such as $\hat{E}^2$ in the Schwinger model are trivial to circuitize without approximation because Trotterizing their Pauli decomposition incurs no commutation errors:
$[I,Z]=0$.
The off-diagonal hopping terms, however, are not as trivial to implement exactly, in part due to $[X,Y]\neq 0$.
Diagonalizing the involved field operators is not an option since they are nilpotent.
The fields' Hermitian and anti-Hermitian parts are of course diagonalizable, but they are not by themselves gauge-covariant and they fail to commute.
Reference \cite{Shaw:2020udc} went forward with the Hermitian and antihermitian fields, postponing the construction of fully gauge-invariant Trotter steps.

\begin{figure}
  \includegraphics[width=0.35\textwidth]{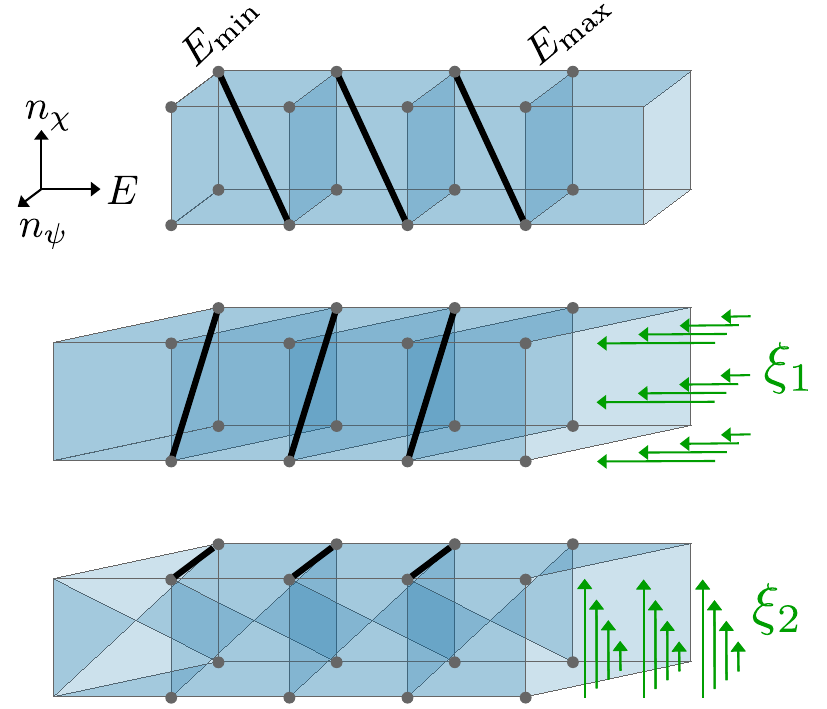}
  \caption{
    \label{fig:hoppingShears}
    Graphical representation of the transitions induced by hopping terms.
    Top: the transitions before any change of basis.
    Middle: the transitions after $\hat{\xi}_1$.
    Bottom: the transitions after $\hat{\xi}_2 \hat{\xi}_1$.
  }
\end{figure}

Insight into the computational nature of gauge-invariant transitions can be gained by graphically examining the structure of hopping terms as a whole within the space of quantum numbers.
The top panel of Fig. \ref{fig:hoppingShears} depicts the transitions induced by $\hat{T}_{\text{hop}}$ and $\hat{T}_{\text{hop}}^\dagger$ as edges in the $(n_\psi, n_\chi, E)$ space of quantum numbers (shown for $\eta=2$).
An edge connecting $ ( n_{\psi}^{i}, n_{\chi}^{i}, E^{i} ) $ and $ ( n_{\psi}^{f} , n_{\chi}^{f} , E^{f} ) $ corresponds to a nonzero matrix element $ \langle n_{\psi}^{f} , n_{\chi}^{f} , E^{f} | (\hat{T}_{\text{hop}} + \hat{T}_{\text{hop}}^\dagger) | n_{\psi}^{i}, n_{\chi}^{i}, E^{i} \rangle $.
The transitions form what is known in graph theory as a geometric graph or spatial network.
This particular spatial network consists of parallel edges whose slant represents simultaneous changes to all three relevant quantum numbers---as given, hopping terms are off-diagonal on all three field registers.

If the edges were instead oriented parallel to one of the axes, the interaction would be off-diagonal on a single register only.
Such a block-diagonalization would make it easier to avoid most gauge-violating transitions by simply excluding all $X$ and $Y$ operations on the other registers' qubits.
It is clear from Fig. \ref{fig:hoppingShears} that this scenario is entirely attainable by appropriately shearing the graph.

The middle section of Fig. \ref{fig:hoppingShears} show the result of a cyclic shear transformation, which is mathematically given by
\begin{align}
  \hat{\xi}_1 &= \delta_{ \hat{n}_\psi , 0 } + \delta_{ \hat{n}_\psi , 1 } \hat{\lambda}^- \ , \label{eq:hoppingShear1} \\
  \hat{\xi}_1 \hat{T}_{\text{hop}} \hat{\xi}_1^\dagger &= x \, \sigma^-_\psi \sigma^+_\chi ( 1 - \delta_{ \hat{E} , E_{\text{max}} } ) \ . \label{eq:hoppingShear2} 
\end{align}
Above, $\delta$ symbols serve as shorthand for projection operators: $\delta_{\hat{n}_\psi ,0} = (\ket{0} \bra{0})_\psi$, $\delta_{\hat{E},E_{\text{max}}} = \ket{E_{\text{max}}} \bra{E_{\text{max}}}$, etc.
We also introduce cyclic incrementers $ \hat{\lambda}^{\pm} = \sum_{\mathcal{E}=0}^{2^\eta-1} \ket{\mathcal{E}\pm 1} \bra{ \mathcal{E}}$.
The cyclicity of the shear transformation thus means that when a node gets sheared beyond the extent of its range, it wraps around to the opposite boundary and continues being sheared.
In this way, the shears provide unitary mappings that can be applied to computational basis states.

Under $\hat{\xi}_1$, $\hat{T}_{\text{hop}}$ becomes off-diagonal on two registers only.
This can then be combined with another shear,
\begin{align}
  \hat{\xi}_2 &= \delta_{ \hat{n}_\psi , 0 } + \delta_{ \hat{n}_\psi , 1 } X_\chi \ , \\
  \hat{\xi}_2 ( \hat{\xi}_1 \hat{T}_{\text{hop}} \hat{\xi}_1^\dagger ) \hat{\xi}_2^\dagger &= x \, \delta_{ \hat{n}_\chi , 1 } \sigma^-_\psi ( 1 - \delta_{ \hat{E} , E_{\text{max}} } ) \ . \nonumber
\end{align}
After the combined transformation $\hat{\xi}_2 \hat{\xi}_1$, $\hat{T}_{\text{hop}}$ becomes off-diagonal on the space of a single qubit.
The projector $( 1 - \delta_{ \hat{E} , E_{\text{max}} } )$ that has arisen is necessary to prevent a reduced set of possible gauge-violating errors: those that correspond to wrap-around effects at the cutoffs.
This is the work left to be done by hand, as far as charge conservation is concerned.

\begin{figure}
  \includegraphics{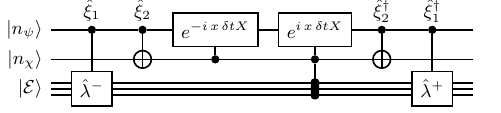}
  \caption{Circuit to simulate a gauge-fermion hopping term, $x ( \hat{\psi}^\dagger \hat{\chi} \hat{U} - \hat{\psi} \hat{\chi}^\dagger \hat{U}^\dagger ) $.
    The multiqubit electric register could be any number of qubits.
}
  \label{fig:schwingerHop}
\end{figure}

Fortuitously, in this case of gauge-fermion hopping terms, the work left to be done by hand is negligible.
The Hermitian operator to be simulated, $ \delta_{ \hat{n}_\chi , 1 } X_\psi ( 1 - \delta_{ \hat{E} , E_{\text{max}} } ) $, can be implemented with two controlled $X$-rotations.
Figure \ref{fig:schwingerHop} shows these gates sandwiched between the shears.
This circuit, which was only tailor-made to conserve charge, is actually an exact gate decomposition for the hopping propagator.

\section{Example: Compact U(1) plaquettes}
The shearing of plaquettes is explained below by first considering a toy model, two-link ``plaquette'':
two links joined end to end.
The toy plaquette highlights every key feature of solving plaquette operators using shears, but is more easily visualized than the case of four links.
The generalization to the four-link (true) plaquette is summarized thereafter.

\begin{figure}[t]
  \includegraphics[width=.48\textwidth]{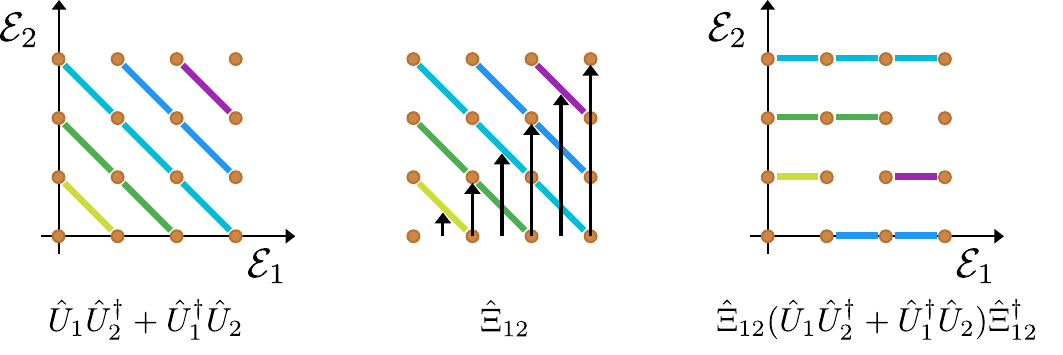}
  \caption{\label{fig:toyShear}
    Left: applications of $ \hat{U}_1 \hat{U}_2^\dagger $ and $ \hat{U}_1^\dagger \hat{U}_2  $ can only move a state by $ \pm ( 1, -1 ) $ in the 12-plane, forming chains of reachable states.
    The available chains are each displayed in a distinct color (shown here for $\eta=2$).
    Right: In the cyclically sheared electric coordinates, allowed moves generated by $ \hat{U}_1 \hat{U}_2^\dagger  $ and $ \hat{U}_1^\dagger \hat{U}_2  $ are parallel to the $ \mathcal{E}_1 $-axis.
  }
\end{figure}

Consider two links labeled 1 and 2 with their truncated plaquette operator given by
\begin{align}
  \hat{U}_1 \hat{U}_2^\dagger &= \hat{\lambda}_1^+ [ 1 - \delta_{ \hat{E}_1 , E_{\text{max}} } ] \hat{\lambda}_2^- [ 1 - \delta_{ \hat{E}_{2} , E_{\text{min}} } ]  \nonumber \\
  &= \hat{\lambda}_1^+ \hat{\lambda}_2^- [ 1 - \delta_{ \hat{\mathcal{E}}_1 , -1 } ] [ 1 - \delta_{ \hat{\mathcal{E}}_{2} , 0 } ] \ . \label{eq:factoredToyPlaq}
\end{align}
A consequence of $ [ \hat{U}_1 \hat{U}_2^\dagger , \hat{E}_1 + \hat{E}_2 ] = 0 $ is that the electric configurations that can be mixed by $ \hat{U}_1 \hat{U}_2^\dagger $ lie on various one-dimensional lines in the space of electric quantum numbers.
As illustrated in Fig.~\ref{fig:toyShear}, these transitions can be aligned with one axis by a shear in the 12-plane,
\begin{align}
  \hat{\Xi}_{12} &\equiv \sum_{j=0}^{2^\eta - 1} \delta_{\hat{\mathcal{E}}_1, j} ( \hat{\lambda}_2^+ )^{j} \ .
\end{align}
Computationally, $\hat{\Xi}_{12}$ is nothing but addition modulo $ 2^\eta $:
$\ket{ \mathcal{E}_1 } \ket{ \mathcal{E}_2 } \overset{ \Xi_{12} }{ \rightarrow } \ket{ \mathcal{E}_1 } \ket{ \mathcal{E}_2 + \mathcal{E}_1 } $.
Modular addition of two $\eta$-bit integers is an elementary routine called for throughout quantum computation, with two possible implementations being the ripple-carry adder \cite{Cuccaro:2004xxx} and carry-lookahead adder \cite{Draper:2004ayj}, and the particular choice can be tailored to the available resources.
Applying $\hat{\Xi}_{12}$ to the operators appearing in (\ref{eq:factoredToyPlaq}), we find that $ \hat{\lambda}_2^- $ and $ \hat{\mathcal{E}}_1 $ are invariant, whereas $ \hat{\lambda}_1^+ \overset{ \Xi_{12} }{ \rightarrow }\hat{\lambda}_1^+ \hat{\lambda}_2^+ $ and $ \hat{\mathcal{E}}_2 \overset{ \Xi_{12} }{ \rightarrow } \hat{\mathcal{E}}_2 - \hat{\mathcal{E}}_1 $.
Therefore
\begin{align}
  \hat{\Xi}_{12} \hat{U}_1 \hat{U}_2^\dagger \hat{\Xi}_{12}^\dagger &= \hat{U}_1 [ 1 - \delta_{ \hat{\mathcal{E}}_2 - \hat{\mathcal{E}}_1 , 0 } ] \ , \label{eq:shearedToyPlaq}
\end{align}
which is off-diagonal on register 1 alone.

\begin{figure}[t!]
  \includegraphics[width=.48\textwidth]{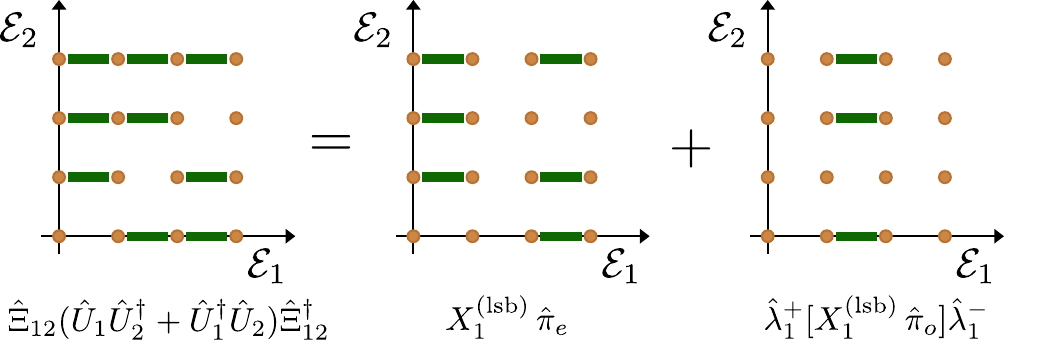}
  \caption{\label{fig:shearedEdgeGrouping}
    The edges representing $ \hat{U}_1 \hat{U}_2^\dagger + \text{H.c.} $ can be grouped by the even-odd parity of the column they are in, corresponding to the terms $ \hat{h}_e $ and $ \hat{h}_o $.
}
\end{figure}

To simulate the right-hand side of (\ref{eq:shearedToyPlaq}), we can upcycle a strategy used in \cite{Shaw:2020udc}.
The identity $ \hat{\lambda}^{\pm} = \sigma_{\mp}^{\text{(lsb)}} + \hat{\lambda}^+ \sigma_{\mp}^{\text{(lsb)}} \hat{\lambda}^- $, where $ \sigma^{\text{(lsb)}}_{\mp} $ acts on the least significant (qu)bit of $\mathcal{E}$, is used in (\ref{eq:shearedToyPlaq}) to obtain
\begin{align}
  \hat{U}_1 \hat{U}_2^\dagger  + \hat{U}_1^\dagger \hat{U}_2  & \overset{ \Xi_{12} } { \longrightarrow } \hat{h}_{e} + \hat{h}_{o} \, , \nonumber \\
  \hat{h}_{e} &= \hat{\pi}_e X_{1}^{\text{(lsb)}} , \\
  \hat{\pi}_e &\equiv ( 1 - \delta_{ \hat{\mathcal{E}}_2, 2 \lfloor \hat{\mathcal{E}}_1 / 2 \rfloor } ) , \nonumber \\
  \hat{h}_{o} &= \lambda_1^+ \hat{\pi}_o X_{1}^{\text{(lsb)}} \lambda_1^- , \\
  \hat{\pi}_o &\equiv ( 1 - \delta_{ 2 \lfloor \hat{\mathcal{E}}_1 / 2 \rfloor + 1 , -1 } ) ( 1 - \delta_{ \hat{\mathcal{E}}_2, 2 \lfloor \hat{\mathcal{E}}_1 / 2 \rfloor + 1 } ) . \nonumber
\end{align}
The projection operator $\hat{\pi}_e$ ($\hat{\pi}_o$) is named as such because it explicitly depends on $ \mathcal{E}_1 $ rounded down (up) to the nearest even (odd) integer;
in particular, $\hat{\pi}_e$ and $\hat{\pi}_o$ each commute with $X_{1}^{\text{(lsb)}}$.
Figure \ref{fig:shearedEdgeGrouping} provides intuition for the identity $ \hat{\Xi}_{12} ( \hat{U}_1 \hat{U}_2^\dagger + \hat{U}_1^\dagger \hat{U}_2   ) \hat{\Xi}_{12}^\dagger = \hat{h}_e + \hat{h}_o $ by interpreting the graph as a linear combination of pairwise mixings.
The two terms correspond to a so-called ``coloring'' of the graph, for which the chromatic number is 2.
The isolation of the terms $\hat{h}_e$ and $\hat{h}_o$ has the effect of breaking up chains of states connected by powers of the sheared plaquette into pairwise transitions in two-state subspaces;
in $\hat{h}_e$ they are bit-flips of the least significant bit, and the same can be said for $\hat{h}_o$ if the basis is shifted by one unit.
The projectors $\hat{\pi}_e$ and $\hat{\pi}_o$ that arise serve to prevent unphysical wrap-around effects at the cutoffs.

\begin{figure*}
  \includegraphics{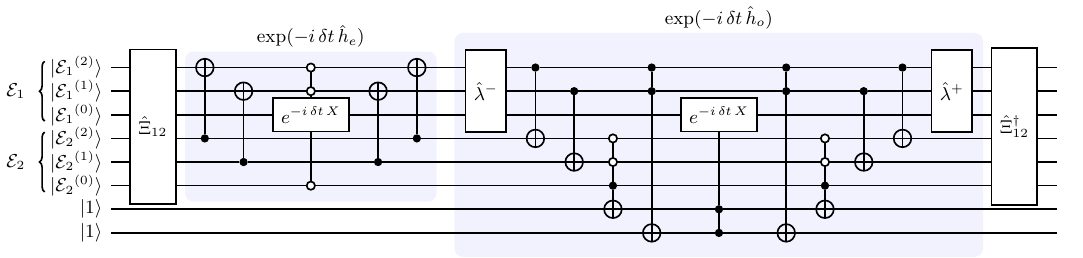}
  \caption{\label{fig:toyPlaqTrott}
    Digital quantum circuit for gauge-invariant, first-order Trotter-Suzuki approximation to $ \exp [ -i \, \delta t ( \hat{U}_1 \hat{U}_2^\dagger + \hat{U}_1^\dagger \hat{U}_2  ) ] $ on a single toy plaquette (shown for $\eta=3$).
    Filled and empty circles are used for conditional operations activated by the $\ket{1}$ or $\ket{0}$ states, respectively, of the control qubit(s) \cite{Nielsen:2012yss}.
}
\end{figure*}

Next, we consider a first-order Trotterization
$ e^{ -i \, \delta t ( \hat{U}_1 \hat{U}_2^\dagger + \hat{U}_1^\dagger \hat{U}_2  ) } \simeq \hat{\Xi}_{12}^{\dagger} e^{ -i \, \delta t \, \hat{h}_o } e^{ -i \, \delta t \, \hat{h}_e } \hat{\Xi}_{12} $, which incurs error due to $ [ \hat{h}_e , \hat{h}_o ] \neq 0 $, but respects gauge invariance.
Now all that is needed is circuits for the Trotter factors
\begin{align}
  \exp [ -i \, \delta t \, \hat{h}_o ] &= \hat{\lambda}_1^+ \exp [ -i \, \delta t \, \hat{\pi}_o X_{1}^{\text{(lsb)}} ] \hat{\lambda}_1^- \ , \\
  \exp [ -i \, \delta t \, \hat{h}_e ] &= \exp [ -i \, \delta t \, \hat{\pi}_e X_{1}^{\text{(lsb)}} ] \ .
\end{align}
At the heart of each factor is an $X$-rotation on register 1's least significant bit, controlled by the other $\eta - 1$ bits as dictated by $ \hat{\pi}_e $ or $ \hat{\pi}_o $.
Figure \ref{fig:toyPlaqTrott} shows a possible circuit implementation of the Trotterized hopping propagator for the case of $\eta=3$.
Protocols for $\hat{\lambda}^{\pm}$ and $\hat{\Xi}_{12}$ are left up to the user.
The routine calls for two ancilla qubits.

Moving on, we now generalize to the case of four links 0, 1, 2, and 3.
The truncated plaquette operator is
\begin{align}
  \hat{U}_{\square} = \hat{U}_0 \hat{U}_1 \hat{U}_2^\dagger \hat{U}_3^\dagger  &= \hat{\lambda}_0^+ \hat{\lambda}_1^+ \hat{\lambda}_2^- \hat{\lambda}_3^- [ 1 - \delta_{ \hat{\mathcal{E}}_0 , -1 } ] [ 1 - \delta_{ \hat{\mathcal{E}}_1 , -1 } ] \times \nonumber \\
  &\quad \ \ [ 1 - \delta_{ \hat{\mathcal{E}}_2 , 0 } ] [ 1 - \delta_{ \hat{\mathcal{E}}_3 , 0 } ] \ .
\end{align}
Like before, the gauge-invariant moves by $ \pm ( 1, 1, -1, -1) $ in the  $ ( E_0, E_1, E_2, E_3 ) $ hypercube induced by plaquettes form parallel chains in the space of quantum numbers that can be cyclically sheared to align with a single axis.

First, the $ E_0 $ and $ E_3 $ components of the moves can by eliminated by shears in the 01- and 23-planes,
\begin{align}
  \hat{\Xi}_{01} &\equiv \sum_{j=0}^{2^\eta - 1} \delta_{ \hat{\mathcal{E}}_1 , j} ( \hat{\lambda}_0^- ) ^j \ , \\
  \hat{\Xi}_{01} \hat{U}_0 \hat{U}_1 \hat{\Xi}_{01}^\dagger &= \hat{U}_1 (1 - \delta_{ \hat{\mathcal{E}}_0 + \hat{\mathcal{E}}_1 , -1} ) \ , \nonumber \\
  \hat{\Xi}_{23} &\equiv \sum_{k=0}^{2^\eta-1} \delta_{ \hat{\mathcal{E}}_2, k }( \hat{\lambda}_3^- ) ^k \ , \\
  \hat{\Xi}_{23} \hat{U}_2^\dagger \hat{U}_3^\dagger \hat{\Xi}_{23}^\dagger &= \hat{U}_2^\dagger ( 1 - \delta_{ \hat{\mathcal{E}}_3 + \hat{\mathcal{E}}_2 , 0} ) \ . \nonumber
\end{align}
Combining the two, $ \hat{U}_{\square} $ is transformed into $ \hat{U}_1 \hat{U}_2^\dagger (1 - \delta_{ \hat{\mathcal{E}}_0 + \hat{\mathcal{E}}_1 , -1} ) ( 1 - \delta_{ \hat{\mathcal{E}}_3 + \hat{\mathcal{E}}_2 , 0} ) $: this is essentially the two-link toy plaquette, but with additional cutoff-enforcing projectors.
The $ E_2 $ components can then be eliminated just like before, by porting over $\hat{\Xi}_{12}$ unchanged.
Recalling how $ \hat{\Xi}_{12} $ transforms the operators $ \hat{U}_1 \hat{U}_2^\dagger $, $ \hat{\mathcal{E}}_1 $, and $ \hat{\mathcal{E}}_2 $, we arrive at a sheared plaquette that is off-diagonal on register 1 alone,
\begin{align}
  \hat{U}_{\square} \overset{\Xi_{12} \Xi_{23} \Xi_{01}}{\xrightarrow{\hspace*{28pt}}}&= \hat{U}_1 ( 1 - \delta_{ \hat{\mathcal{E}}_2 - \hat{\mathcal{E}}_1 , 0 } ) (1 - \delta_{ \hat{\mathcal{E}}_0 + \hat{\mathcal{E}}_1 , -1} ) \times \nonumber \\
  & \quad ( 1 - \delta_{ \hat{\mathcal{E}}_3 + \hat{\mathcal{E}}_2 - \hat{\mathcal{E}}_1 , 0} ) \ .
\end{align}

Looking toward Trotterization, let us express the sheared plaquette as $ \hat{\lambda}_1^+ \hat{\Pi} $, where
\begin{align}
  \hat{\Pi} &\equiv [ 1 - \delta_{ \hat{\mathcal{E}}_0 + \hat{\mathcal{E}}_1 , -1} ] [ 1 - \delta_{ \hat{\mathcal{E}}_1 , -1 } ] \times \nonumber \\
  &\qquad [ 1 - \delta_{ \hat{\mathcal{E}}_2 - \hat{\mathcal{E}}_1 , 0 } ] [ 1 - \delta_{ \hat{\mathcal{E}}_3 + \hat{\mathcal{E}}_2 - \hat{\mathcal{E}}_1 , 0} ] \, . \nonumber
\end{align}
$\hat{\Pi}$ is a projector on to the states in the electric hypercube not destroyed by $ \hat{U}_{\square} $, after it has been sheared.
Substituting for $ \hat{\lambda}^{\pm} $ and simplifying as in the toy example, we find that the Hamiltonian term that needs to be simulated in the sheared basis is given by
\begin{align}
  \hat{\Xi}_{12} \hat{\Xi}_{23} & \hat{\Xi}_{01} ( \hat{U}_{\square} + \hat{U}_{\square}^\dagger ) \hat{\Xi}_{01}^\dagger \hat{\Xi}_{23}^\dagger \hat{\Xi}_{12}^\dagger = \hat{H}_e + \hat{H}_o \, , \\
  \hat{H}_e &= \hat{\Pi}_e X_{1}^{\text{(lsb)}} \, , \\
  \hat{\Pi}_e &\equiv [ 1 - \delta_{ \hat{\mathcal{E}}_0 + 2 \lfloor \hat{\mathcal{E}}_1/2 \rfloor , -1} \, ] \times \nonumber \\
  &\qquad [ 1 - \delta_{ \hat{\mathcal{E}}_2 - 2 \lfloor \hat{\mathcal{E}}_1/2 \rfloor , 0 } \, ] [ 1 - \delta_{ \hat{\mathcal{E}}_3 + \hat{\mathcal{E}}_2 - 2 \lfloor \hat{\mathcal{E}}_1/2 \rfloor , 0} \, ] \, , \nonumber \\
  \hat{H}_o &= \hat{\lambda}_1^+ \hat{\Pi}_o X_{1}^{\text{(lsb)}} \hat{\lambda}_1^- \, , \\
  \hat{\Pi}_o &\equiv [ 1 - \delta_{ \hat{\mathcal{E}}_0 + 2 \lfloor \hat{\mathcal{E}}_1/2 \rfloor + 1  , -1} \, ] [ 1 - \delta_{ 2 \lfloor \hat{\mathcal{E}}_1/2 \rfloor + 1  , -1 } \, ] \times \nonumber \\
  &\qquad [ 1 - \delta_{ \hat{\mathcal{E}}_2 - 2 \lfloor \hat{\mathcal{E}}_1/2 \rfloor - 1  , 0 } \, ] [ 1 - \delta_{ \hat{\mathcal{E}}_3 + \hat{\mathcal{E}}_2 - 2 \lfloor \hat{\mathcal{E}}_1/2 \rfloor - 1  , 0} \, ] \, . \nonumber
\end{align}
Like in the two-link model, $ \hat{H}_e $ and $ \hat{H}_o $ are essentially controlled $X$-rotations on a single qubit, and they can be simulated separately in a Trotter-Suzuki expansion without violating gauge invariance.
A quantum circuit for this (the counterpart to Fig. \ref{fig:toyPlaqTrott}) is provided in Fig.~\ref{fig:fullPlaqCircuit} of the Appendix.

\section{Discussion}
The objective of this work was to furnish a new strategy for constructing digital quantum simulation algorithms for Abelian gauge theories that do not compromise gauge invariance.
What makes gauge invariance nontrivial is the necessity of approximating time evolution with gates that generally are not gauge-invariant, while at the same time enacting tightly correlated changes across many qubits.
A key observation is that the particular transitions induced by a given gauge-invariant interaction can be thought of as parallel edges of a spatial network in the space of involved charge quantum numbers.
It follows that appropriately chosen cyclic shears can be used to align the edges parallel to the axis of a single quantum number, essentially converting the correlated changes across multiple qubit registers into controlled increments on a single register.

One obvious question remains:
How practical is it to make the circuits gauge-invariant?
To this end, we will compare the sheared hopping propagator to its counterpart developed in Ref.~\cite{Shaw:2020udc}.
As detailed in the Appendix, an equitable comparison of the gate counting (in the approximation that $\hat{U}$ is allowed to wrap around at the cutoff) leads to an upper bound of $ 4 \eta^2 + 20 $ on the number of CNOTs called for within the shearing approach described in this work, as compared with the earlier CNOT upper bound of $ 4 \eta^2 - 4 \eta + 18 $.
The subleading cost increase (in $\eta$) can, however, be avoided by an alternative choice of shears (also in the Appendix), giving a final count of $ 4 \eta^2 - 4 \eta + 20$ --- just two more than the original algorithm.
Given that the original circuit can induce unphysical transitions with probability as high as 0.8 if $ x \, \delta t $ grows to 1.8, and $x\rightarrow \infty$ corresponds to the continuum limit of the Schwinger model, it would seem that two more CNOTs is a small price to pay to achieve the exact hopping propagator.

The operators studied in this work are important for compact U(1) gauge theories, but the shearing method's applicability is in fact broader.
Key features that made the gauge-invariant transitions tractable were the simultaneously diagonalizable constraints (guaranteeing basis states that can each be identified as allowed or unallowed nodes), and the Cartesian structure of the truncated quantum numbers (which unitarily maps into itself under cyclic shears).
Kogut-Susskind-based \cite{Kogut:1974ag} formulations of non-Abelian, SU(2) lattice gauge theory share neither of these:
Gauss's law has noncommuting color components, and angular momentum quantum numbers $\{j,m_L,m_R\}$ on links have a staggered pyramidal structure that does not naturally truncate to a boxlike grid.
The loop-string-hadron formulation of SU(2) lattice gauge theory \cite{Raychowdhury:2019iki}, however, is ideally suited for applications of shears because the same gauge-invariant states are encoded using strictly Abelian flux constraints and naturally Cartesian quantum numbers.
Following the original version of this manuscript, Ref.~\cite{Davoudi:2022xmb} has been published, which demonstrates the usage of shears in such an application to SU(2) lattice gauge theory.

Local gauge symmetries are most celebrated for their key role in defining the Standard Model of particle physics.
The potential for simulating the Standard Model Hamiltonian efficiently, and with all known symmetries intact, will be an important clue as to whether or not our own universe could be a simulation \cite{Beane:2012rz} and further our insight into its intrinsic informational complexity.
Could it be that optimized gauge-invariant simulation protocols ultimately cost fewer resources than optimized gauge-violating protocols? 
Intuition would suggest the opposite---that imposing the extra constraint ought to increase costs---yet empirical findings suggest that the principle of gauge invariance can act as a guide to finding more efficient protocols \cite{Davoudi:2022xmb}.

\begin{acknowledgments}
  J.R.S. thanks Zohreh Davoudi and Alexander Shaw for essential conversations that inspired this work and for valuable input on the manuscript.
  This research was supported by the U.S. Department of Energy (DOE)'s Office of Science Early Career Award DE-SC0020271, and by the US DOE's Office of Science, Office of Advanced Scientific Computing Research, Accelerated Research in Quantum Computing program award DE-SC0020312.
\end{acknowledgments}

\appendix
\section*{Appendix}

\subsection{Alternate hopping-term circuit\label{subsec:altHopping}}
\begin{figure}[t]
  \includegraphics[]{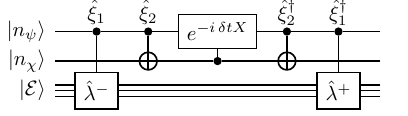}
  \caption{Circuit to simulate a Schwinger model hopping term $\hat{\psi}^\dagger \hat{\chi} \hat{U} + \text{H.c.}$ when the link operator is allowed to ``wrap around'' at the electric cutoffs.
}
  \label{fig:schwingerHop-wrapped}
\end{figure}

The hopping circuit given in Fig. \ref{fig:schwingerHop} implements the hopping propagator without approximation, but it apparently comes at an increased cost relative to the approximate circuit originally developed in Ref.~\cite{Shaw:2020udc}.
To keep the comparison as straightforward as possible, we will allow the link operator to wrap around at the cutoff ($\hat{U} \ket{E_{\text{max}}} = \ket{E_{\text{min}}}$) as was done in Ref.~\cite{Shaw:2020udc}.
In a computational framework in which CNOTs were considered to be the ``expensive'' resource, the cost of a single hopping term propagator was found to be up to $ 4 \eta ( \eta - 1) + 18 $ CNOTs, where $\eta$ is the number of qubits allocated to each electric link register.
(For details, see Sec.~3.3 of Ref.~\cite{Shaw:2020udc}).
Adapting that counting to the cutoff-wrapped hopping propagator in Fig. \ref{fig:schwingerHop-wrapped}, we have:
\begin{enumerate}[(i)]
  \item 2 explicit CNOTs
  \item 2 CNOTs used to implement the controlled $ e^{-i \, x \, \delta t X} $
  \item $ 2 \eta^2 $ CNOTS embedded in the controlled $\hat{\lambda}^-$: $ \eta ( \eta - 1 ) $ for a quantum Fourier transform, plus another $ \eta ( \eta - 1 ) $ for its inverse, plus an additional $ 2 \eta $ involved with adding the $\chi$-register control to $ \eta $ single-qubit rotations applied in Fourier space
  \item $ 2 \eta^2 $ similarly for the controlled $\hat{\lambda}^+$
\end{enumerate}
This gives $ 4 \eta^2 + 4 $ CNOTs in total.
This is to be compared with $ 4 \eta^2 - 4 \eta + 18 $ CNOTs for the gauge-violating algorithm.
Based on the available information, the shear-based circuit in Fig.~\ref{fig:schwingerHop-wrapped} increases the CNOT cost at subleading order in $\eta$, which can be attributed to the addition of a control qubit to each basis incrementer $\hat{\lambda}^{\pm}$.

\begin{figure}
  \includegraphics[width=1.04\columnwidth]{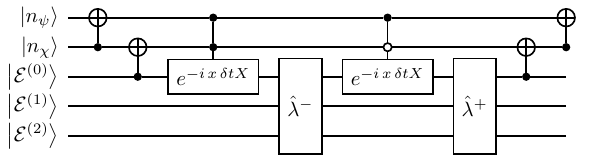}
  \caption{Circuit to simulate a Schwinger model hopping term $x \, \hat{\psi}^\dagger \hat{\chi} \hat{U} + \text{H.c.}$, shown here for a three-qubit electric register.
    The change of basis in this circuit involves two \textsc{CNOT}s, and while the interpretation of it as a shear of the quantum numbers is lost, the main guiding idea still applies: Each \textsc{CNOT} serves to help bring the edges parallel to a single axis.
}
  \label{fig:schwingerHop-Alt}
\end{figure}

If the shears in \eqref{eq:hoppingShear1} and \eqref{eq:hoppingShear2} increase the CNOT cost by calling for a control on the basis incrementers, one would like to know whether or not this is avoidable within the shearing approach.
Note that the purpose of the controlled $\hat{\lambda}^{\pm}$ gates was to induce a shear that removed the component of the graph edges along the $E$ axis.
This suggests that it may be advantageous to instead keep the edges' nonzero $E$ components, taking the $E$-axis as the one the edges are aligned to.
Two alternative cyclic shears that will do this are
\begin{align}
  \hat{\xi}_1' &= \delta_{\hat{n}_\chi , 0} + X_{\psi} \delta_{\hat{n}_\chi , 1} \ , \\
  \hat{\xi}_2' &= \sum_{j=0}^{2^\eta - 1} \delta_{\hat{\mathcal{E}}, j} (X_{\chi})^j \nonumber \\
  &= \delta_{\hat{\mathcal{E}} \text{(mod 2)}, 0} + X_{\chi} \delta_{\hat{\mathcal{E}} \text{(mod 2)}, 1} \ ,
\end{align}
where $\xi_1'$ removes the edges' components along the $n_{\psi}$ axis, and then $\xi_2'$ removes the components along the $n_{\chi}$ direction.
One then has
\begin{align}
  &\xi_2' \xi_1' ( T_{\text{hop}} + T_{\text{hop}}^\dagger ) \xi_1^{\prime \dagger} \xi_2^{\prime \dagger} \nonumber \\
  &= x \, \delta_{\hat{n}_\psi, 1 } \big[ \delta_{\hat{n}_\chi , 0 } \hat{\lambda}^+ X^{\text{(lsb)}} ( 1 - (\delta_{\hat{\mathcal{E}},-1} - \delta_{\hat{\mathcal{E}},-2}) ) \hat{\lambda}^- \nonumber \\
  & \qquad \qquad \ + \delta_{\hat{n}_\chi , 1 } X^{\text{(lsb)}} \big] .
\end{align}
In the cutoff-wrapped version, this term is simplified as
\begin{align}
  &( T_{\text{hop}} + T_{\text{hop}}^\dagger ) \nonumber \\
  &\overset{\text{wrapped $U$}}{\xrightarrow{\hspace*{24pt}}}
  \overset{\xi_2' \xi_1'}{\xrightarrow{\hspace*{16pt}}}
  \ x \, \delta_{\hat{n}_\psi, 1 } \big[ \delta_{\hat{n}_\chi , 0 } \hat{\lambda}^+ X^{\text{(lsb)}} \hat{\lambda}^- + \delta_{\hat{n}_\chi , 1 } X^{\text{(lsb)}} \big]  .
\end{align}
\begin{figure*}
  \hspace{0cm}\includegraphics[]{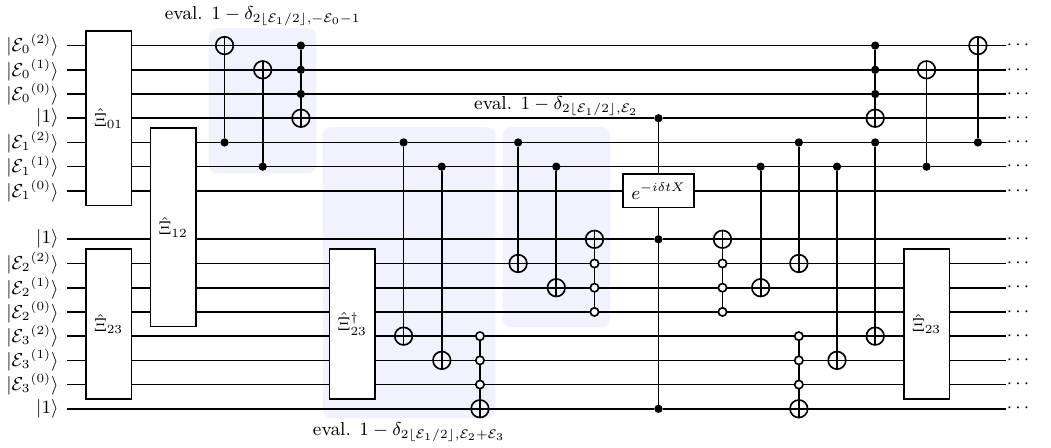}
  \hspace{1cm}\includegraphics[]{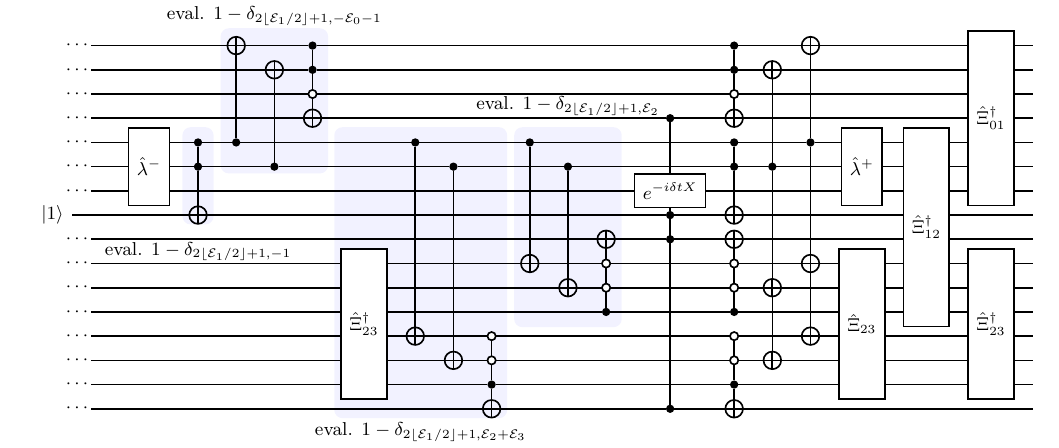}
  \caption{\label{fig:fullPlaqCircuit}
    Digital quantum circuit for gauge-invariant, first-order Trotter-Suzuki approximation to $ \exp [ -i \, \delta t ( \hat{U}_0 \hat{U}_1 \hat{U}_2^\dagger \hat{U}_3^\dagger + \hat{U}_0^\dagger \hat{U}_1^\dagger \hat{U}_2 \hat{U}_3 ) ] $ for a true, four-link plaquette (shown for $\eta=3$).
    Filled and empty circles are used for conditional operations activated by the $\ket{1}$ or $\ket{0}$ states, respectively, of the control qubit(s) \cite{Nielsen:2012yss}.}
\end{figure*}
As desired, this term does not require controls on the incrementers.
An exact circuit for it is shown in Fig. \ref{fig:schwingerHop-Alt}.
The cost goes as follows:
\begin{enumerate}[(i)]
  \item 4 explicit CNOTs
  \item 8 CNOTs for the doubly controlled $e^{-i \, x \, \delta t X}$ gate
  \item 8 CNOTs for the doubly controlled $e^{+i \, x \, \delta t X}$ gate
  \item $2 \eta (\eta - 1)$ CNOTs for the $\hat{\lambda}^-$ as done in Ref.~\cite{Shaw:2020udc}
  \item $2 \eta (\eta - 1)$ CNOTs for the $\hat{\lambda}^+$
\end{enumerate}
This gives $ 4 \eta^2 - 4 \eta + 20 $ CNOTs in total, a constant increase of two relative to the old circuit.
Additionally, the old circuit is accidentally exact for $\eta<3$;
when we reach $\eta=3$, 42 CNOTs are called for by the old circuit, which is much larger than the additional two needed to make the exact, gauge-conserving circuit.

Since it apparently costs more CNOTs to implement the new hopping term circuits, it is appropriate to ask what is gained in terms of the simulated physics.
One way to answer this as follows:
Suppose $\ket{\psi}$ is some initial state in a definite charge sector, $\mathcal{H}_{\text{phys}}$ is the space of states that can mix with $\ket{\psi}$ under evolution by the exact hopping propagator, and $\mathcal{H}_{\text{unphys}}$ is the orthogonal complement of $\mathcal{H}_{\text{phys}}$.
Given that the approximate propagator circuit causes leakage into $\mathcal{H}_{\text{unphys}}$ for $\eta \geq 3$, what is the worst-case probability (as a function of Trotter step time) of finding the evolved state in $\mathcal{H}_{\text{unphys}}$?
A numerical check reveals that this probability can be as high as $0.8$ if $x \, \delta t$ grows as large as 1.8.
In the continuum limit of asymptotically free theories like QCD, $x \rightarrow \infty$, in which case $\delta t$ must approach zero to keep the unphysical transitions suppressed.

\subsection{Four-link (full plaquette) circuit}

The circuit in Fig.~\ref{fig:fullPlaqCircuit} is a generalization of Fig.~\ref{fig:toyPlaqTrott} to the ordinary number of gauge links and represents the Trotterized propagator associated to a plaquette operator in a compact U(1) lattice gauge theory (discussed in the main text).
In this case the plaquette operator is off-diagonal on four electric quantum numbers $E_i$ ($i=0,1,2,3$).
The gauge-invariant moves induced by $ U_{\square} + U_{\square}^{\dagger} $ can be thought of as edges running in the $(1,1,-1,-1)$ direction of a four-dimensional grid.
Shears in the 01- and 23-planes at the beginning of the circuit realign the edges to the $(0,1,-1,0)$ direction.
From that point forward, the circuit is essentially following the same steps as were taken for the two-link toy plaquette:
propagating by the analogue of $\hat{h}_e$ and then $\hat{h}_o$, with appropriate arithmetic and controls throughout serving to effect the electric cutoffs in a four-dimensional space.

\bibliography{shearing}

\newcommand{\noopsort}[1]{}
\begin{thebibliography}{142}%
\makeatletter
\providecommand \@ifxundefined [1]{%
 \@ifx{#1\undefined}
}%
\providecommand \@ifnum [1]{%
 \ifnum #1\expandafter \@firstoftwo
 \else \expandafter \@secondoftwo
 \fi
}%
\providecommand \@ifx [1]{%
 \ifx #1\expandafter \@firstoftwo
 \else \expandafter \@secondoftwo
 \fi
}%
\providecommand \natexlab [1]{#1}%
\providecommand \enquote  [1]{``#1''}%
\providecommand \bibnamefont  [1]{#1}%
\providecommand \bibfnamefont [1]{#1}%
\providecommand \citenamefont [1]{#1}%
\providecommand \href@noop [0]{\@secondoftwo}%
\providecommand \href [0]{\begingroup \@sanitize@url \@href}%
\providecommand \@href[1]{\@@startlink{#1}\@@href}%
\providecommand \@@href[1]{\endgroup#1\@@endlink}%
\providecommand \@sanitize@url [0]{\catcode `\\12\catcode `\$12\catcode
  `\&12\catcode `\#12\catcode `\^12\catcode `\_12\catcode `\%12\relax}%
\providecommand \@@startlink[1]{}%
\providecommand \@@endlink[0]{}%
\providecommand \url  [0]{\begingroup\@sanitize@url \@url }%
\providecommand \@url [1]{\endgroup\@href {#1}{\urlprefix }}%
\providecommand \urlprefix  [0]{URL }%
\providecommand \Eprint [0]{\href }%
\providecommand \doibase [0]{https://doi.org/}%
\providecommand \selectlanguage [0]{\@gobble}%
\providecommand \bibinfo  [0]{\@secondoftwo}%
\providecommand \bibfield  [0]{\@secondoftwo}%
\providecommand \translation [1]{[#1]}%
\providecommand \BibitemOpen [0]{}%
\providecommand \bibitemStop [0]{}%
\providecommand \bibitemNoStop [0]{.\EOS\space}%
\providecommand \EOS [0]{\spacefactor3000\relax}%
\providecommand \BibitemShut  [1]{\csname bibitem#1\endcsname}%
\let\auto@bib@innerbib\@empty
\bibitem [{\citenamefont {Luscher}(1986)}]{Luscher:1986pf}%
  \BibitemOpen
  \bibfield  {author} {\bibinfo {author} {\bibfnamefont {M.}~\bibnamefont
  {Luscher}},\ }\bibfield  {title} {\bibinfo {title} {{Volume Dependence of the
  Energy Spectrum in Massive Quantum Field Theories. 2. Scattering States}},\
  }\href {https://doi.org/10.1007/BF01211097} {\bibfield  {journal} {\bibinfo
  {journal} {Commun. Math. Phys.}\ }\textbf {\bibinfo {volume} {105}},\
  \bibinfo {pages} {153} (\bibinfo {year} {1986})}\BibitemShut {NoStop}%
\bibitem [{\citenamefont {Luscher}(1991)}]{Luscher:1990ux}%
  \BibitemOpen
  \bibfield  {author} {\bibinfo {author} {\bibfnamefont {M.}~\bibnamefont
  {Luscher}},\ }\bibfield  {title} {\bibinfo {title} {{Two particle states on a
  torus and their relation to the scattering matrix}},\ }\href
  {https://doi.org/10.1016/0550-3213(91)90366-6} {\bibfield  {journal}
  {\bibinfo  {journal} {Nucl. Phys. B}\ }\textbf {\bibinfo {volume} {354}},\
  \bibinfo {pages} {531} (\bibinfo {year} {1991})}\BibitemShut {NoStop}%
\bibitem [{\citenamefont {Martinez}\ \emph {et~al.}(2016)\citenamefont
  {Martinez} \emph {et~al.}}]{Martinez:2016yna}%
  \BibitemOpen
  \bibfield  {author} {\bibinfo {author} {\bibfnamefont {E.~A.}\ \bibnamefont
  {Martinez}} \emph {et~al.},\ }\bibfield  {title} {\bibinfo {title}
  {{Real-time dynamics of lattice gauge theories with a few-qubit quantum
  computer}},\ }\href {https://doi.org/10.1038/nature18318} {\bibfield
  {journal} {\bibinfo  {journal} {Nature}\ }\textbf {\bibinfo {volume} {534}},\
  \bibinfo {pages} {516} (\bibinfo {year} {2016})},\ \Eprint
  {https://arxiv.org/abs/1605.04570} {arXiv:1605.04570 [quant-ph]} \BibitemShut
  {NoStop}%
\bibitem [{\citenamefont {Klco}\ \emph {et~al.}(2018)\citenamefont {Klco},
  \citenamefont {Dumitrescu}, \citenamefont {McCaskey}, \citenamefont {Morris},
  \citenamefont {Pooser}, \citenamefont {Sanz}, \citenamefont {Solano},
  \citenamefont {Lougovski},\ and\ \citenamefont {Savage}}]{Klco:2018kyo}%
  \BibitemOpen
  \bibfield  {author} {\bibinfo {author} {\bibfnamefont {N.}~\bibnamefont
  {Klco}}, \bibinfo {author} {\bibfnamefont {E.~F.}\ \bibnamefont
  {Dumitrescu}}, \bibinfo {author} {\bibfnamefont {A.~J.}\ \bibnamefont
  {McCaskey}}, \bibinfo {author} {\bibfnamefont {T.~D.}\ \bibnamefont
  {Morris}}, \bibinfo {author} {\bibfnamefont {R.~C.}\ \bibnamefont {Pooser}},
  \bibinfo {author} {\bibfnamefont {M.}~\bibnamefont {Sanz}}, \bibinfo {author}
  {\bibfnamefont {E.}~\bibnamefont {Solano}}, \bibinfo {author} {\bibfnamefont
  {P.}~\bibnamefont {Lougovski}},\ and\ \bibinfo {author} {\bibfnamefont
  {M.~J.}\ \bibnamefont {Savage}},\ }\bibfield  {title} {\bibinfo {title}
  {{Quantum-classical computation of Schwinger model dynamics using quantum
  computers}},\ }\href {https://doi.org/10.1103/PhysRevA.98.032331} {\bibfield
  {journal} {\bibinfo  {journal} {Phys. Rev. A}\ }\textbf {\bibinfo {volume}
  {98}},\ \bibinfo {pages} {032331} (\bibinfo {year} {2018})},\ \Eprint
  {https://arxiv.org/abs/1803.03326} {arXiv:1803.03326 [quant-ph]} \BibitemShut
  {NoStop}%
\bibitem [{\citenamefont {Kokail}\ \emph {et~al.}(2019)\citenamefont {Kokail}
  \emph {et~al.}}]{Kokail:2018eiw}%
  \BibitemOpen
  \bibfield  {author} {\bibinfo {author} {\bibfnamefont {C.}~\bibnamefont
  {Kokail}} \emph {et~al.},\ }\bibfield  {title} {\bibinfo {title}
  {{Self-verifying variational quantum simulation of lattice models}},\ }\href
  {https://doi.org/10.1038/s41586-019-1177-4} {\bibfield  {journal} {\bibinfo
  {journal} {Nature}\ }\textbf {\bibinfo {volume} {569}},\ \bibinfo {pages}
  {355} (\bibinfo {year} {2019})},\ \Eprint {https://arxiv.org/abs/1810.03421}
  {arXiv:1810.03421 [quant-ph]} \BibitemShut {NoStop}%
\bibitem [{\citenamefont {Schweizer}\ \emph {et~al.}(2019)\citenamefont
  {Schweizer}, \citenamefont {Grusdt}, \citenamefont {Berngruber},
  \citenamefont {Barbiero}, \citenamefont {Demler}, \citenamefont {Goldman},
  \citenamefont {Bloch},\ and\ \citenamefont
  {Aidelsburger}}]{Schweizer:2019lwx}%
  \BibitemOpen
  \bibfield  {author} {\bibinfo {author} {\bibfnamefont {C.}~\bibnamefont
  {Schweizer}}, \bibinfo {author} {\bibfnamefont {F.}~\bibnamefont {Grusdt}},
  \bibinfo {author} {\bibfnamefont {M.}~\bibnamefont {Berngruber}}, \bibinfo
  {author} {\bibfnamefont {L.}~\bibnamefont {Barbiero}}, \bibinfo {author}
  {\bibfnamefont {E.}~\bibnamefont {Demler}}, \bibinfo {author} {\bibfnamefont
  {N.}~\bibnamefont {Goldman}}, \bibinfo {author} {\bibfnamefont
  {I.}~\bibnamefont {Bloch}},\ and\ \bibinfo {author} {\bibfnamefont
  {M.}~\bibnamefont {Aidelsburger}},\ }\bibfield  {title} {\bibinfo {title}
  {{Floquet approach to $\mathbb{Z}_{2}$ lattice gauge theories with ultracold
  atoms in optical lattices}},\ }\href
  {https://doi.org/10.1038/s41567-019-0649-7} {\bibfield  {journal} {\bibinfo
  {journal} {Nature Phys.}\ }\textbf {\bibinfo {volume} {15}},\ \bibinfo
  {pages} {1168} (\bibinfo {year} {2019})},\ \Eprint
  {https://arxiv.org/abs/1901.07103} {arXiv:1901.07103 [cond-mat.quant-gas]}
  \BibitemShut {NoStop}%
\bibitem [{\citenamefont {Klco}\ \emph {et~al.}(2020)\citenamefont {Klco},
  \citenamefont {Stryker},\ and\ \citenamefont {Savage}}]{Klco:2019evd}%
  \BibitemOpen
  \bibfield  {author} {\bibinfo {author} {\bibfnamefont {N.}~\bibnamefont
  {Klco}}, \bibinfo {author} {\bibfnamefont {J.~R.}\ \bibnamefont {Stryker}},\
  and\ \bibinfo {author} {\bibfnamefont {M.~J.}\ \bibnamefont {Savage}},\
  }\bibfield  {title} {\bibinfo {title} {{SU(2) non-Abelian gauge field theory
  in one dimension on digital quantum computers}},\ }\href
  {https://doi.org/10.1103/PhysRevD.101.074512} {\bibfield  {journal} {\bibinfo
   {journal} {Phys. Rev. D}\ }\textbf {\bibinfo {volume} {101}},\ \bibinfo
  {pages} {074512} (\bibinfo {year} {2020})},\ \Eprint
  {https://arxiv.org/abs/1908.06935} {arXiv:1908.06935 [quant-ph]} \BibitemShut
  {NoStop}%
\bibitem [{\citenamefont {Yang}\ \emph {et~al.}(2020)\citenamefont {Yang},
  \citenamefont {Sun}, \citenamefont {Ott}, \citenamefont {Wang}, \citenamefont
  {Zache}, \citenamefont {Halimeh}, \citenamefont {Yuan}, \citenamefont
  {Hauke},\ and\ \citenamefont {Pan}}]{Yang:2020yer}%
  \BibitemOpen
  \bibfield  {author} {\bibinfo {author} {\bibfnamefont {B.}~\bibnamefont
  {Yang}}, \bibinfo {author} {\bibfnamefont {H.}~\bibnamefont {Sun}}, \bibinfo
  {author} {\bibfnamefont {R.}~\bibnamefont {Ott}}, \bibinfo {author}
  {\bibfnamefont {H.-Y.}\ \bibnamefont {Wang}}, \bibinfo {author}
  {\bibfnamefont {T.~V.}\ \bibnamefont {Zache}}, \bibinfo {author}
  {\bibfnamefont {J.~C.}\ \bibnamefont {Halimeh}}, \bibinfo {author}
  {\bibfnamefont {Z.-S.}\ \bibnamefont {Yuan}}, \bibinfo {author}
  {\bibfnamefont {P.}~\bibnamefont {Hauke}},\ and\ \bibinfo {author}
  {\bibfnamefont {J.-W.}\ \bibnamefont {Pan}},\ }\bibfield  {title} {\bibinfo
  {title} {{Observation of gauge invariance in a 71-site
  Bose\textendash{}Hubbard quantum simulator}},\ }\href
  {https://doi.org/10.1038/s41586-020-2910-8} {\bibfield  {journal} {\bibinfo
  {journal} {Nature}\ }\textbf {\bibinfo {volume} {587}},\ \bibinfo {pages}
  {392} (\bibinfo {year} {2020})},\ \Eprint {https://arxiv.org/abs/2003.08945}
  {arXiv:2003.08945 [cond-mat.quant-gas]} \BibitemShut {NoStop}%
\bibitem [{\citenamefont {Kreshchuk}\ \emph {et~al.}(2021)\citenamefont
  {Kreshchuk}, \citenamefont {Jia}, \citenamefont {Kirby}, \citenamefont
  {Goldstein}, \citenamefont {Vary},\ and\ \citenamefont
  {Love}}]{Kreshchuk:2020kcz}%
  \BibitemOpen
  \bibfield  {author} {\bibinfo {author} {\bibfnamefont {M.}~\bibnamefont
  {Kreshchuk}}, \bibinfo {author} {\bibfnamefont {S.}~\bibnamefont {Jia}},
  \bibinfo {author} {\bibfnamefont {W.~M.}\ \bibnamefont {Kirby}}, \bibinfo
  {author} {\bibfnamefont {G.}~\bibnamefont {Goldstein}}, \bibinfo {author}
  {\bibfnamefont {J.~P.}\ \bibnamefont {Vary}},\ and\ \bibinfo {author}
  {\bibfnamefont {P.~J.}\ \bibnamefont {Love}},\ }\bibfield  {title} {\bibinfo
  {title} {{Light-Front Field Theory on Current Quantum Computers}},\ }\href
  {https://doi.org/10.3390/e23050597} {\bibfield  {journal} {\bibinfo
  {journal} {Entropy}\ }\textbf {\bibinfo {volume} {23}},\ \bibinfo {pages}
  {597} (\bibinfo {year} {2021})},\ \Eprint {https://arxiv.org/abs/2009.07885}
  {arXiv:2009.07885 [quant-ph]} \BibitemShut {NoStop}%
\bibitem [{\citenamefont {Klco}\ and\ \citenamefont
  {Savage}(2020{\natexlab{a}})}]{Klco:2019xro}%
  \BibitemOpen
  \bibfield  {author} {\bibinfo {author} {\bibfnamefont {N.}~\bibnamefont
  {Klco}}\ and\ \bibinfo {author} {\bibfnamefont {M.~J.}\ \bibnamefont
  {Savage}},\ }\bibfield  {title} {\bibinfo {title} {{Minimally entangled state
  preparation of localized wave functions on quantum computers}},\ }\href
  {https://doi.org/10.1103/PhysRevA.102.012612} {\bibfield  {journal} {\bibinfo
   {journal} {Phys. Rev. A}\ }\textbf {\bibinfo {volume} {102}},\ \bibinfo
  {pages} {012612} (\bibinfo {year} {2020}{\natexlab{a}})},\ \Eprint
  {https://arxiv.org/abs/1904.10440} {arXiv:1904.10440 [quant-ph]} \BibitemShut
  {NoStop}%
\bibitem [{\citenamefont {Gustafson}\ \emph {et~al.}(2021)\citenamefont
  {Gustafson}, \citenamefont {Zhu}, \citenamefont {Dreher}, \citenamefont
  {Linke},\ and\ \citenamefont {Meurice}}]{Gustafson:2021imb}%
  \BibitemOpen
  \bibfield  {author} {\bibinfo {author} {\bibfnamefont {E.}~\bibnamefont
  {Gustafson}}, \bibinfo {author} {\bibfnamefont {Y.}~\bibnamefont {Zhu}},
  \bibinfo {author} {\bibfnamefont {P.}~\bibnamefont {Dreher}}, \bibinfo
  {author} {\bibfnamefont {N.~M.}\ \bibnamefont {Linke}},\ and\ \bibinfo
  {author} {\bibfnamefont {Y.}~\bibnamefont {Meurice}},\ }\bibfield  {title}
  {\bibinfo {title} {{Real-time quantum calculations of phase shifts using wave
  packet time delays}},\ }\href {https://doi.org/10.1103/PhysRevD.104.054507}
  {\bibfield  {journal} {\bibinfo  {journal} {Phys. Rev. D}\ }\textbf {\bibinfo
  {volume} {104}},\ \bibinfo {pages} {054507} (\bibinfo {year} {2021})},\
  \Eprint {https://arxiv.org/abs/2103.06848} {arXiv:2103.06848 [hep-lat]}
  \BibitemShut {NoStop}%
\bibitem [{\citenamefont {Bauer}\ \emph
  {et~al.}(2021{\natexlab{a}})\citenamefont {Bauer}, \citenamefont {Freytsis},\
  and\ \citenamefont {Nachman}}]{Bauer:2021gup}%
  \BibitemOpen
  \bibfield  {author} {\bibinfo {author} {\bibfnamefont {C.~W.}\ \bibnamefont
  {Bauer}}, \bibinfo {author} {\bibfnamefont {M.}~\bibnamefont {Freytsis}},\
  and\ \bibinfo {author} {\bibfnamefont {B.}~\bibnamefont {Nachman}},\
  }\bibfield  {title} {\bibinfo {title} {{Simulating Collider Physics on
  Quantum Computers Using Effective Field Theories}},\ }\href
  {https://doi.org/10.1103/PhysRevLett.127.212001} {\bibfield  {journal}
  {\bibinfo  {journal} {Phys. Rev. Lett.}\ }\textbf {\bibinfo {volume} {127}},\
  \bibinfo {pages} {212001} (\bibinfo {year} {2021}{\natexlab{a}})},\ \Eprint
  {https://arxiv.org/abs/2102.05044} {arXiv:2102.05044 [hep-ph]} \BibitemShut
  {NoStop}%
\bibitem [{\citenamefont {Atas}\ \emph {et~al.}(2021)\citenamefont {Atas},
  \citenamefont {Zhang}, \citenamefont {Lewis}, \citenamefont {Jahanpour},
  \citenamefont {Haase},\ and\ \citenamefont {Muschik}}]{Atas:2021ext}%
  \BibitemOpen
  \bibfield  {author} {\bibinfo {author} {\bibfnamefont {Y.~Y.}\ \bibnamefont
  {Atas}}, \bibinfo {author} {\bibfnamefont {J.}~\bibnamefont {Zhang}},
  \bibinfo {author} {\bibfnamefont {R.}~\bibnamefont {Lewis}}, \bibinfo
  {author} {\bibfnamefont {A.}~\bibnamefont {Jahanpour}}, \bibinfo {author}
  {\bibfnamefont {J.~F.}\ \bibnamefont {Haase}},\ and\ \bibinfo {author}
  {\bibfnamefont {C.~A.}\ \bibnamefont {Muschik}},\ }\bibfield  {title}
  {\bibinfo {title} {{SU(2) hadrons on a quantum computer via a variational
  approach}},\ }\href {https://doi.org/10.1038/s41467-021-26825-4} {\bibfield
  {journal} {\bibinfo  {journal} {Nature Commun.}\ }\textbf {\bibinfo {volume}
  {12}},\ \bibinfo {pages} {6499} (\bibinfo {year} {2021})},\ \Eprint
  {https://arxiv.org/abs/2102.08920} {arXiv:2102.08920 [quant-ph]} \BibitemShut
  {NoStop}%
\bibitem [{\citenamefont {A~Rahman}\ \emph {et~al.}(2021)\citenamefont
  {A~Rahman}, \citenamefont {Lewis}, \citenamefont {Mendicelli},\ and\
  \citenamefont {Powell}}]{ARahman:2021ktn}%
  \BibitemOpen
  \bibfield  {author} {\bibinfo {author} {\bibfnamefont {S.}~\bibnamefont
  {A~Rahman}}, \bibinfo {author} {\bibfnamefont {R.}~\bibnamefont {Lewis}},
  \bibinfo {author} {\bibfnamefont {E.}~\bibnamefont {Mendicelli}},\ and\
  \bibinfo {author} {\bibfnamefont {S.}~\bibnamefont {Powell}},\ }\bibfield
  {title} {\bibinfo {title} {{SU(2) lattice gauge theory on a quantum
  annealer}},\ }\href {https://doi.org/10.1103/PhysRevD.104.034501} {\bibfield
  {journal} {\bibinfo  {journal} {Phys. Rev. D}\ }\textbf {\bibinfo {volume}
  {104}},\ \bibinfo {pages} {034501} (\bibinfo {year} {2021})},\ \Eprint
  {https://arxiv.org/abs/2103.08661} {arXiv:2103.08661 [hep-lat]} \BibitemShut
  {NoStop}%
\bibitem [{\citenamefont {Ciavarella}\ \emph {et~al.}(2021)\citenamefont
  {Ciavarella}, \citenamefont {Klco},\ and\ \citenamefont
  {Savage}}]{Ciavarella:2021nmj}%
  \BibitemOpen
  \bibfield  {author} {\bibinfo {author} {\bibfnamefont {A.}~\bibnamefont
  {Ciavarella}}, \bibinfo {author} {\bibfnamefont {N.}~\bibnamefont {Klco}},\
  and\ \bibinfo {author} {\bibfnamefont {M.~J.}\ \bibnamefont {Savage}},\
  }\bibfield  {title} {\bibinfo {title} {{Trailhead for quantum simulation of
  SU(3) Yang-Mills lattice gauge theory in the local multiplet basis}},\ }\href
  {https://doi.org/10.1103/PhysRevD.103.094501} {\bibfield  {journal} {\bibinfo
   {journal} {Phys. Rev. D}\ }\textbf {\bibinfo {volume} {103}},\ \bibinfo
  {pages} {094501} (\bibinfo {year} {2021})},\ \Eprint
  {https://arxiv.org/abs/2101.10227} {arXiv:2101.10227 [quant-ph]} \BibitemShut
  {NoStop}%
\bibitem [{\citenamefont {Huffman}\ \emph {et~al.}(2022)\citenamefont
  {Huffman}, \citenamefont {Garc\'\i{}a~Vera},\ and\ \citenamefont
  {Banerjee}}]{Huffman:2021gsi}%
  \BibitemOpen
  \bibfield  {author} {\bibinfo {author} {\bibfnamefont {E.}~\bibnamefont
  {Huffman}}, \bibinfo {author} {\bibfnamefont {M.}~\bibnamefont
  {Garc\'\i{}a~Vera}},\ and\ \bibinfo {author} {\bibfnamefont {D.}~\bibnamefont
  {Banerjee}},\ }\bibfield  {title} {\bibinfo {title} {{Toward the real-time
  evolution of gauge-invariant $\mathbb Z_2$ and $U(1)$ quantum link models on
  noisy intermediate-scale quantum hardware with error mitigation}},\ }\href
  {https://doi.org/10.1103/PhysRevD.106.094502} {\bibfield  {journal} {\bibinfo
   {journal} {Phys. Rev. D}\ }\textbf {\bibinfo {volume} {106}},\ \bibinfo
  {pages} {094502} (\bibinfo {year} {2022})},\ \Eprint
  {https://arxiv.org/abs/2109.15065} {arXiv:2109.15065 [quant-ph]} \BibitemShut
  {NoStop}%
\bibitem [{\citenamefont {Xu}\ and\ \citenamefont {Xue}(2022)}]{Xu:2021tey}%
  \BibitemOpen
  \bibfield  {author} {\bibinfo {author} {\bibfnamefont {B.}~\bibnamefont
  {Xu}}\ and\ \bibinfo {author} {\bibfnamefont {W.}~\bibnamefont {Xue}},\
  }\bibfield  {title} {\bibinfo {title} {{(3+1)-dimensional Schwinger pair
  production with quantum computers}},\ }\href
  {https://doi.org/10.1103/PhysRevD.106.116007} {\bibfield  {journal} {\bibinfo
   {journal} {Phys. Rev. D}\ }\textbf {\bibinfo {volume} {106}},\ \bibinfo
  {pages} {116007} (\bibinfo {year} {2022})},\ \Eprint
  {https://arxiv.org/abs/2112.06863} {arXiv:2112.06863 [quant-ph]} \BibitemShut
  {NoStop}%
\bibitem [{\citenamefont {Ciavarella}\ and\ \citenamefont
  {Chernyshev}(2022)}]{Ciavarella:2021lel}%
  \BibitemOpen
  \bibfield  {author} {\bibinfo {author} {\bibfnamefont {A.~N.}\ \bibnamefont
  {Ciavarella}}\ and\ \bibinfo {author} {\bibfnamefont {I.~A.}\ \bibnamefont
  {Chernyshev}},\ }\bibfield  {title} {\bibinfo {title} {{Preparation of the
  SU(3) lattice Yang-Mills vacuum with variational quantum methods}},\ }\href
  {https://doi.org/10.1103/PhysRevD.105.074504} {\bibfield  {journal} {\bibinfo
   {journal} {Phys. Rev. D}\ }\textbf {\bibinfo {volume} {105}},\ \bibinfo
  {pages} {074504} (\bibinfo {year} {2022})},\ \Eprint
  {https://arxiv.org/abs/2112.09083} {arXiv:2112.09083 [quant-ph]} \BibitemShut
  {NoStop}%
\bibitem [{\citenamefont {Nguyen}\ \emph {et~al.}(2022)\citenamefont {Nguyen},
  \citenamefont {Tran}, \citenamefont {Zhu}, \citenamefont {Green},
  \citenamefont {Alderete}, \citenamefont {Davoudi},\ and\ \citenamefont
  {Linke}}]{Nguyen:2021hyk}%
  \BibitemOpen
  \bibfield  {author} {\bibinfo {author} {\bibfnamefont {N.~H.}\ \bibnamefont
  {Nguyen}}, \bibinfo {author} {\bibfnamefont {M.~C.}\ \bibnamefont {Tran}},
  \bibinfo {author} {\bibfnamefont {Y.}~\bibnamefont {Zhu}}, \bibinfo {author}
  {\bibfnamefont {A.~M.}\ \bibnamefont {Green}}, \bibinfo {author}
  {\bibfnamefont {C.~H.}\ \bibnamefont {Alderete}}, \bibinfo {author}
  {\bibfnamefont {Z.}~\bibnamefont {Davoudi}},\ and\ \bibinfo {author}
  {\bibfnamefont {N.~M.}\ \bibnamefont {Linke}},\ }\bibfield  {title} {\bibinfo
  {title} {{Digital Quantum Simulation of the Schwinger Model and Symmetry
  Protection with Trapped Ions}},\ }\href
  {https://doi.org/10.1103/PRXQuantum.3.020324} {\bibfield  {journal} {\bibinfo
   {journal} {PRX Quantum}\ }\textbf {\bibinfo {volume} {3}},\ \bibinfo {pages}
  {020324} (\bibinfo {year} {2022})},\ \Eprint
  {https://arxiv.org/abs/2112.14262} {arXiv:2112.14262 [quant-ph]} \BibitemShut
  {NoStop}%
\bibitem [{\citenamefont {Illa}\ and\ \citenamefont
  {Savage}(2022)}]{Illa:2022jqb}%
  \BibitemOpen
  \bibfield  {author} {\bibinfo {author} {\bibfnamefont {M.}~\bibnamefont
  {Illa}}\ and\ \bibinfo {author} {\bibfnamefont {M.~J.}\ \bibnamefont
  {Savage}},\ }\bibfield  {title} {\bibinfo {title} {{Basic elements for
  simulations of standard-model physics with quantum annealers: Multigrid and
  clock states}},\ }\href {https://doi.org/10.1103/PhysRevA.106.052605}
  {\bibfield  {journal} {\bibinfo  {journal} {Phys. Rev. A}\ }\textbf {\bibinfo
  {volume} {106}},\ \bibinfo {pages} {052605} (\bibinfo {year} {2022})},\
  \Eprint {https://arxiv.org/abs/2202.12340} {arXiv:2202.12340 [quant-ph]}
  \BibitemShut {NoStop}%
\bibitem [{\citenamefont {Mildenberger}\ \emph {et~al.}(2025)\citenamefont
  {Mildenberger}, \citenamefont {Mruczkiewicz}, \citenamefont {Halimeh},
  \citenamefont {Jiang},\ and\ \citenamefont {Hauke}}]{Mildenberger:2022jqr}%
  \BibitemOpen
  \bibfield  {author} {\bibinfo {author} {\bibfnamefont {J.}~\bibnamefont
  {Mildenberger}}, \bibinfo {author} {\bibfnamefont {W.}~\bibnamefont
  {Mruczkiewicz}}, \bibinfo {author} {\bibfnamefont {J.~C.}\ \bibnamefont
  {Halimeh}}, \bibinfo {author} {\bibfnamefont {Z.}~\bibnamefont {Jiang}},\
  and\ \bibinfo {author} {\bibfnamefont {P.}~\bibnamefont {Hauke}},\ }\bibfield
   {title} {\bibinfo {title} {{Confinement in a ${{\mathbb{Z}}}_{2}$ lattice
  gauge theory on a quantum computer}},\ }\href
  {https://doi.org/10.1038/s41567-024-02723-6} {\bibfield  {journal} {\bibinfo
  {journal} {Nature Phys.}\ }\textbf {\bibinfo {volume} {21}},\ \bibinfo
  {pages} {312} (\bibinfo {year} {2025})},\ \Eprint
  {https://arxiv.org/abs/2203.08905} {arXiv:2203.08905 [quant-ph]} \BibitemShut
  {NoStop}%
\bibitem [{\citenamefont {Pardo}\ \emph {et~al.}(2023)\citenamefont {Pardo},
  \citenamefont {Greenberg}, \citenamefont {Fortinsky}, \citenamefont {Katz},\
  and\ \citenamefont {Zohar}}]{Pardo:2022hrp}%
  \BibitemOpen
  \bibfield  {author} {\bibinfo {author} {\bibfnamefont {G.}~\bibnamefont
  {Pardo}}, \bibinfo {author} {\bibfnamefont {T.}~\bibnamefont {Greenberg}},
  \bibinfo {author} {\bibfnamefont {A.}~\bibnamefont {Fortinsky}}, \bibinfo
  {author} {\bibfnamefont {N.}~\bibnamefont {Katz}},\ and\ \bibinfo {author}
  {\bibfnamefont {E.}~\bibnamefont {Zohar}},\ }\bibfield  {title} {\bibinfo
  {title} {{Resource-efficient quantum simulation of lattice gauge theories in
  arbitrary dimensions: Solving for Gauss's law and fermion elimination}},\
  }\href {https://doi.org/10.1103/PhysRevResearch.5.023077} {\bibfield
  {journal} {\bibinfo  {journal} {Phys. Rev. Res.}\ }\textbf {\bibinfo {volume}
  {5}},\ \bibinfo {pages} {023077} (\bibinfo {year} {2023})},\ \Eprint
  {https://arxiv.org/abs/2206.00685} {arXiv:2206.00685 [quant-ph]} \BibitemShut
  {NoStop}%
\bibitem [{\citenamefont {Farrell}\ \emph
  {et~al.}(2023{\natexlab{a}})\citenamefont {Farrell}, \citenamefont
  {Chernyshev}, \citenamefont {Powell}, \citenamefont {Zemlevskiy},
  \citenamefont {Illa},\ and\ \citenamefont {Savage}}]{Farrell:2022wyt}%
  \BibitemOpen
  \bibfield  {author} {\bibinfo {author} {\bibfnamefont {R.~C.}\ \bibnamefont
  {Farrell}}, \bibinfo {author} {\bibfnamefont {I.~A.}\ \bibnamefont
  {Chernyshev}}, \bibinfo {author} {\bibfnamefont {S.~J.~M.}\ \bibnamefont
  {Powell}}, \bibinfo {author} {\bibfnamefont {N.~A.}\ \bibnamefont
  {Zemlevskiy}}, \bibinfo {author} {\bibfnamefont {M.}~\bibnamefont {Illa}},\
  and\ \bibinfo {author} {\bibfnamefont {M.~J.}\ \bibnamefont {Savage}},\
  }\bibfield  {title} {\bibinfo {title} {{Preparations for quantum simulations
  of quantum chromodynamics in 1+1 dimensions. I. Axial gauge}},\ }\href
  {https://doi.org/10.1103/PhysRevD.107.054512} {\bibfield  {journal} {\bibinfo
   {journal} {Phys. Rev. D}\ }\textbf {\bibinfo {volume} {107}},\ \bibinfo
  {pages} {054512} (\bibinfo {year} {2023}{\natexlab{a}})},\ \Eprint
  {https://arxiv.org/abs/2207.01731} {arXiv:2207.01731 [quant-ph]} \BibitemShut
  {NoStop}%
\bibitem [{\citenamefont {Farrell}\ \emph
  {et~al.}(2023{\natexlab{b}})\citenamefont {Farrell}, \citenamefont
  {Chernyshev}, \citenamefont {Powell}, \citenamefont {Zemlevskiy},
  \citenamefont {Illa},\ and\ \citenamefont {Savage}}]{Farrell:2022vyh}%
  \BibitemOpen
  \bibfield  {author} {\bibinfo {author} {\bibfnamefont {R.~C.}\ \bibnamefont
  {Farrell}}, \bibinfo {author} {\bibfnamefont {I.~A.}\ \bibnamefont
  {Chernyshev}}, \bibinfo {author} {\bibfnamefont {S.~J.~M.}\ \bibnamefont
  {Powell}}, \bibinfo {author} {\bibfnamefont {N.~A.}\ \bibnamefont
  {Zemlevskiy}}, \bibinfo {author} {\bibfnamefont {M.}~\bibnamefont {Illa}},\
  and\ \bibinfo {author} {\bibfnamefont {M.~J.}\ \bibnamefont {Savage}},\
  }\bibfield  {title} {\bibinfo {title} {{Preparations for quantum simulations
  of quantum chromodynamics in 1+1 dimensions. II. Single-baryon
  \ensuremath{\beta}-decay in real time}},\ }\href
  {https://doi.org/10.1103/PhysRevD.107.054513} {\bibfield  {journal} {\bibinfo
   {journal} {Phys. Rev. D}\ }\textbf {\bibinfo {volume} {107}},\ \bibinfo
  {pages} {054513} (\bibinfo {year} {2023}{\natexlab{b}})},\ \Eprint
  {https://arxiv.org/abs/2209.10781} {arXiv:2209.10781 [quant-ph]} \BibitemShut
  {NoStop}%
\bibitem [{\citenamefont {Gupta}\ \emph {et~al.}(2024)\citenamefont {Gupta},
  \citenamefont {Javanmard}, \citenamefont {Osborne},\ and\ \citenamefont
  {Santos}}]{Gupta:2024gnw}%
  \BibitemOpen
  \bibfield  {author} {\bibinfo {author} {\bibfnamefont {S.}~\bibnamefont
  {Gupta}}, \bibinfo {author} {\bibfnamefont {Y.}~\bibnamefont {Javanmard}},
  \bibinfo {author} {\bibfnamefont {T.~J.}\ \bibnamefont {Osborne}},\ and\
  \bibinfo {author} {\bibfnamefont {L.}~\bibnamefont {Santos}},\ }\bibfield
  {title} {\bibinfo {title} {{Simulation of a Rohksar\textendash{}Kivelson
  ladder on a NISQ device}},\ }\href
  {https://doi.org/10.1038/s41598-024-79480-2} {\bibfield  {journal} {\bibinfo
  {journal} {Sci. Rep.}\ }\textbf {\bibinfo {volume} {14}},\ \bibinfo {pages}
  {29276} (\bibinfo {year} {2024})},\ \Eprint
  {https://arxiv.org/abs/2401.16326} {arXiv:2401.16326 [quant-ph]} \BibitemShut
  {NoStop}%
\bibitem [{\citenamefont {Davoudi}\ \emph {et~al.}(2024)\citenamefont
  {Davoudi}, \citenamefont {Hsieh},\ and\ \citenamefont
  {Kadam}}]{Davoudi:2024wyv}%
  \BibitemOpen
  \bibfield  {author} {\bibinfo {author} {\bibfnamefont {Z.}~\bibnamefont
  {Davoudi}}, \bibinfo {author} {\bibfnamefont {C.-C.}\ \bibnamefont {Hsieh}},\
  and\ \bibinfo {author} {\bibfnamefont {S.~V.}\ \bibnamefont {Kadam}},\
  }\bibfield  {title} {\bibinfo {title} {{Scattering wave packets of hadrons in
  gauge theories: Preparation on a quantum computer}},\ }\href
  {https://doi.org/10.22331/q-2024-11-11-1520} {\bibfield  {journal} {\bibinfo
  {journal} {Quantum}\ }\textbf {\bibinfo {volume} {8}},\ \bibinfo {pages}
  {1520} (\bibinfo {year} {2024})},\ \Eprint {https://arxiv.org/abs/2402.00840}
  {arXiv:2402.00840 [quant-ph]} \BibitemShut {NoStop}%
\bibitem [{\citenamefont {Ciavarella}\ and\ \citenamefont
  {Bauer}(2024)}]{Ciavarella:2024fzw}%
  \BibitemOpen
  \bibfield  {author} {\bibinfo {author} {\bibfnamefont {A.~N.}\ \bibnamefont
  {Ciavarella}}\ and\ \bibinfo {author} {\bibfnamefont {C.~W.}\ \bibnamefont
  {Bauer}},\ }\bibfield  {title} {\bibinfo {title} {{Quantum Simulation of
  SU(3) Lattice Yang-Mills Theory at Leading Order in Large-Nc Expansion}},\
  }\href {https://doi.org/10.1103/PhysRevLett.133.111901} {\bibfield  {journal}
  {\bibinfo  {journal} {Phys. Rev. Lett.}\ }\textbf {\bibinfo {volume} {133}},\
  \bibinfo {pages} {111901} (\bibinfo {year} {2024})},\ \Eprint
  {https://arxiv.org/abs/2402.10265} {arXiv:2402.10265 [hep-ph]} \BibitemShut
  {NoStop}%
\bibitem [{\citenamefont {Crippa}\ \emph {et~al.}(2024)\citenamefont {Crippa},
  \citenamefont {Jansen},\ and\ \citenamefont {Rinaldi}}]{Crippa:2024hso}%
  \BibitemOpen
  \bibfield  {author} {\bibinfo {author} {\bibfnamefont {A.}~\bibnamefont
  {Crippa}}, \bibinfo {author} {\bibfnamefont {K.}~\bibnamefont {Jansen}},\
  and\ \bibinfo {author} {\bibfnamefont {E.}~\bibnamefont {Rinaldi}},\
  }\bibfield  {title} {\bibinfo {title} {{Analysis of the confinement string in
  (2 + 1)-dimensional Quantum Electrodynamics with a trapped-ion quantum
  computer}},\ }\href@noop {} {\  (\bibinfo {year} {2024})},\ \Eprint
  {https://arxiv.org/abs/2411.05628} {arXiv:2411.05628 [hep-lat]} \BibitemShut
  {NoStop}%
\bibitem [{\citenamefont {Zhang}\ \emph {et~al.}(2025)\citenamefont {Zhang},
  \citenamefont {Guo}, \citenamefont {Wang},\ and\ \citenamefont
  {Xing}}]{Zhang:2024fgv}%
  \BibitemOpen
  \bibfield  {author} {\bibinfo {author} {\bibfnamefont {G.}~\bibnamefont
  {Zhang}}, \bibinfo {author} {\bibfnamefont {X.}~\bibnamefont {Guo}}, \bibinfo
  {author} {\bibfnamefont {E.}~\bibnamefont {Wang}},\ and\ \bibinfo {author}
  {\bibfnamefont {H.}~\bibnamefont {Xing}} (\bibinfo {collaboration} {QuNu}),\
  }\bibfield  {title} {\bibinfo {title} {{Quantum computing of chirality
  imbalance in SU(2) gauge theory}},\ }\href
  {https://doi.org/10.1103/PhysRevD.111.056031} {\bibfield  {journal} {\bibinfo
   {journal} {Phys. Rev. D}\ }\textbf {\bibinfo {volume} {111}},\ \bibinfo
  {pages} {056031} (\bibinfo {year} {2025})},\ \Eprint
  {https://arxiv.org/abs/2411.18869} {arXiv:2411.18869 [hep-ph]} \BibitemShut
  {NoStop}%
\bibitem [{\citenamefont {Than}\ \emph {et~al.}(2024)\citenamefont {Than} \emph
  {et~al.}}]{Than:2024zaj}%
  \BibitemOpen
  \bibfield  {author} {\bibinfo {author} {\bibfnamefont {A.~T.}\ \bibnamefont
  {Than}} \emph {et~al.},\ }\bibfield  {title} {\bibinfo {title} {{The phase
  diagram of quantum chromodynamics in one dimension on a quantum computer}},\
  }\href@noop {} {\  (\bibinfo {year} {2024})},\ \Eprint
  {https://arxiv.org/abs/2501.00579} {arXiv:2501.00579 [quant-ph]} \BibitemShut
  {NoStop}%
\bibitem [{\citenamefont {Schuhmacher}\ \emph {et~al.}(2025)\citenamefont
  {Schuhmacher}, \citenamefont {Su}, \citenamefont {Osborne}, \citenamefont
  {Gandon}, \citenamefont {Halimeh},\ and\ \citenamefont
  {Tavernelli}}]{Schuhmacher:2025ehh}%
  \BibitemOpen
  \bibfield  {author} {\bibinfo {author} {\bibfnamefont {J.}~\bibnamefont
  {Schuhmacher}}, \bibinfo {author} {\bibfnamefont {G.-X.}\ \bibnamefont {Su}},
  \bibinfo {author} {\bibfnamefont {J.~J.}\ \bibnamefont {Osborne}}, \bibinfo
  {author} {\bibfnamefont {A.}~\bibnamefont {Gandon}}, \bibinfo {author}
  {\bibfnamefont {J.~C.}\ \bibnamefont {Halimeh}},\ and\ \bibinfo {author}
  {\bibfnamefont {I.}~\bibnamefont {Tavernelli}},\ }\bibfield  {title}
  {\bibinfo {title} {{Observation of hadron scattering in a lattice gauge
  theory on a quantum computer}},\ }\href@noop {} {\  (\bibinfo {year}
  {2025})},\ \Eprint {https://arxiv.org/abs/2505.20387} {arXiv:2505.20387
  [quant-ph]} \BibitemShut {NoStop}%
\bibitem [{\citenamefont {Davoudi}\ \emph {et~al.}(2025)\citenamefont
  {Davoudi}, \citenamefont {Hsieh},\ and\ \citenamefont
  {Kadam}}]{Davoudi:2025rdv}%
  \BibitemOpen
  \bibfield  {author} {\bibinfo {author} {\bibfnamefont {Z.}~\bibnamefont
  {Davoudi}}, \bibinfo {author} {\bibfnamefont {C.-C.}\ \bibnamefont {Hsieh}},\
  and\ \bibinfo {author} {\bibfnamefont {S.~V.}\ \bibnamefont {Kadam}},\
  }\bibfield  {title} {\bibinfo {title} {{Quantum computation of hadron
  scattering in a lattice gauge theory}},\ }\href@noop {} {\  (\bibinfo {year}
  {2025})},\ \Eprint {https://arxiv.org/abs/2505.20408} {arXiv:2505.20408
  [quant-ph]} \BibitemShut {NoStop}%
\bibitem [{\citenamefont {Jordan}\ \emph {et~al.}(2012)\citenamefont {Jordan},
  \citenamefont {Lee},\ and\ \citenamefont {Preskill}}]{Jordan:2012xnu}%
  \BibitemOpen
  \bibfield  {author} {\bibinfo {author} {\bibfnamefont {S.~P.}\ \bibnamefont
  {Jordan}}, \bibinfo {author} {\bibfnamefont {K.~S.~M.}\ \bibnamefont {Lee}},\
  and\ \bibinfo {author} {\bibfnamefont {J.}~\bibnamefont {Preskill}},\
  }\bibfield  {title} {\bibinfo {title} {{Quantum Algorithms for Quantum Field
  Theories}},\ }\href {https://doi.org/10.1126/science.1217069} {\bibfield
  {journal} {\bibinfo  {journal} {Science}\ }\textbf {\bibinfo {volume}
  {336}},\ \bibinfo {pages} {1130} (\bibinfo {year} {2012})},\ \Eprint
  {https://arxiv.org/abs/1111.3633} {arXiv:1111.3633 [quant-ph]} \BibitemShut
  {NoStop}%
\bibitem [{\citenamefont {Jordan}\ \emph
  {et~al.}(2014{\natexlab{a}})\citenamefont {Jordan}, \citenamefont {Lee},\
  and\ \citenamefont {Preskill}}]{Jordan:2011ci}%
  \BibitemOpen
  \bibfield  {author} {\bibinfo {author} {\bibfnamefont {S.~P.}\ \bibnamefont
  {Jordan}}, \bibinfo {author} {\bibfnamefont {K.~S.~M.}\ \bibnamefont {Lee}},\
  and\ \bibinfo {author} {\bibfnamefont {J.}~\bibnamefont {Preskill}},\
  }\bibfield  {title} {\bibinfo {title} {{Quantum Computation of Scattering in
  Scalar Quantum Field Theories}},\ }\href@noop {} {\bibfield  {journal}
  {\bibinfo  {journal} {Quant. Inf. Comput.}\ }\textbf {\bibinfo {volume}
  {14}},\ \bibinfo {pages} {1014} (\bibinfo {year} {2014}{\natexlab{a}})},\
  \Eprint {https://arxiv.org/abs/1112.4833} {arXiv:1112.4833 [hep-th]}
  \BibitemShut {NoStop}%
\bibitem [{\citenamefont {Bhattacharyya}\ \emph {et~al.}(2018)\citenamefont
  {Bhattacharyya}, \citenamefont {Shekar},\ and\ \citenamefont
  {Sinha}}]{Bhattacharyya:2018bbv}%
  \BibitemOpen
  \bibfield  {author} {\bibinfo {author} {\bibfnamefont {A.}~\bibnamefont
  {Bhattacharyya}}, \bibinfo {author} {\bibfnamefont {A.}~\bibnamefont
  {Shekar}},\ and\ \bibinfo {author} {\bibfnamefont {A.}~\bibnamefont
  {Sinha}},\ }\bibfield  {title} {\bibinfo {title} {{Circuit complexity in
  interacting QFTs and RG flows}},\ }\href
  {https://doi.org/10.1007/JHEP10(2018)140} {\bibfield  {journal} {\bibinfo
  {journal} {JHEP}\ }\textbf {\bibinfo {volume} {10}},\ \bibinfo {pages}
  {140}},\ \Eprint {https://arxiv.org/abs/1808.03105} {arXiv:1808.03105
  [hep-th]} \BibitemShut {NoStop}%
\bibitem [{\citenamefont {Klco}\ and\ \citenamefont
  {Savage}(2019)}]{Klco:2018zqz}%
  \BibitemOpen
  \bibfield  {author} {\bibinfo {author} {\bibfnamefont {N.}~\bibnamefont
  {Klco}}\ and\ \bibinfo {author} {\bibfnamefont {M.~J.}\ \bibnamefont
  {Savage}},\ }\bibfield  {title} {\bibinfo {title} {{Digitization of scalar
  fields for quantum computing}},\ }\href
  {https://doi.org/10.1103/PhysRevA.99.052335} {\bibfield  {journal} {\bibinfo
  {journal} {Phys. Rev. A}\ }\textbf {\bibinfo {volume} {99}},\ \bibinfo
  {pages} {052335} (\bibinfo {year} {2019})},\ \Eprint
  {https://arxiv.org/abs/1808.10378} {arXiv:1808.10378 [quant-ph]} \BibitemShut
  {NoStop}%
\bibitem [{\citenamefont {Klco}\ and\ \citenamefont
  {Savage}(2020{\natexlab{b}})}]{Klco:2019yrb}%
  \BibitemOpen
  \bibfield  {author} {\bibinfo {author} {\bibfnamefont {N.}~\bibnamefont
  {Klco}}\ and\ \bibinfo {author} {\bibfnamefont {M.~J.}\ \bibnamefont
  {Savage}},\ }\bibfield  {title} {\bibinfo {title} {{Systematically
  Localizable Operators for Quantum Simulations of Quantum Field Theories}},\
  }\href {https://doi.org/10.1103/PhysRevA.102.012619} {\bibfield  {journal}
  {\bibinfo  {journal} {Phys. Rev. A}\ }\textbf {\bibinfo {volume} {102}},\
  \bibinfo {pages} {012619} (\bibinfo {year} {2020}{\natexlab{b}})},\ \Eprint
  {https://arxiv.org/abs/1912.03577} {arXiv:1912.03577 [quant-ph]} \BibitemShut
  {NoStop}%
\bibitem [{\citenamefont {Barata}\ \emph {et~al.}(2021)\citenamefont {Barata},
  \citenamefont {Mueller}, \citenamefont {Tarasov},\ and\ \citenamefont
  {Venugopalan}}]{Barata:2020jtq}%
  \BibitemOpen
  \bibfield  {author} {\bibinfo {author} {\bibfnamefont {J.~a.}\ \bibnamefont
  {Barata}}, \bibinfo {author} {\bibfnamefont {N.}~\bibnamefont {Mueller}},
  \bibinfo {author} {\bibfnamefont {A.}~\bibnamefont {Tarasov}},\ and\ \bibinfo
  {author} {\bibfnamefont {R.}~\bibnamefont {Venugopalan}},\ }\bibfield
  {title} {\bibinfo {title} {{Single-particle digitization strategy for quantum
  computation of a $\phi^4$ scalar field theory}},\ }\href
  {https://doi.org/10.1103/PhysRevA.103.042410} {\bibfield  {journal} {\bibinfo
   {journal} {Phys. Rev. A}\ }\textbf {\bibinfo {volume} {103}},\ \bibinfo
  {pages} {042410} (\bibinfo {year} {2021})},\ \Eprint
  {https://arxiv.org/abs/2012.00020} {arXiv:2012.00020 [hep-th]} \BibitemShut
  {NoStop}%
\bibitem [{\citenamefont {Ignacio~Cirac}\ \emph {et~al.}(2010)\citenamefont
  {Ignacio~Cirac}, \citenamefont {Maraner},\ and\ \citenamefont
  {Pachos}}]{IgnacioCirac:2010us}%
  \BibitemOpen
  \bibfield  {author} {\bibinfo {author} {\bibfnamefont {J.}~\bibnamefont
  {Ignacio~Cirac}}, \bibinfo {author} {\bibfnamefont {P.}~\bibnamefont
  {Maraner}},\ and\ \bibinfo {author} {\bibfnamefont {J.~K.}\ \bibnamefont
  {Pachos}},\ }\bibfield  {title} {\bibinfo {title} {{Cold atom simulation of
  interacting relativistic quantum field theories}},\ }\href
  {https://doi.org/10.1103/PhysRevLett.105.190403} {\bibfield  {journal}
  {\bibinfo  {journal} {Phys. Rev. Lett.}\ }\textbf {\bibinfo {volume} {105}},\
  \bibinfo {pages} {190403} (\bibinfo {year} {2010})},\ \Eprint
  {https://arxiv.org/abs/1006.2975} {arXiv:1006.2975 [cond-mat.str-el]}
  \BibitemShut {NoStop}%
\bibitem [{\citenamefont {Casanova}\ \emph {et~al.}(2011)\citenamefont
  {Casanova}, \citenamefont {Lamata}, \citenamefont {Egusquiza}, \citenamefont
  {Gerritsma}, \citenamefont {Roos}, \citenamefont {Garcia-Ripoll},\ and\
  \citenamefont {Solano}}]{Casanova:2011wh}%
  \BibitemOpen
  \bibfield  {author} {\bibinfo {author} {\bibfnamefont {J.}~\bibnamefont
  {Casanova}}, \bibinfo {author} {\bibfnamefont {L.}~\bibnamefont {Lamata}},
  \bibinfo {author} {\bibfnamefont {I.~L.}\ \bibnamefont {Egusquiza}}, \bibinfo
  {author} {\bibfnamefont {R.}~\bibnamefont {Gerritsma}}, \bibinfo {author}
  {\bibfnamefont {C.~F.}\ \bibnamefont {Roos}}, \bibinfo {author}
  {\bibfnamefont {J.~J.}\ \bibnamefont {Garcia-Ripoll}},\ and\ \bibinfo
  {author} {\bibfnamefont {E.}~\bibnamefont {Solano}},\ }\bibfield  {title}
  {\bibinfo {title} {{Quantum Simulation of Quantum Field Theories in Trapped
  Ions}},\ }\href {https://doi.org/10.1103/PhysRevLett.107.260501} {\bibfield
  {journal} {\bibinfo  {journal} {Phys. Rev. Lett.}\ }\textbf {\bibinfo
  {volume} {107}},\ \bibinfo {pages} {260501} (\bibinfo {year} {2011})},\
  \Eprint {https://arxiv.org/abs/1107.5233} {arXiv:1107.5233 [quant-ph]}
  \BibitemShut {NoStop}%
\bibitem [{\citenamefont {Jordan}\ \emph
  {et~al.}(2014{\natexlab{b}})\citenamefont {Jordan}, \citenamefont {Lee},\
  and\ \citenamefont {Preskill}}]{Jordan:2014tma}%
  \BibitemOpen
  \bibfield  {author} {\bibinfo {author} {\bibfnamefont {S.~P.}\ \bibnamefont
  {Jordan}}, \bibinfo {author} {\bibfnamefont {K.~S.~M.}\ \bibnamefont {Lee}},\
  and\ \bibinfo {author} {\bibfnamefont {J.}~\bibnamefont {Preskill}},\
  }\bibfield  {title} {\bibinfo {title} {{Quantum Algorithms for Fermionic
  Quantum Field Theories}},\ }\href@noop {} {\  (\bibinfo {year}
  {2014}{\natexlab{b}})},\ \Eprint {https://arxiv.org/abs/1404.7115}
  {arXiv:1404.7115 [hep-th]} \BibitemShut {NoStop}%
\bibitem [{\citenamefont {Hamed~Moosavian}\ and\ \citenamefont
  {Jordan}(2018)}]{HamedMoosavian:2017koz}%
  \BibitemOpen
  \bibfield  {author} {\bibinfo {author} {\bibfnamefont {A.}~\bibnamefont
  {Hamed~Moosavian}}\ and\ \bibinfo {author} {\bibfnamefont {S.}~\bibnamefont
  {Jordan}},\ }\bibfield  {title} {\bibinfo {title} {{Faster Quantum Algorithm
  to simulate Fermionic Quantum Field Theory}},\ }\href
  {https://doi.org/10.1103/PhysRevA.98.012332} {\bibfield  {journal} {\bibinfo
  {journal} {Phys. Rev. A}\ }\textbf {\bibinfo {volume} {98}},\ \bibinfo
  {pages} {012332} (\bibinfo {year} {2018})},\ \Eprint
  {https://arxiv.org/abs/1711.04006} {arXiv:1711.04006 [quant-ph]} \BibitemShut
  {NoStop}%
\bibitem [{\citenamefont {Lamm}\ \emph
  {et~al.}(2020{\natexlab{a}})\citenamefont {Lamm}, \citenamefont {Lawrence},\
  and\ \citenamefont {Yamauchi}}]{Lamm:2019uyc}%
  \BibitemOpen
  \bibfield  {author} {\bibinfo {author} {\bibfnamefont {H.}~\bibnamefont
  {Lamm}}, \bibinfo {author} {\bibfnamefont {S.}~\bibnamefont {Lawrence}},\
  and\ \bibinfo {author} {\bibfnamefont {Y.}~\bibnamefont {Yamauchi}} (\bibinfo
  {collaboration} {NuQS}),\ }\bibfield  {title} {\bibinfo {title} {{Parton
  physics on a quantum computer}},\ }\href
  {https://doi.org/10.1103/PhysRevResearch.2.013272} {\bibfield  {journal}
  {\bibinfo  {journal} {Phys. Rev. Res.}\ }\textbf {\bibinfo {volume} {2}},\
  \bibinfo {pages} {013272} (\bibinfo {year} {2020}{\natexlab{a}})},\ \Eprint
  {https://arxiv.org/abs/1908.10439} {arXiv:1908.10439 [hep-lat]} \BibitemShut
  {NoStop}%
\bibitem [{\citenamefont {Mueller}\ \emph {et~al.}(2020)\citenamefont
  {Mueller}, \citenamefont {Tarasov},\ and\ \citenamefont
  {Venugopalan}}]{Mueller:2019qqj}%
  \BibitemOpen
  \bibfield  {author} {\bibinfo {author} {\bibfnamefont {N.}~\bibnamefont
  {Mueller}}, \bibinfo {author} {\bibfnamefont {A.}~\bibnamefont {Tarasov}},\
  and\ \bibinfo {author} {\bibfnamefont {R.}~\bibnamefont {Venugopalan}},\
  }\bibfield  {title} {\bibinfo {title} {{Deeply inelastic scattering structure
  functions on a hybrid quantum computer}},\ }\href
  {https://doi.org/10.1103/PhysRevD.102.016007} {\bibfield  {journal} {\bibinfo
   {journal} {Phys. Rev. D}\ }\textbf {\bibinfo {volume} {102}},\ \bibinfo
  {pages} {016007} (\bibinfo {year} {2020})},\ \Eprint
  {https://arxiv.org/abs/1908.07051} {arXiv:1908.07051 [hep-th]} \BibitemShut
  {NoStop}%
\bibitem [{\citenamefont {Harmalkar}\ \emph {et~al.}(2020)\citenamefont
  {Harmalkar}, \citenamefont {Lamm},\ and\ \citenamefont
  {Lawrence}}]{Harmalkar:2020mpd}%
  \BibitemOpen
  \bibfield  {author} {\bibinfo {author} {\bibfnamefont {S.}~\bibnamefont
  {Harmalkar}}, \bibinfo {author} {\bibfnamefont {H.}~\bibnamefont {Lamm}},\
  and\ \bibinfo {author} {\bibfnamefont {S.}~\bibnamefont {Lawrence}} (\bibinfo
  {collaboration} {NuQS}),\ }\bibfield  {title} {\bibinfo {title} {{Quantum
  Simulation of Field Theories Without State Preparation}},\ }\href@noop {} {\
  (\bibinfo {year} {2020})},\ \Eprint {https://arxiv.org/abs/2001.11490}
  {arXiv:2001.11490 [hep-lat]} \BibitemShut {NoStop}%
\bibitem [{\citenamefont {Kharzeev}\ and\ \citenamefont
  {Kikuchi}(2020)}]{Kharzeev:2020kgc}%
  \BibitemOpen
  \bibfield  {author} {\bibinfo {author} {\bibfnamefont {D.~E.}\ \bibnamefont
  {Kharzeev}}\ and\ \bibinfo {author} {\bibfnamefont {Y.}~\bibnamefont
  {Kikuchi}},\ }\bibfield  {title} {\bibinfo {title} {{Real-time chiral
  dynamics from a digital quantum simulation}},\ }\href
  {https://doi.org/10.1103/PhysRevResearch.2.023342} {\bibfield  {journal}
  {\bibinfo  {journal} {Phys. Rev. Res.}\ }\textbf {\bibinfo {volume} {2}},\
  \bibinfo {pages} {023342} (\bibinfo {year} {2020})},\ \Eprint
  {https://arxiv.org/abs/2001.00698} {arXiv:2001.00698 [hep-ph]} \BibitemShut
  {NoStop}%
\bibitem [{\citenamefont {Bauer}\ \emph
  {et~al.}(2021{\natexlab{b}})\citenamefont {Bauer}, \citenamefont {de~Jong},
  \citenamefont {Nachman},\ and\ \citenamefont {Provasoli}}]{Bauer:2019qxa}%
  \BibitemOpen
  \bibfield  {author} {\bibinfo {author} {\bibfnamefont {C.~W.}\ \bibnamefont
  {Bauer}}, \bibinfo {author} {\bibfnamefont {W.~A.}\ \bibnamefont {de~Jong}},
  \bibinfo {author} {\bibfnamefont {B.}~\bibnamefont {Nachman}},\ and\ \bibinfo
  {author} {\bibfnamefont {D.}~\bibnamefont {Provasoli}},\ }\bibfield  {title}
  {\bibinfo {title} {{Quantum Algorithm for High Energy Physics Simulations}},\
  }\href {https://doi.org/10.1103/PhysRevLett.126.062001} {\bibfield  {journal}
  {\bibinfo  {journal} {Phys. Rev. Lett.}\ }\textbf {\bibinfo {volume} {126}},\
  \bibinfo {pages} {062001} (\bibinfo {year} {2021}{\natexlab{b}})},\ \Eprint
  {https://arxiv.org/abs/1904.03196} {arXiv:1904.03196 [hep-ph]} \BibitemShut
  {NoStop}%
\bibitem [{\citenamefont {Davoudi}\ \emph
  {et~al.}(2021{\natexlab{a}})\citenamefont {Davoudi}, \citenamefont {Linke},\
  and\ \citenamefont {Pagano}}]{Davoudi:2021ney}%
  \BibitemOpen
  \bibfield  {author} {\bibinfo {author} {\bibfnamefont {Z.}~\bibnamefont
  {Davoudi}}, \bibinfo {author} {\bibfnamefont {N.~M.}\ \bibnamefont {Linke}},\
  and\ \bibinfo {author} {\bibfnamefont {G.}~\bibnamefont {Pagano}},\
  }\bibfield  {title} {\bibinfo {title} {{Toward simulating quantum field
  theories with controlled phonon-ion dynamics: A hybrid analog-digital
  approach}},\ }\href {https://doi.org/10.1103/PhysRevResearch.3.043072}
  {\bibfield  {journal} {\bibinfo  {journal} {Phys. Rev. Res.}\ }\textbf
  {\bibinfo {volume} {3}},\ \bibinfo {pages} {043072} (\bibinfo {year}
  {2021}{\natexlab{a}})},\ \Eprint {https://arxiv.org/abs/2104.09346}
  {arXiv:2104.09346 [quant-ph]} \BibitemShut {NoStop}%
\bibitem [{\citenamefont {Ba\~nuls}\ \emph {et~al.}(2020)\citenamefont
  {Ba\~nuls} \emph {et~al.}}]{Banuls:2019bmf}%
  \BibitemOpen
  \bibfield  {author} {\bibinfo {author} {\bibfnamefont {M.~C.}\ \bibnamefont
  {Ba\~nuls}} \emph {et~al.},\ }\bibfield  {title} {\bibinfo {title}
  {{Simulating Lattice Gauge Theories within Quantum Technologies}},\ }\href
  {https://doi.org/10.1140/epjd/e2020-100571-8} {\bibfield  {journal} {\bibinfo
   {journal} {Eur. Phys. J. D}\ }\textbf {\bibinfo {volume} {74}},\ \bibinfo
  {pages} {165} (\bibinfo {year} {2020})},\ \Eprint
  {https://arxiv.org/abs/1911.00003} {arXiv:1911.00003 [quant-ph]} \BibitemShut
  {NoStop}%
\bibitem [{\citenamefont {Byrnes}\ and\ \citenamefont
  {Yamamoto}(2006)}]{Byrnes:2005qx}%
  \BibitemOpen
  \bibfield  {author} {\bibinfo {author} {\bibfnamefont {T.}~\bibnamefont
  {Byrnes}}\ and\ \bibinfo {author} {\bibfnamefont {Y.}~\bibnamefont
  {Yamamoto}},\ }\bibfield  {title} {\bibinfo {title} {{Simulating lattice
  gauge theories on a quantum computer}},\ }\href
  {https://doi.org/10.1103/PhysRevA.73.022328} {\bibfield  {journal} {\bibinfo
  {journal} {Phys. Rev. A}\ }\textbf {\bibinfo {volume} {73}},\ \bibinfo
  {pages} {022328} (\bibinfo {year} {2006})},\ \Eprint
  {https://arxiv.org/abs/quant-ph/0510027} {arXiv:quant-ph/0510027}
  \BibitemShut {NoStop}%
\bibitem [{\citenamefont {Stryker}(2019)}]{Stryker:2018efp}%
  \BibitemOpen
  \bibfield  {author} {\bibinfo {author} {\bibfnamefont {J.~R.}\ \bibnamefont
  {Stryker}},\ }\bibfield  {title} {\bibinfo {title} {{Oracles for Gauss's law
  on digital quantum computers}},\ }\href
  {https://doi.org/10.1103/PhysRevA.99.042301} {\bibfield  {journal} {\bibinfo
  {journal} {Phys. Rev. A}\ }\textbf {\bibinfo {volume} {99}},\ \bibinfo
  {pages} {042301} (\bibinfo {year} {2019})},\ \Eprint
  {https://arxiv.org/abs/1812.01617} {arXiv:1812.01617 [quant-ph]} \BibitemShut
  {NoStop}%
\bibitem [{\citenamefont {Lamm}\ \emph {et~al.}(2019)\citenamefont {Lamm},
  \citenamefont {Lawrence},\ and\ \citenamefont {Yamauchi}}]{Lamm:2019bik}%
  \BibitemOpen
  \bibfield  {author} {\bibinfo {author} {\bibfnamefont {H.}~\bibnamefont
  {Lamm}}, \bibinfo {author} {\bibfnamefont {S.}~\bibnamefont {Lawrence}},\
  and\ \bibinfo {author} {\bibfnamefont {Y.}~\bibnamefont {Yamauchi}} (\bibinfo
  {collaboration} {NuQS}),\ }\bibfield  {title} {\bibinfo {title} {{General
  Methods for Digital Quantum Simulation of Gauge Theories}},\ }\href
  {https://doi.org/10.1103/PhysRevD.100.034518} {\bibfield  {journal} {\bibinfo
   {journal} {Phys. Rev. D}\ }\textbf {\bibinfo {volume} {100}},\ \bibinfo
  {pages} {034518} (\bibinfo {year} {2019})},\ \Eprint
  {https://arxiv.org/abs/1903.08807} {arXiv:1903.08807 [hep-lat]} \BibitemShut
  {NoStop}%
\bibitem [{\citenamefont {Alexandru}\ \emph {et~al.}(2019)\citenamefont
  {Alexandru}, \citenamefont {Bedaque}, \citenamefont {Harmalkar},
  \citenamefont {Lamm}, \citenamefont {Lawrence},\ and\ \citenamefont
  {Warrington}}]{Alexandru:2019nsa}%
  \BibitemOpen
  \bibfield  {author} {\bibinfo {author} {\bibfnamefont {A.}~\bibnamefont
  {Alexandru}}, \bibinfo {author} {\bibfnamefont {P.~F.}\ \bibnamefont
  {Bedaque}}, \bibinfo {author} {\bibfnamefont {S.}~\bibnamefont {Harmalkar}},
  \bibinfo {author} {\bibfnamefont {H.}~\bibnamefont {Lamm}}, \bibinfo {author}
  {\bibfnamefont {S.}~\bibnamefont {Lawrence}},\ and\ \bibinfo {author}
  {\bibfnamefont {N.~C.}\ \bibnamefont {Warrington}} (\bibinfo {collaboration}
  {NuQS}),\ }\bibfield  {title} {\bibinfo {title} {{Gluon Field Digitization
  for Quantum Computers}},\ }\href
  {https://doi.org/10.1103/PhysRevD.100.114501} {\bibfield  {journal} {\bibinfo
   {journal} {Phys. Rev. D}\ }\textbf {\bibinfo {volume} {100}},\ \bibinfo
  {pages} {114501} (\bibinfo {year} {2019})},\ \Eprint
  {https://arxiv.org/abs/1906.11213} {arXiv:1906.11213 [hep-lat]} \BibitemShut
  {NoStop}%
\bibitem [{\citenamefont {Zohar}\ and\ \citenamefont
  {Cirac}(2019)}]{Zohar:2019ygc}%
  \BibitemOpen
  \bibfield  {author} {\bibinfo {author} {\bibfnamefont {E.}~\bibnamefont
  {Zohar}}\ and\ \bibinfo {author} {\bibfnamefont {J.~I.}\ \bibnamefont
  {Cirac}},\ }\bibfield  {title} {\bibinfo {title} {{Removing Staggered
  Fermionic Matter in $U(N)$ and $SU(N)$ Lattice Gauge Theories}},\ }\href
  {https://doi.org/10.1103/PhysRevD.99.114511} {\bibfield  {journal} {\bibinfo
  {journal} {Phys. Rev. D}\ }\textbf {\bibinfo {volume} {99}},\ \bibinfo
  {pages} {114511} (\bibinfo {year} {2019})},\ \Eprint
  {https://arxiv.org/abs/1905.00652} {arXiv:1905.00652 [quant-ph]} \BibitemShut
  {NoStop}%
\bibitem [{\citenamefont {Shaw}\ \emph {et~al.}(2020)\citenamefont {Shaw},
  \citenamefont {Lougovski}, \citenamefont {Stryker},\ and\ \citenamefont
  {Wiebe}}]{Shaw:2020udc}%
  \BibitemOpen
  \bibfield  {author} {\bibinfo {author} {\bibfnamefont {A.~F.}\ \bibnamefont
  {Shaw}}, \bibinfo {author} {\bibfnamefont {P.}~\bibnamefont {Lougovski}},
  \bibinfo {author} {\bibfnamefont {J.~R.}\ \bibnamefont {Stryker}},\ and\
  \bibinfo {author} {\bibfnamefont {N.}~\bibnamefont {Wiebe}},\ }\bibfield
  {title} {\bibinfo {title} {{Quantum Algorithms for Simulating the Lattice
  Schwinger Model}},\ }\href {https://doi.org/10.22331/q-2020-08-10-306}
  {\bibfield  {journal} {\bibinfo  {journal} {Quantum}\ }\textbf {\bibinfo
  {volume} {4}},\ \bibinfo {pages} {306} (\bibinfo {year} {2020})},\ \Eprint
  {https://arxiv.org/abs/2002.11146} {arXiv:2002.11146 [quant-ph]} \BibitemShut
  {NoStop}%
\bibitem [{\citenamefont {Chakraborty}\ \emph {et~al.}(2022)\citenamefont
  {Chakraborty}, \citenamefont {Honda}, \citenamefont {Izubuchi}, \citenamefont
  {Kikuchi},\ and\ \citenamefont {Tomiya}}]{Chakraborty:2020uhf}%
  \BibitemOpen
  \bibfield  {author} {\bibinfo {author} {\bibfnamefont {B.}~\bibnamefont
  {Chakraborty}}, \bibinfo {author} {\bibfnamefont {M.}~\bibnamefont {Honda}},
  \bibinfo {author} {\bibfnamefont {T.}~\bibnamefont {Izubuchi}}, \bibinfo
  {author} {\bibfnamefont {Y.}~\bibnamefont {Kikuchi}},\ and\ \bibinfo {author}
  {\bibfnamefont {A.}~\bibnamefont {Tomiya}},\ }\bibfield  {title} {\bibinfo
  {title} {{Classically emulated digital quantum simulation of the Schwinger
  model with a topological term via adiabatic state preparation}},\ }\href
  {https://doi.org/10.1103/PhysRevD.105.094503} {\bibfield  {journal} {\bibinfo
   {journal} {Phys. Rev. D}\ }\textbf {\bibinfo {volume} {105}},\ \bibinfo
  {pages} {094503} (\bibinfo {year} {2022})},\ \Eprint
  {https://arxiv.org/abs/2001.00485} {arXiv:2001.00485 [hep-lat]} \BibitemShut
  {NoStop}%
\bibitem [{\citenamefont {Liu}\ and\ \citenamefont {Xin}(2020)}]{Liu:2020eoa}%
  \BibitemOpen
  \bibfield  {author} {\bibinfo {author} {\bibfnamefont {J.}~\bibnamefont
  {Liu}}\ and\ \bibinfo {author} {\bibfnamefont {Y.}~\bibnamefont {Xin}},\
  }\bibfield  {title} {\bibinfo {title} {{Quantum simulation of quantum field
  theories as quantum chemistry}},\ }\href
  {https://doi.org/10.1007/JHEP12(2020)011} {\bibfield  {journal} {\bibinfo
  {journal} {JHEP}\ }\textbf {\bibinfo {volume} {12}},\ \bibinfo {pages}
  {011}},\ \Eprint {https://arxiv.org/abs/2004.13234} {arXiv:2004.13234
  [hep-th]} \BibitemShut {NoStop}%
\bibitem [{\citenamefont {Paulson}\ \emph {et~al.}(2021)\citenamefont {Paulson}
  \emph {et~al.}}]{Paulson:2020zjd}%
  \BibitemOpen
  \bibfield  {author} {\bibinfo {author} {\bibfnamefont {D.}~\bibnamefont
  {Paulson}} \emph {et~al.},\ }\bibfield  {title} {\bibinfo {title}
  {{Simulating 2D Effects in Lattice Gauge Theories on a Quantum Computer}},\
  }\href {https://doi.org/10.1103/PRXQuantum.2.030334} {\bibfield  {journal}
  {\bibinfo  {journal} {PRX Quantum}\ }\textbf {\bibinfo {volume} {2}},\
  \bibinfo {pages} {030334} (\bibinfo {year} {2021})},\ \Eprint
  {https://arxiv.org/abs/2008.09252} {arXiv:2008.09252 [quant-ph]} \BibitemShut
  {NoStop}%
\bibitem [{\citenamefont {Ji}\ \emph {et~al.}(2020)\citenamefont {Ji},
  \citenamefont {Lamm},\ and\ \citenamefont {Zhu}}]{Ji:2020kjk}%
  \BibitemOpen
  \bibfield  {author} {\bibinfo {author} {\bibfnamefont {Y.}~\bibnamefont
  {Ji}}, \bibinfo {author} {\bibfnamefont {H.}~\bibnamefont {Lamm}},\ and\
  \bibinfo {author} {\bibfnamefont {S.}~\bibnamefont {Zhu}} (\bibinfo
  {collaboration} {NuQS}),\ }\bibfield  {title} {\bibinfo {title} {{Gluon Field
  Digitization via Group Space Decimation for Quantum Computers}},\ }\href
  {https://doi.org/10.1103/PhysRevD.102.114513} {\bibfield  {journal} {\bibinfo
   {journal} {Phys. Rev. D}\ }\textbf {\bibinfo {volume} {102}},\ \bibinfo
  {pages} {114513} (\bibinfo {year} {2020})},\ \Eprint
  {https://arxiv.org/abs/2005.14221} {arXiv:2005.14221 [hep-lat]} \BibitemShut
  {NoStop}%
\bibitem [{\citenamefont {Davoudi}\ \emph
  {et~al.}(2021{\natexlab{b}})\citenamefont {Davoudi}, \citenamefont
  {Raychowdhury},\ and\ \citenamefont {Shaw}}]{Davoudi:2020yln}%
  \BibitemOpen
  \bibfield  {author} {\bibinfo {author} {\bibfnamefont {Z.}~\bibnamefont
  {Davoudi}}, \bibinfo {author} {\bibfnamefont {I.}~\bibnamefont
  {Raychowdhury}},\ and\ \bibinfo {author} {\bibfnamefont {A.}~\bibnamefont
  {Shaw}},\ }\bibfield  {title} {\bibinfo {title} {{Search for efficient
  formulations for Hamiltonian simulation of non-Abelian lattice gauge
  theories}},\ }\href {https://doi.org/10.1103/PhysRevD.104.074505} {\bibfield
  {journal} {\bibinfo  {journal} {Phys. Rev. D}\ }\textbf {\bibinfo {volume}
  {104}},\ \bibinfo {pages} {074505} (\bibinfo {year} {2021}{\natexlab{b}})},\
  \Eprint {https://arxiv.org/abs/2009.11802} {arXiv:2009.11802 [hep-lat]}
  \BibitemShut {NoStop}%
\bibitem [{\citenamefont {Bender}\ and\ \citenamefont
  {Zohar}(2020)}]{Bender:2020ztu}%
  \BibitemOpen
  \bibfield  {author} {\bibinfo {author} {\bibfnamefont {J.}~\bibnamefont
  {Bender}}\ and\ \bibinfo {author} {\bibfnamefont {E.}~\bibnamefont {Zohar}},\
  }\bibfield  {title} {\bibinfo {title} {{Gauge redundancy-free formulation of
  compact QED with dynamical matter for quantum and classical computations}},\
  }\href {https://doi.org/10.1103/PhysRevD.102.114517} {\bibfield  {journal}
  {\bibinfo  {journal} {Phys. Rev. D}\ }\textbf {\bibinfo {volume} {102}},\
  \bibinfo {pages} {114517} (\bibinfo {year} {2020})},\ \Eprint
  {https://arxiv.org/abs/2008.01349} {arXiv:2008.01349 [quant-ph]} \BibitemShut
  {NoStop}%
\bibitem [{\citenamefont {Haase}\ \emph {et~al.}(2021)\citenamefont {Haase},
  \citenamefont {Dellantonio}, \citenamefont {Celi}, \citenamefont {Paulson},
  \citenamefont {Kan}, \citenamefont {Jansen},\ and\ \citenamefont
  {Muschik}}]{Haase:2020kaj}%
  \BibitemOpen
  \bibfield  {author} {\bibinfo {author} {\bibfnamefont {J.~F.}\ \bibnamefont
  {Haase}}, \bibinfo {author} {\bibfnamefont {L.}~\bibnamefont {Dellantonio}},
  \bibinfo {author} {\bibfnamefont {A.}~\bibnamefont {Celi}}, \bibinfo {author}
  {\bibfnamefont {D.}~\bibnamefont {Paulson}}, \bibinfo {author} {\bibfnamefont
  {A.}~\bibnamefont {Kan}}, \bibinfo {author} {\bibfnamefont {K.}~\bibnamefont
  {Jansen}},\ and\ \bibinfo {author} {\bibfnamefont {C.~A.}\ \bibnamefont
  {Muschik}},\ }\bibfield  {title} {\bibinfo {title} {{A resource efficient
  approach for quantum and classical simulations of gauge theories in particle
  physics}},\ }\href {https://doi.org/10.22331/q-2021-02-04-393} {\bibfield
  {journal} {\bibinfo  {journal} {Quantum}\ }\textbf {\bibinfo {volume} {5}},\
  \bibinfo {pages} {393} (\bibinfo {year} {2021})},\ \Eprint
  {https://arxiv.org/abs/2006.14160} {arXiv:2006.14160 [quant-ph]} \BibitemShut
  {NoStop}%
\bibitem [{\citenamefont {Kan}\ \emph {et~al.}(2021)\citenamefont {Kan},
  \citenamefont {Funcke}, \citenamefont {K\"uhn}, \citenamefont {Dellantonio},
  \citenamefont {Zhang}, \citenamefont {Haase}, \citenamefont {Muschik},\ and\
  \citenamefont {Jansen}}]{Kan:2021nyu}%
  \BibitemOpen
  \bibfield  {author} {\bibinfo {author} {\bibfnamefont {A.}~\bibnamefont
  {Kan}}, \bibinfo {author} {\bibfnamefont {L.}~\bibnamefont {Funcke}},
  \bibinfo {author} {\bibfnamefont {S.}~\bibnamefont {K\"uhn}}, \bibinfo
  {author} {\bibfnamefont {L.}~\bibnamefont {Dellantonio}}, \bibinfo {author}
  {\bibfnamefont {J.}~\bibnamefont {Zhang}}, \bibinfo {author} {\bibfnamefont
  {J.~F.}\ \bibnamefont {Haase}}, \bibinfo {author} {\bibfnamefont {C.~A.}\
  \bibnamefont {Muschik}},\ and\ \bibinfo {author} {\bibfnamefont
  {K.}~\bibnamefont {Jansen}},\ }\bibfield  {title} {\bibinfo {title}
  {{Investigating a (3+1)D topological \ensuremath{\theta}-term in the
  Hamiltonian formulation of lattice gauge theories for quantum and classical
  simulations}},\ }\href {https://doi.org/10.1103/PhysRevD.104.034504}
  {\bibfield  {journal} {\bibinfo  {journal} {Phys. Rev. D}\ }\textbf {\bibinfo
  {volume} {104}},\ \bibinfo {pages} {034504} (\bibinfo {year} {2021})},\
  \Eprint {https://arxiv.org/abs/2105.06019} {arXiv:2105.06019 [hep-lat]}
  \BibitemShut {NoStop}%
\bibitem [{\citenamefont {Zohar}\ and\ \citenamefont
  {Reznik}(2011)}]{Zohar:2011cw}%
  \BibitemOpen
  \bibfield  {author} {\bibinfo {author} {\bibfnamefont {E.}~\bibnamefont
  {Zohar}}\ and\ \bibinfo {author} {\bibfnamefont {B.}~\bibnamefont {Reznik}},\
  }\bibfield  {title} {\bibinfo {title} {{Confinement and lattice QED electric
  flux-tubes simulated with ultracold atoms}},\ }\href
  {https://doi.org/10.1103/PhysRevLett.107.275301} {\bibfield  {journal}
  {\bibinfo  {journal} {Phys. Rev. Lett.}\ }\textbf {\bibinfo {volume} {107}},\
  \bibinfo {pages} {275301} (\bibinfo {year} {2011})},\ \Eprint
  {https://arxiv.org/abs/1108.1562} {arXiv:1108.1562 [quant-ph]} \BibitemShut
  {NoStop}%
\bibitem [{\citenamefont {Zohar}\ \emph
  {et~al.}(2013{\natexlab{a}})\citenamefont {Zohar}, \citenamefont {Cirac},\
  and\ \citenamefont {Reznik}}]{Zohar:2013zla}%
  \BibitemOpen
  \bibfield  {author} {\bibinfo {author} {\bibfnamefont {E.}~\bibnamefont
  {Zohar}}, \bibinfo {author} {\bibfnamefont {J.~I.}\ \bibnamefont {Cirac}},\
  and\ \bibinfo {author} {\bibfnamefont {B.}~\bibnamefont {Reznik}},\
  }\bibfield  {title} {\bibinfo {title} {{Quantum simulations of gauge theories
  with ultracold atoms: local gauge invariance from angular momentum
  conservation}},\ }\href {https://doi.org/10.1103/PhysRevA.88.023617}
  {\bibfield  {journal} {\bibinfo  {journal} {Phys. Rev. A}\ }\textbf {\bibinfo
  {volume} {88}},\ \bibinfo {pages} {023617} (\bibinfo {year}
  {2013}{\natexlab{a}})},\ \Eprint {https://arxiv.org/abs/1303.5040}
  {arXiv:1303.5040 [quant-ph]} \BibitemShut {NoStop}%
\bibitem [{\citenamefont {Zohar}\ and\ \citenamefont
  {Burrello}(2015)}]{Zohar:2014qma}%
  \BibitemOpen
  \bibfield  {author} {\bibinfo {author} {\bibfnamefont {E.}~\bibnamefont
  {Zohar}}\ and\ \bibinfo {author} {\bibfnamefont {M.}~\bibnamefont
  {Burrello}},\ }\bibfield  {title} {\bibinfo {title} {{Formulation of lattice
  gauge theories for quantum simulations}},\ }\href
  {https://doi.org/10.1103/PhysRevD.91.054506} {\bibfield  {journal} {\bibinfo
  {journal} {Phys. Rev. D}\ }\textbf {\bibinfo {volume} {91}},\ \bibinfo
  {pages} {054506} (\bibinfo {year} {2015})},\ \Eprint
  {https://arxiv.org/abs/1409.3085} {arXiv:1409.3085 [quant-ph]} \BibitemShut
  {NoStop}%
\bibitem [{\citenamefont {Zohar}\ \emph
  {et~al.}(2017{\natexlab{a}})\citenamefont {Zohar}, \citenamefont {Farace},
  \citenamefont {Reznik},\ and\ \citenamefont {Cirac}}]{Zohar:2016wmo}%
  \BibitemOpen
  \bibfield  {author} {\bibinfo {author} {\bibfnamefont {E.}~\bibnamefont
  {Zohar}}, \bibinfo {author} {\bibfnamefont {A.}~\bibnamefont {Farace}},
  \bibinfo {author} {\bibfnamefont {B.}~\bibnamefont {Reznik}},\ and\ \bibinfo
  {author} {\bibfnamefont {J.~I.}\ \bibnamefont {Cirac}},\ }\bibfield  {title}
  {\bibinfo {title} {{Digital quantum simulation of $\mathbb{Z}_2$ lattice
  gauge theories with dynamical fermionic matter}},\ }\href
  {https://doi.org/10.1103/PhysRevLett.118.070501} {\bibfield  {journal}
  {\bibinfo  {journal} {Phys. Rev. Lett.}\ }\textbf {\bibinfo {volume} {118}},\
  \bibinfo {pages} {070501} (\bibinfo {year} {2017}{\natexlab{a}})},\ \Eprint
  {https://arxiv.org/abs/1607.03656} {arXiv:1607.03656 [quant-ph]} \BibitemShut
  {NoStop}%
\bibitem [{\citenamefont {Bender}\ \emph {et~al.}(2018)\citenamefont {Bender},
  \citenamefont {Zohar}, \citenamefont {Farace},\ and\ \citenamefont
  {Cirac}}]{Bender:2018rdp}%
  \BibitemOpen
  \bibfield  {author} {\bibinfo {author} {\bibfnamefont {J.}~\bibnamefont
  {Bender}}, \bibinfo {author} {\bibfnamefont {E.}~\bibnamefont {Zohar}},
  \bibinfo {author} {\bibfnamefont {A.}~\bibnamefont {Farace}},\ and\ \bibinfo
  {author} {\bibfnamefont {J.~I.}\ \bibnamefont {Cirac}},\ }\bibfield  {title}
  {\bibinfo {title} {{Digital quantum simulation of lattice gauge theories in
  three spatial dimensions}},\ }\href
  {https://doi.org/10.1088/1367-2630/aadb71} {\bibfield  {journal} {\bibinfo
  {journal} {New J. Phys.}\ }\textbf {\bibinfo {volume} {20}},\ \bibinfo
  {pages} {093001} (\bibinfo {year} {2018})},\ \Eprint
  {https://arxiv.org/abs/1804.02082} {arXiv:1804.02082 [quant-ph]} \BibitemShut
  {NoStop}%
\bibitem [{\citenamefont {Davoudi}\ \emph {et~al.}(2020)\citenamefont
  {Davoudi}, \citenamefont {Hafezi}, \citenamefont {Monroe}, \citenamefont
  {Pagano}, \citenamefont {Seif},\ and\ \citenamefont
  {Shaw}}]{Davoudi:2019bhy}%
  \BibitemOpen
  \bibfield  {author} {\bibinfo {author} {\bibfnamefont {Z.}~\bibnamefont
  {Davoudi}}, \bibinfo {author} {\bibfnamefont {M.}~\bibnamefont {Hafezi}},
  \bibinfo {author} {\bibfnamefont {C.}~\bibnamefont {Monroe}}, \bibinfo
  {author} {\bibfnamefont {G.}~\bibnamefont {Pagano}}, \bibinfo {author}
  {\bibfnamefont {A.}~\bibnamefont {Seif}},\ and\ \bibinfo {author}
  {\bibfnamefont {A.}~\bibnamefont {Shaw}},\ }\bibfield  {title} {\bibinfo
  {title} {{Towards analog quantum simulations of lattice gauge theories with
  trapped ions}},\ }\href {https://doi.org/10.1103/PhysRevResearch.2.023015}
  {\bibfield  {journal} {\bibinfo  {journal} {Phys. Rev. Res.}\ }\textbf
  {\bibinfo {volume} {2}},\ \bibinfo {pages} {023015} (\bibinfo {year}
  {2020})},\ \Eprint {https://arxiv.org/abs/1908.03210} {arXiv:1908.03210
  [quant-ph]} \BibitemShut {NoStop}%
\bibitem [{\citenamefont {Mil}\ \emph {et~al.}(2020)\citenamefont {Mil},
  \citenamefont {Zache}, \citenamefont {Hegde}, \citenamefont {Xia},
  \citenamefont {Bhatt}, \citenamefont {Oberthaler}, \citenamefont {Hauke},
  \citenamefont {Berges},\ and\ \citenamefont {Jendrzejewski}}]{Mil:2019pbt}%
  \BibitemOpen
  \bibfield  {author} {\bibinfo {author} {\bibfnamefont {A.}~\bibnamefont
  {Mil}}, \bibinfo {author} {\bibfnamefont {T.~V.}\ \bibnamefont {Zache}},
  \bibinfo {author} {\bibfnamefont {A.}~\bibnamefont {Hegde}}, \bibinfo
  {author} {\bibfnamefont {A.}~\bibnamefont {Xia}}, \bibinfo {author}
  {\bibfnamefont {R.~P.}\ \bibnamefont {Bhatt}}, \bibinfo {author}
  {\bibfnamefont {M.~K.}\ \bibnamefont {Oberthaler}}, \bibinfo {author}
  {\bibfnamefont {P.}~\bibnamefont {Hauke}}, \bibinfo {author} {\bibfnamefont
  {J.}~\bibnamefont {Berges}},\ and\ \bibinfo {author} {\bibfnamefont
  {F.}~\bibnamefont {Jendrzejewski}},\ }\bibfield  {title} {\bibinfo {title}
  {{A scalable realization of local U(1) gauge invariance in cold atomic
  mixtures}},\ }\href {https://doi.org/10.1126/science.aaz5312} {\bibfield
  {journal} {\bibinfo  {journal} {Science}\ }\textbf {\bibinfo {volume}
  {367}},\ \bibinfo {pages} {1128} (\bibinfo {year} {2020})},\ \Eprint
  {https://arxiv.org/abs/1909.07641} {arXiv:1909.07641 [cond-mat.quant-gas]}
  \BibitemShut {NoStop}%
\bibitem [{\citenamefont {Banerjee}\ \emph {et~al.}(2012)\citenamefont
  {Banerjee}, \citenamefont {Dalmonte}, \citenamefont {Muller}, \citenamefont
  {Rico}, \citenamefont {Stebler}, \citenamefont {Wiese},\ and\ \citenamefont
  {Zoller}}]{Banerjee:2012pg}%
  \BibitemOpen
  \bibfield  {author} {\bibinfo {author} {\bibfnamefont {D.}~\bibnamefont
  {Banerjee}}, \bibinfo {author} {\bibfnamefont {M.}~\bibnamefont {Dalmonte}},
  \bibinfo {author} {\bibfnamefont {M.}~\bibnamefont {Muller}}, \bibinfo
  {author} {\bibfnamefont {E.}~\bibnamefont {Rico}}, \bibinfo {author}
  {\bibfnamefont {P.}~\bibnamefont {Stebler}}, \bibinfo {author} {\bibfnamefont
  {U.~J.}\ \bibnamefont {Wiese}},\ and\ \bibinfo {author} {\bibfnamefont
  {P.}~\bibnamefont {Zoller}},\ }\bibfield  {title} {\bibinfo {title} {{Atomic
  Quantum Simulation of Dynamical Gauge Fields coupled to Fermionic Matter:
  From String Breaking to Evolution after a Quench}},\ }\href
  {https://doi.org/10.1103/PhysRevLett.109.175302} {\bibfield  {journal}
  {\bibinfo  {journal} {Phys. Rev. Lett.}\ }\textbf {\bibinfo {volume} {109}},\
  \bibinfo {pages} {175302} (\bibinfo {year} {2012})},\ \Eprint
  {https://arxiv.org/abs/1205.6366} {arXiv:1205.6366 [cond-mat.quant-gas]}
  \BibitemShut {NoStop}%
\bibitem [{\citenamefont {Tagliacozzo}\ \emph {et~al.}(2013)\citenamefont
  {Tagliacozzo}, \citenamefont {Celi}, \citenamefont {Zamora},\ and\
  \citenamefont {Lewenstein}}]{Tagliacozzo:2012vg}%
  \BibitemOpen
  \bibfield  {author} {\bibinfo {author} {\bibfnamefont {L.}~\bibnamefont
  {Tagliacozzo}}, \bibinfo {author} {\bibfnamefont {A.}~\bibnamefont {Celi}},
  \bibinfo {author} {\bibfnamefont {A.}~\bibnamefont {Zamora}},\ and\ \bibinfo
  {author} {\bibfnamefont {M.}~\bibnamefont {Lewenstein}},\ }\bibfield  {title}
  {\bibinfo {title} {{Optical Abelian Lattice Gauge Theories}},\ }\href
  {https://doi.org/10.1016/j.aop.2012.11.009} {\bibfield  {journal} {\bibinfo
  {journal} {Annals Phys.}\ }\textbf {\bibinfo {volume} {330}},\ \bibinfo
  {pages} {160} (\bibinfo {year} {2013})},\ \Eprint
  {https://arxiv.org/abs/1205.0496} {arXiv:1205.0496 [cond-mat.quant-gas]}
  \BibitemShut {NoStop}%
\bibitem [{\citenamefont {Mezzacapo}\ \emph {et~al.}(2015)\citenamefont
  {Mezzacapo}, \citenamefont {Rico}, \citenamefont {Sab\'\i{}n}, \citenamefont
  {Egusquiza}, \citenamefont {Lamata},\ and\ \citenamefont
  {Solano}}]{Mezzacapo:2015bra}%
  \BibitemOpen
  \bibfield  {author} {\bibinfo {author} {\bibfnamefont {A.}~\bibnamefont
  {Mezzacapo}}, \bibinfo {author} {\bibfnamefont {E.}~\bibnamefont {Rico}},
  \bibinfo {author} {\bibfnamefont {C.}~\bibnamefont {Sab\'\i{}n}}, \bibinfo
  {author} {\bibfnamefont {I.~L.}\ \bibnamefont {Egusquiza}}, \bibinfo {author}
  {\bibfnamefont {L.}~\bibnamefont {Lamata}},\ and\ \bibinfo {author}
  {\bibfnamefont {E.}~\bibnamefont {Solano}},\ }\bibfield  {title} {\bibinfo
  {title} {{Non-Abelian $SU(2)$ Lattice Gauge Theories in Superconducting
  Circuits}},\ }\href {https://doi.org/10.1103/PhysRevLett.115.240502}
  {\bibfield  {journal} {\bibinfo  {journal} {Phys. Rev. Lett.}\ }\textbf
  {\bibinfo {volume} {115}},\ \bibinfo {pages} {240502} (\bibinfo {year}
  {2015})},\ \Eprint {https://arxiv.org/abs/1505.04720} {arXiv:1505.04720
  [quant-ph]} \BibitemShut {NoStop}%
\bibitem [{\citenamefont {Zache}\ \emph {et~al.}(2018)\citenamefont {Zache},
  \citenamefont {Hebenstreit}, \citenamefont {Jendrzejewski}, \citenamefont
  {Oberthaler}, \citenamefont {Berges},\ and\ \citenamefont
  {Hauke}}]{Zache:2018jbt}%
  \BibitemOpen
  \bibfield  {author} {\bibinfo {author} {\bibfnamefont {T.~V.}\ \bibnamefont
  {Zache}}, \bibinfo {author} {\bibfnamefont {F.}~\bibnamefont {Hebenstreit}},
  \bibinfo {author} {\bibfnamefont {F.}~\bibnamefont {Jendrzejewski}}, \bibinfo
  {author} {\bibfnamefont {M.~K.}\ \bibnamefont {Oberthaler}}, \bibinfo
  {author} {\bibfnamefont {J.}~\bibnamefont {Berges}},\ and\ \bibinfo {author}
  {\bibfnamefont {P.}~\bibnamefont {Hauke}},\ }\bibfield  {title} {\bibinfo
  {title} {{Quantum simulation of lattice gauge theories using Wilson
  fermions}},\ }\href {https://doi.org/10.1088/2058-9565/aac33b} {\bibfield
  {journal} {\bibinfo  {journal} {Quantum Sci. Technol.}\ }\textbf {\bibinfo
  {volume} {3}},\ \bibinfo {pages} {034010} (\bibinfo {year} {2018})},\ \Eprint
  {https://arxiv.org/abs/1802.06704} {arXiv:1802.06704 [cond-mat.quant-gas]}
  \BibitemShut {NoStop}%
\bibitem [{\citenamefont {Zohar}\ \emph
  {et~al.}(2013{\natexlab{b}})\citenamefont {Zohar}, \citenamefont {Cirac},\
  and\ \citenamefont {Reznik}}]{Zohar:2012xf}%
  \BibitemOpen
  \bibfield  {author} {\bibinfo {author} {\bibfnamefont {E.}~\bibnamefont
  {Zohar}}, \bibinfo {author} {\bibfnamefont {J.~I.}\ \bibnamefont {Cirac}},\
  and\ \bibinfo {author} {\bibfnamefont {B.}~\bibnamefont {Reznik}},\
  }\bibfield  {title} {\bibinfo {title} {{Cold-Atom Quantum Simulator for SU(2)
  Yang-Mills Lattice Gauge Theory}},\ }\href
  {https://doi.org/10.1103/PhysRevLett.110.125304} {\bibfield  {journal}
  {\bibinfo  {journal} {Phys. Rev. Lett.}\ }\textbf {\bibinfo {volume} {110}},\
  \bibinfo {pages} {125304} (\bibinfo {year} {2013}{\natexlab{b}})},\ \Eprint
  {https://arxiv.org/abs/1211.2241} {arXiv:1211.2241 [quant-ph]} \BibitemShut
  {NoStop}%
\bibitem [{\citenamefont {Mathur}\ and\ \citenamefont
  {Sreeraj}(2016)}]{Mathur:2016cko}%
  \BibitemOpen
  \bibfield  {author} {\bibinfo {author} {\bibfnamefont {M.}~\bibnamefont
  {Mathur}}\ and\ \bibinfo {author} {\bibfnamefont {T.~P.}\ \bibnamefont
  {Sreeraj}},\ }\bibfield  {title} {\bibinfo {title} {{Lattice Gauge Theories
  and Spin Models}},\ }\href {https://doi.org/10.1103/PhysRevD.94.085029}
  {\bibfield  {journal} {\bibinfo  {journal} {Phys. Rev. D}\ }\textbf {\bibinfo
  {volume} {94}},\ \bibinfo {pages} {085029} (\bibinfo {year} {2016})},\
  \Eprint {https://arxiv.org/abs/1604.00315} {arXiv:1604.00315 [hep-lat]}
  \BibitemShut {NoStop}%
\bibitem [{\citenamefont {Raychowdhury}\ and\ \citenamefont
  {Stryker}(2020{\natexlab{a}})}]{Raychowdhury:2019iki}%
  \BibitemOpen
  \bibfield  {author} {\bibinfo {author} {\bibfnamefont {I.}~\bibnamefont
  {Raychowdhury}}\ and\ \bibinfo {author} {\bibfnamefont {J.~R.}\ \bibnamefont
  {Stryker}},\ }\bibfield  {title} {\bibinfo {title} {{Loop, string, and hadron
  dynamics in SU(2) Hamiltonian lattice gauge theories}},\ }\href
  {https://doi.org/10.1103/PhysRevD.101.114502} {\bibfield  {journal} {\bibinfo
   {journal} {Phys. Rev. D}\ }\textbf {\bibinfo {volume} {101}},\ \bibinfo
  {pages} {114502} (\bibinfo {year} {2020}{\natexlab{a}})},\ \Eprint
  {https://arxiv.org/abs/1912.06133} {arXiv:1912.06133 [hep-lat]} \BibitemShut
  {NoStop}%
\bibitem [{\citenamefont {Raychowdhury}\ and\ \citenamefont
  {Stryker}(2020{\natexlab{b}})}]{Raychowdhury:2018osk}%
  \BibitemOpen
  \bibfield  {author} {\bibinfo {author} {\bibfnamefont {I.}~\bibnamefont
  {Raychowdhury}}\ and\ \bibinfo {author} {\bibfnamefont {J.~R.}\ \bibnamefont
  {Stryker}},\ }\bibfield  {title} {\bibinfo {title} {{Solving Gauss's Law on
  Digital Quantum Computers with Loop-String-Hadron Digitization}},\ }\href
  {https://doi.org/10.1103/PhysRevResearch.2.033039} {\bibfield  {journal}
  {\bibinfo  {journal} {Phys. Rev. Res.}\ }\textbf {\bibinfo {volume} {2}},\
  \bibinfo {pages} {033039} (\bibinfo {year} {2020}{\natexlab{b}})},\ \Eprint
  {https://arxiv.org/abs/1812.07554} {arXiv:1812.07554 [hep-lat]} \BibitemShut
  {NoStop}%
\bibitem [{\citenamefont {Dasgupta}\ and\ \citenamefont
  {Raychowdhury}(2022)}]{Dasgupta:2020itb}%
  \BibitemOpen
  \bibfield  {author} {\bibinfo {author} {\bibfnamefont {R.}~\bibnamefont
  {Dasgupta}}\ and\ \bibinfo {author} {\bibfnamefont {I.}~\bibnamefont
  {Raychowdhury}},\ }\bibfield  {title} {\bibinfo {title} {{Cold-atom quantum
  simulator for string and hadron dynamics in non-Abelian lattice gauge
  theory}},\ }\href {https://doi.org/10.1103/PhysRevA.105.023322} {\bibfield
  {journal} {\bibinfo  {journal} {Phys. Rev. A}\ }\textbf {\bibinfo {volume}
  {105}},\ \bibinfo {pages} {023322} (\bibinfo {year} {2022})},\ \Eprint
  {https://arxiv.org/abs/2009.13969} {arXiv:2009.13969 [hep-lat]} \BibitemShut
  {NoStop}%
\bibitem [{\citenamefont {Kreshchuk}\ \emph {et~al.}(2022)\citenamefont
  {Kreshchuk}, \citenamefont {Kirby}, \citenamefont {Goldstein}, \citenamefont
  {Beauchemin},\ and\ \citenamefont {Love}}]{Kreshchuk:2020dla}%
  \BibitemOpen
  \bibfield  {author} {\bibinfo {author} {\bibfnamefont {M.}~\bibnamefont
  {Kreshchuk}}, \bibinfo {author} {\bibfnamefont {W.~M.}\ \bibnamefont
  {Kirby}}, \bibinfo {author} {\bibfnamefont {G.}~\bibnamefont {Goldstein}},
  \bibinfo {author} {\bibfnamefont {H.}~\bibnamefont {Beauchemin}},\ and\
  \bibinfo {author} {\bibfnamefont {P.~J.}\ \bibnamefont {Love}},\ }\bibfield
  {title} {\bibinfo {title} {{Quantum simulation of quantum field theory in the
  light-front formulation}},\ }\href
  {https://doi.org/10.1103/PhysRevA.105.032418} {\bibfield  {journal} {\bibinfo
   {journal} {Phys. Rev. A}\ }\textbf {\bibinfo {volume} {105}},\ \bibinfo
  {pages} {032418} (\bibinfo {year} {2022})},\ \Eprint
  {https://arxiv.org/abs/2002.04016} {arXiv:2002.04016 [quant-ph]} \BibitemShut
  {NoStop}%
\bibitem [{\citenamefont {Buser}\ \emph {et~al.}(2021)\citenamefont {Buser},
  \citenamefont {Gharibyan}, \citenamefont {Hanada}, \citenamefont {Honda},\
  and\ \citenamefont {Liu}}]{Buser:2020cvn}%
  \BibitemOpen
  \bibfield  {author} {\bibinfo {author} {\bibfnamefont {A.~J.}\ \bibnamefont
  {Buser}}, \bibinfo {author} {\bibfnamefont {H.}~\bibnamefont {Gharibyan}},
  \bibinfo {author} {\bibfnamefont {M.}~\bibnamefont {Hanada}}, \bibinfo
  {author} {\bibfnamefont {M.}~\bibnamefont {Honda}},\ and\ \bibinfo {author}
  {\bibfnamefont {J.}~\bibnamefont {Liu}},\ }\bibfield  {title} {\bibinfo
  {title} {{Quantum simulation of gauge theory via orbifold lattice}},\ }\href
  {https://doi.org/10.1007/JHEP09(2021)034} {\bibfield  {journal} {\bibinfo
  {journal} {JHEP}\ }\textbf {\bibinfo {volume} {09}},\ \bibinfo {pages}
  {034}},\ \Eprint {https://arxiv.org/abs/2011.06576} {arXiv:2011.06576
  [hep-th]} \BibitemShut {NoStop}%
\bibitem [{\citenamefont {Kan}\ and\ \citenamefont {Nam}(2023)}]{Kan:2022esj}%
  \BibitemOpen
  \bibfield  {author} {\bibinfo {author} {\bibfnamefont {A.}~\bibnamefont
  {Kan}}\ and\ \bibinfo {author} {\bibfnamefont {Y.}~\bibnamefont {Nam}},\
  }\bibfield  {title} {\bibinfo {title} {{Simulating lattice quantum
  electrodynamics on a quantum computer}},\ }\href
  {https://doi.org/10.1088/2058-9565/aca0b8} {\bibfield  {journal} {\bibinfo
  {journal} {Quantum Sci. Technol.}\ }\textbf {\bibinfo {volume} {8}},\
  \bibinfo {pages} {015008} (\bibinfo {year} {2023})}\BibitemShut {NoStop}%
\bibitem [{\citenamefont {D'Andrea}\ \emph {et~al.}(2024)\citenamefont
  {D'Andrea}, \citenamefont {Bauer}, \citenamefont {Grabowska},\ and\
  \citenamefont {Freytsis}}]{DAndrea:2023qnr}%
  \BibitemOpen
  \bibfield  {author} {\bibinfo {author} {\bibfnamefont {I.}~\bibnamefont
  {D'Andrea}}, \bibinfo {author} {\bibfnamefont {C.~W.}\ \bibnamefont {Bauer}},
  \bibinfo {author} {\bibfnamefont {D.~M.}\ \bibnamefont {Grabowska}},\ and\
  \bibinfo {author} {\bibfnamefont {M.}~\bibnamefont {Freytsis}},\ }\bibfield
  {title} {\bibinfo {title} {{New basis for Hamiltonian SU(2) simulations}},\
  }\href {https://doi.org/10.1103/PhysRevD.109.074501} {\bibfield  {journal}
  {\bibinfo  {journal} {Phys. Rev. D}\ }\textbf {\bibinfo {volume} {109}},\
  \bibinfo {pages} {074501} (\bibinfo {year} {2024})},\ \Eprint
  {https://arxiv.org/abs/2307.11829} {arXiv:2307.11829 [hep-ph]} \BibitemShut
  {NoStop}%
\bibitem [{\citenamefont {Sukeno}\ and\ \citenamefont
  {Okuda}(2024)}]{Sukeno:2023uhx}%
  \BibitemOpen
  \bibfield  {author} {\bibinfo {author} {\bibfnamefont {H.}~\bibnamefont
  {Sukeno}}\ and\ \bibinfo {author} {\bibfnamefont {T.}~\bibnamefont {Okuda}},\
  }\bibfield  {title} {\bibinfo {title} {{Measurement-based quantum simulation
  of Abelian lattice gauge theories}},\ }\href
  {https://doi.org/10.22323/1.453.0232} {\bibfield  {journal} {\bibinfo
  {journal} {PoS}\ }\textbf {\bibinfo {volume} {LATTICE2023}},\ \bibinfo
  {pages} {232} (\bibinfo {year} {2024})},\ \Eprint
  {https://arxiv.org/abs/2401.10259} {arXiv:2401.10259 [hep-lat]} \BibitemShut
  {NoStop}%
\bibitem [{\citenamefont {Bauer}\ \emph {et~al.}(2023)\citenamefont {Bauer},
  \citenamefont {Davoudi}, \citenamefont {Klco},\ and\ \citenamefont
  {Savage}}]{Bauer:2023qgm}%
  \BibitemOpen
  \bibfield  {author} {\bibinfo {author} {\bibfnamefont {C.~W.}\ \bibnamefont
  {Bauer}}, \bibinfo {author} {\bibfnamefont {Z.}~\bibnamefont {Davoudi}},
  \bibinfo {author} {\bibfnamefont {N.}~\bibnamefont {Klco}},\ and\ \bibinfo
  {author} {\bibfnamefont {M.~J.}\ \bibnamefont {Savage}},\ }\bibfield  {title}
  {\bibinfo {title} {{Quantum simulation of fundamental particles and
  forces}},\ }\href {https://doi.org/10.1038/s42254-023-00599-8} {\bibfield
  {journal} {\bibinfo  {journal} {Nature Rev. Phys.}\ }\textbf {\bibinfo
  {volume} {5}},\ \bibinfo {pages} {420} (\bibinfo {year} {2023})},\ \Eprint
  {https://arxiv.org/abs/2404.06298} {arXiv:2404.06298 [hep-ph]} \BibitemShut
  {NoStop}%
\bibitem [{\citenamefont {Feynman}(1982)}]{Feynman:1981tf}%
  \BibitemOpen
  \bibfield  {author} {\bibinfo {author} {\bibfnamefont {R.~P.}\ \bibnamefont
  {Feynman}},\ }\bibfield  {title} {\bibinfo {title} {{Simulating physics with
  computers}},\ }\href {https://doi.org/10.1007/BF02650179} {\bibfield
  {journal} {\bibinfo  {journal} {Int. J. Theor. Phys.}\ }\textbf {\bibinfo
  {volume} {21}},\ \bibinfo {pages} {467} (\bibinfo {year} {1982})}\BibitemShut
  {NoStop}%
\bibitem [{\citenamefont {Lloyd}(1996)}]{Lloyd:1996aai}%
  \BibitemOpen
  \bibfield  {author} {\bibinfo {author} {\bibfnamefont {S.}~\bibnamefont
  {Lloyd}},\ }\bibfield  {title} {\bibinfo {title} {{Universal Quantum
  Simulators}},\ }\href {https://doi.org/10.1126/science.273.5278.1073}
  {\bibfield  {journal} {\bibinfo  {journal} {Science}\ }\textbf {\bibinfo
  {volume} {273}},\ \bibinfo {pages} {1073} (\bibinfo {year}
  {1996})}\BibitemShut {NoStop}%
\bibitem [{\citenamefont {Arute}\ \emph {et~al.}(2019)\citenamefont {Arute}
  \emph {et~al.}}]{Arute:2019zxq}%
  \BibitemOpen
  \bibfield  {author} {\bibinfo {author} {\bibfnamefont {F.}~\bibnamefont
  {Arute}} \emph {et~al.},\ }\bibfield  {title} {\bibinfo {title} {{Quantum
  supremacy using a programmable superconducting processor}},\ }\href
  {https://doi.org/10.1038/s41586-019-1666-5} {\bibfield  {journal} {\bibinfo
  {journal} {Nature}\ }\textbf {\bibinfo {volume} {574}},\ \bibinfo {pages}
  {505} (\bibinfo {year} {2019})},\ \Eprint {https://arxiv.org/abs/1910.11333}
  {arXiv:1910.11333 [quant-ph]} \BibitemShut {NoStop}%
\bibitem [{\citenamefont {Chow}\ \emph {et~al.}(2012)\citenamefont {Chow} \emph
  {et~al.}}]{Chow:2012zps}%
  \BibitemOpen
  \bibfield  {author} {\bibinfo {author} {\bibfnamefont {J.~M.}\ \bibnamefont
  {Chow}} \emph {et~al.},\ }\bibfield  {title} {\bibinfo {title} {{Universal
  Quantum Gate Set Approaching Fault-Tolerant Thresholds with Superconducting
  Qubits}},\ }\href {https://doi.org/10.1103/PhysRevLett.109.060501} {\bibfield
   {journal} {\bibinfo  {journal} {Phys. Rev. Lett.}\ }\textbf {\bibinfo
  {volume} {109}},\ \bibinfo {pages} {060501} (\bibinfo {year}
  {2012})}\BibitemShut {NoStop}%
\bibitem [{\citenamefont {Debnath}\ \emph {et~al.}(2016)\citenamefont
  {Debnath}, \citenamefont {Linke}, \citenamefont {Figgatt}, \citenamefont
  {Landsman}, \citenamefont {Wright},\ and\ \citenamefont
  {Monroe}}]{Debnath:2016xdi}%
  \BibitemOpen
  \bibfield  {author} {\bibinfo {author} {\bibfnamefont {S.}~\bibnamefont
  {Debnath}}, \bibinfo {author} {\bibfnamefont {N.~M.}\ \bibnamefont {Linke}},
  \bibinfo {author} {\bibfnamefont {C.}~\bibnamefont {Figgatt}}, \bibinfo
  {author} {\bibfnamefont {K.~A.}\ \bibnamefont {Landsman}}, \bibinfo {author}
  {\bibfnamefont {K.}~\bibnamefont {Wright}},\ and\ \bibinfo {author}
  {\bibfnamefont {C.}~\bibnamefont {Monroe}},\ }\bibfield  {title} {\bibinfo
  {title} {{Demonstration of a small programmable quantum computer with atomic
  qubits}},\ }\href {https://doi.org/10.1038/nature18648} {\bibfield  {journal}
  {\bibinfo  {journal} {Nature}\ }\textbf {\bibinfo {volume} {536}},\ \bibinfo
  {pages} {63} (\bibinfo {year} {2016})}\BibitemShut {NoStop}%
\bibitem [{\citenamefont {Reagor}\ \emph {et~al.}(2018)\citenamefont {Reagor}
  \emph {et~al.}}]{Reagor:2018csq}%
  \BibitemOpen
  \bibfield  {author} {\bibinfo {author} {\bibfnamefont {M.}~\bibnamefont
  {Reagor}} \emph {et~al.},\ }\bibfield  {title} {\bibinfo {title}
  {{Demonstration of universal parametric entangling gates on a multi-qubit
  lattice}},\ }\href {https://doi.org/10.1126/sciadv.aao3603} {\bibfield
  {journal} {\bibinfo  {journal} {Sci. Adv.}\ }\textbf {\bibinfo {volume}
  {4}},\ \bibinfo {pages} {aao3603} (\bibinfo {year} {2018})}\BibitemShut
  {NoStop}%
\bibitem [{\citenamefont {Wiese}(2014)}]{Wiese:2014rla}%
  \BibitemOpen
  \bibfield  {author} {\bibinfo {author} {\bibfnamefont {U.-J.}\ \bibnamefont
  {Wiese}},\ }\bibfield  {title} {\bibinfo {title} {{Towards Quantum Simulating
  QCD}},\ }\href {https://doi.org/10.1016/j.nuclphysa.2014.09.102} {\bibfield
  {journal} {\bibinfo  {journal} {Nucl. Phys. A}\ }\textbf {\bibinfo {volume}
  {931}},\ \bibinfo {pages} {246} (\bibinfo {year} {2014})},\ \Eprint
  {https://arxiv.org/abs/1409.7414} {arXiv:1409.7414 [hep-th]} \BibitemShut
  {NoStop}%
\bibitem [{\citenamefont {Zohar}\ \emph
  {et~al.}(2017{\natexlab{b}})\citenamefont {Zohar}, \citenamefont {Farace},
  \citenamefont {Reznik},\ and\ \citenamefont {Cirac}}]{Zohar:2016iic}%
  \BibitemOpen
  \bibfield  {author} {\bibinfo {author} {\bibfnamefont {E.}~\bibnamefont
  {Zohar}}, \bibinfo {author} {\bibfnamefont {A.}~\bibnamefont {Farace}},
  \bibinfo {author} {\bibfnamefont {B.}~\bibnamefont {Reznik}},\ and\ \bibinfo
  {author} {\bibfnamefont {J.~I.}\ \bibnamefont {Cirac}},\ }\bibfield  {title}
  {\bibinfo {title} {{Digital lattice gauge theories}},\ }\href
  {https://doi.org/10.1103/PhysRevA.95.023604} {\bibfield  {journal} {\bibinfo
  {journal} {Phys. Rev. A}\ }\textbf {\bibinfo {volume} {95}},\ \bibinfo
  {pages} {023604} (\bibinfo {year} {2017}{\natexlab{b}})},\ \Eprint
  {https://arxiv.org/abs/1607.08121} {arXiv:1607.08121 [quant-ph]} \BibitemShut
  {NoStop}%
\bibitem [{\citenamefont {Kaplan}\ and\ \citenamefont
  {Stryker}(2020)}]{Kaplan:2018vnj}%
  \BibitemOpen
  \bibfield  {author} {\bibinfo {author} {\bibfnamefont {D.~B.}\ \bibnamefont
  {Kaplan}}\ and\ \bibinfo {author} {\bibfnamefont {J.~R.}\ \bibnamefont
  {Stryker}},\ }\bibfield  {title} {\bibinfo {title} {{Gauss\textquoteright{}s
  law, duality, and the Hamiltonian formulation of U(1) lattice gauge
  theory}},\ }\href {https://doi.org/10.1103/PhysRevD.102.094515} {\bibfield
  {journal} {\bibinfo  {journal} {Phys. Rev. D}\ }\textbf {\bibinfo {volume}
  {102}},\ \bibinfo {pages} {094515} (\bibinfo {year} {2020})},\ \Eprint
  {https://arxiv.org/abs/1806.08797} {arXiv:1806.08797 [hep-lat]} \BibitemShut
  {NoStop}%
\bibitem [{\citenamefont {Meurice}\ \emph
  {et~al.}(2022{\natexlab{a}})\citenamefont {Meurice}, \citenamefont {Sakai},\
  and\ \citenamefont {Unmuth-Yockey}}]{Meurice:2020pxc}%
  \BibitemOpen
  \bibfield  {author} {\bibinfo {author} {\bibfnamefont {Y.}~\bibnamefont
  {Meurice}}, \bibinfo {author} {\bibfnamefont {R.}~\bibnamefont {Sakai}},\
  and\ \bibinfo {author} {\bibfnamefont {J.}~\bibnamefont {Unmuth-Yockey}},\
  }\bibfield  {title} {\bibinfo {title} {{Tensor lattice field theory for
  renormalization and quantum computing}},\ }\href
  {https://doi.org/10.1103/RevModPhys.94.025005} {\bibfield  {journal}
  {\bibinfo  {journal} {Rev. Mod. Phys.}\ }\textbf {\bibinfo {volume} {94}},\
  \bibinfo {pages} {025005} (\bibinfo {year} {2022}{\natexlab{a}})},\ \Eprint
  {https://arxiv.org/abs/2010.06539} {arXiv:2010.06539 [hep-lat]} \BibitemShut
  {NoStop}%
\bibitem [{\citenamefont {Meurice}\ \emph
  {et~al.}(2022{\natexlab{b}})\citenamefont {Meurice}, \citenamefont {Bazavov},
  \citenamefont {Dreher}, \citenamefont {Gustafson}, \citenamefont {Hostetler},
  \citenamefont {Sakai}, \citenamefont {Tsai}, \citenamefont {Unmuth-Yockey},\
  and\ \citenamefont {Zhang}}]{Meurice:2021ujn}%
  \BibitemOpen
  \bibfield  {author} {\bibinfo {author} {\bibfnamefont {Y.}~\bibnamefont
  {Meurice}}, \bibinfo {author} {\bibfnamefont {A.}~\bibnamefont {Bazavov}},
  \bibinfo {author} {\bibfnamefont {P.}~\bibnamefont {Dreher}}, \bibinfo
  {author} {\bibfnamefont {E.}~\bibnamefont {Gustafson}}, \bibinfo {author}
  {\bibfnamefont {L.}~\bibnamefont {Hostetler}}, \bibinfo {author}
  {\bibfnamefont {R.}~\bibnamefont {Sakai}}, \bibinfo {author} {\bibfnamefont
  {S.-W.}\ \bibnamefont {Tsai}}, \bibinfo {author} {\bibfnamefont
  {J.}~\bibnamefont {Unmuth-Yockey}},\ and\ \bibinfo {author} {\bibfnamefont
  {J.}~\bibnamefont {Zhang}},\ }\bibfield  {title} {\bibinfo {title} {{From
  tensors to qubits}},\ }\href {https://doi.org/10.22323/1.396.0608} {\bibfield
   {journal} {\bibinfo  {journal} {PoS}\ }\textbf {\bibinfo {volume}
  {LATTICE2021}},\ \bibinfo {pages} {608} (\bibinfo {year}
  {2022}{\natexlab{b}})},\ \Eprint {https://arxiv.org/abs/2112.10005}
  {arXiv:2112.10005 [hep-lat]} \BibitemShut {NoStop}%
\bibitem [{\citenamefont {Ciavarella}\ \emph {et~al.}(2022)\citenamefont
  {Ciavarella}, \citenamefont {Klco},\ and\ \citenamefont
  {Savage}}]{Ciavarella:2022zhe}%
  \BibitemOpen
  \bibfield  {author} {\bibinfo {author} {\bibfnamefont {A.}~\bibnamefont
  {Ciavarella}}, \bibinfo {author} {\bibfnamefont {N.}~\bibnamefont {Klco}},\
  and\ \bibinfo {author} {\bibfnamefont {M.~J.}\ \bibnamefont {Savage}},\
  }\bibfield  {title} {\bibinfo {title} {{Some Conceptual Aspects of Operator
  Design for Quantum Simulations of Non-Abelian Lattice Gauge Theories}}\
  }(\bibinfo {year} {2022})\ \Eprint {https://arxiv.org/abs/2203.11988}
  {arXiv:2203.11988 [quant-ph]} \BibitemShut {NoStop}%
\bibitem [{\citenamefont {Irmejs}\ \emph {et~al.}(2023)\citenamefont {Irmejs},
  \citenamefont {Banuls},\ and\ \citenamefont {Cirac}}]{Irmejs:2022gwv}%
  \BibitemOpen
  \bibfield  {author} {\bibinfo {author} {\bibfnamefont {R.}~\bibnamefont
  {Irmejs}}, \bibinfo {author} {\bibfnamefont {M.~C.}\ \bibnamefont {Banuls}},\
  and\ \bibinfo {author} {\bibfnamefont {J.~I.}\ \bibnamefont {Cirac}},\
  }\bibfield  {title} {\bibinfo {title} {{Quantum simulation of Z2 lattice
  gauge theory with minimal resources}},\ }\href
  {https://doi.org/10.1103/PhysRevD.108.074503} {\bibfield  {journal} {\bibinfo
   {journal} {Phys. Rev. D}\ }\textbf {\bibinfo {volume} {108}},\ \bibinfo
  {pages} {074503} (\bibinfo {year} {2023})},\ \Eprint
  {https://arxiv.org/abs/2206.08909} {arXiv:2206.08909 [quant-ph]} \BibitemShut
  {NoStop}%
\bibitem [{\citenamefont {Kadam}\ \emph {et~al.}(2023)\citenamefont {Kadam},
  \citenamefont {Raychowdhury},\ and\ \citenamefont {Stryker}}]{Kadam:2022ipf}%
  \BibitemOpen
  \bibfield  {author} {\bibinfo {author} {\bibfnamefont {S.~V.}\ \bibnamefont
  {Kadam}}, \bibinfo {author} {\bibfnamefont {I.}~\bibnamefont
  {Raychowdhury}},\ and\ \bibinfo {author} {\bibfnamefont {J.~R.}\ \bibnamefont
  {Stryker}},\ }\bibfield  {title} {\bibinfo {title} {{Loop-string-hadron
  formulation of an SU(3) gauge theory with dynamical quarks}},\ }\href
  {https://doi.org/10.1103/PhysRevD.107.094513} {\bibfield  {journal} {\bibinfo
   {journal} {Phys. Rev. D}\ }\textbf {\bibinfo {volume} {107}},\ \bibinfo
  {pages} {094513} (\bibinfo {year} {2023})},\ \Eprint
  {https://arxiv.org/abs/2212.04490} {arXiv:2212.04490 [hep-lat]} \BibitemShut
  {NoStop}%
\bibitem [{\citenamefont {Calaj\`o}\ \emph {et~al.}(2024)\citenamefont
  {Calaj\`o}, \citenamefont {Magnifico}, \citenamefont {Edmunds}, \citenamefont
  {Ringbauer}, \citenamefont {Montangero},\ and\ \citenamefont
  {Silvi}}]{Calajo:2024qrc}%
  \BibitemOpen
  \bibfield  {author} {\bibinfo {author} {\bibfnamefont {G.}~\bibnamefont
  {Calaj\`o}}, \bibinfo {author} {\bibfnamefont {G.}~\bibnamefont {Magnifico}},
  \bibinfo {author} {\bibfnamefont {C.}~\bibnamefont {Edmunds}}, \bibinfo
  {author} {\bibfnamefont {M.}~\bibnamefont {Ringbauer}}, \bibinfo {author}
  {\bibfnamefont {S.}~\bibnamefont {Montangero}},\ and\ \bibinfo {author}
  {\bibfnamefont {P.}~\bibnamefont {Silvi}},\ }\bibfield  {title} {\bibinfo
  {title} {{Digital Quantum Simulation of a (1+1)D SU(2) Lattice Gauge Theory
  with Ion Qudits}},\ }\href {https://doi.org/10.1103/PRXQuantum.5.040309}
  {\bibfield  {journal} {\bibinfo  {journal} {PRX Quantum}\ }\textbf {\bibinfo
  {volume} {5}},\ \bibinfo {pages} {040309} (\bibinfo {year} {2024})},\ \Eprint
  {https://arxiv.org/abs/2402.07987} {arXiv:2402.07987 [quant-ph]} \BibitemShut
  {NoStop}%
\bibitem [{\citenamefont {Mariani}(2024)}]{Mariani:2024osg}%
  \BibitemOpen
  \bibfield  {author} {\bibinfo {author} {\bibfnamefont {A.}~\bibnamefont
  {Mariani}},\ }\bibfield  {title} {\bibinfo {title} {{Almost gauge-invariant
  states and the ground state of Yang-Mills theory}},\ }\href
  {https://doi.org/10.1103/PhysRevD.109.094508} {\bibfield  {journal} {\bibinfo
   {journal} {Phys. Rev. D}\ }\textbf {\bibinfo {volume} {109}},\ \bibinfo
  {pages} {094508} (\bibinfo {year} {2024})},\ \Eprint
  {https://arxiv.org/abs/2402.16743} {arXiv:2402.16743 [hep-lat]} \BibitemShut
  {NoStop}%
\bibitem [{\citenamefont {Carena}\ \emph {et~al.}(2024)\citenamefont {Carena},
  \citenamefont {Lamm}, \citenamefont {Li},\ and\ \citenamefont
  {Liu}}]{Carena:2024dzu}%
  \BibitemOpen
  \bibfield  {author} {\bibinfo {author} {\bibfnamefont {M.}~\bibnamefont
  {Carena}}, \bibinfo {author} {\bibfnamefont {H.}~\bibnamefont {Lamm}},
  \bibinfo {author} {\bibfnamefont {Y.-Y.}\ \bibnamefont {Li}},\ and\ \bibinfo
  {author} {\bibfnamefont {W.}~\bibnamefont {Liu}},\ }\bibfield  {title}
  {\bibinfo {title} {{Quantum error thresholds for gauge-redundant
  digitizations of lattice field theories}},\ }\href
  {https://doi.org/10.1103/PhysRevD.110.054516} {\bibfield  {journal} {\bibinfo
   {journal} {Phys. Rev. D}\ }\textbf {\bibinfo {volume} {110}},\ \bibinfo
  {pages} {054516} (\bibinfo {year} {2024})},\ \Eprint
  {https://arxiv.org/abs/2402.16780} {arXiv:2402.16780 [hep-lat]} \BibitemShut
  {NoStop}%
\bibitem [{\citenamefont {Mathew}\ and\ \citenamefont
  {Raychowdhury}(2024)}]{Mathew:2024bed}%
  \BibitemOpen
  \bibfield  {author} {\bibinfo {author} {\bibfnamefont {E.}~\bibnamefont
  {Mathew}}\ and\ \bibinfo {author} {\bibfnamefont {I.}~\bibnamefont
  {Raychowdhury}},\ }\bibfield  {title} {\bibinfo {title} {{Protecting gauge
  symmetries in the the dynamics of SU(3) lattice gauge theories}}\ }\href
  {https://doi.org/10.48550/arXiv.2404.12158} {10.48550/arXiv.2404.12158}
  (\bibinfo {year} {2024}),\ \Eprint {https://arxiv.org/abs/2404.12158}
  {arXiv:2404.12158 [hep-lat]} \BibitemShut {NoStop}%
\bibitem [{\citenamefont {Li}(2024)}]{Li:2024ide}%
  \BibitemOpen
  \bibfield  {author} {\bibinfo {author} {\bibfnamefont {T.}~\bibnamefont
  {Li}},\ }\bibfield  {title} {\bibinfo {title} {{Quantum simulations of
  quantum electrodynamics in Coulomb gauge}},\ }\href@noop {} {\  (\bibinfo
  {year} {2024})},\ \Eprint {https://arxiv.org/abs/2406.01204}
  {arXiv:2406.01204 [hep-lat]} \BibitemShut {NoStop}%
\bibitem [{\citenamefont {Kadam}\ \emph {et~al.}(2025)\citenamefont {Kadam},
  \citenamefont {Naskar}, \citenamefont {Raychowdhury},\ and\ \citenamefont
  {Stryker}}]{Kadam:2024zkj}%
  \BibitemOpen
  \bibfield  {author} {\bibinfo {author} {\bibfnamefont {S.~V.}\ \bibnamefont
  {Kadam}}, \bibinfo {author} {\bibfnamefont {A.}~\bibnamefont {Naskar}},
  \bibinfo {author} {\bibfnamefont {I.}~\bibnamefont {Raychowdhury}},\ and\
  \bibinfo {author} {\bibfnamefont {J.~R.}\ \bibnamefont {Stryker}},\
  }\bibfield  {title} {\bibinfo {title} {{Loop-string-hadron approach to SU(3)
  lattice Yang-Mills theory: Hilbert space of a trivalent vertex}},\ }\href
  {https://doi.org/10.1103/PhysRevD.111.074516} {\bibfield  {journal} {\bibinfo
   {journal} {Phys. Rev. D}\ }\textbf {\bibinfo {volume} {111}},\ \bibinfo
  {pages} {074516} (\bibinfo {year} {2025})},\ \Eprint
  {https://arxiv.org/abs/2407.19181} {arXiv:2407.19181 [hep-lat]} \BibitemShut
  {NoStop}%
\bibitem [{\citenamefont {Grabowska}\ \emph {et~al.}(2024)\citenamefont
  {Grabowska}, \citenamefont {Kane},\ and\ \citenamefont
  {Bauer}}]{Grabowska:2024emw}%
  \BibitemOpen
  \bibfield  {author} {\bibinfo {author} {\bibfnamefont {D.~M.}\ \bibnamefont
  {Grabowska}}, \bibinfo {author} {\bibfnamefont {C.~F.}\ \bibnamefont
  {Kane}},\ and\ \bibinfo {author} {\bibfnamefont {C.~W.}\ \bibnamefont
  {Bauer}},\ }\bibfield  {title} {\bibinfo {title} {{A Fully Gauge-Fixed SU(2)
  Hamiltonian for Quantum Simulations}},\ }\href@noop {} {\  (\bibinfo {year}
  {2024})},\ \Eprint {https://arxiv.org/abs/2409.10610} {arXiv:2409.10610
  [quant-ph]} \BibitemShut {NoStop}%
\bibitem [{\citenamefont {Ma}(2025)}]{Ma:2025ysk}%
  \BibitemOpen
  \bibfield  {author} {\bibinfo {author} {\bibfnamefont {R.}~\bibnamefont
  {Ma}},\ }\bibfield  {title} {\bibinfo {title} {{Packaged Quantum States for
  Quantum Simulation of Lattice Gauge Theories}},\ }\href@noop {} {\  (\bibinfo
  {year} {2025})},\ \Eprint {https://arxiv.org/abs/2502.14654}
  {arXiv:2502.14654 [quant-ph]} \BibitemShut {NoStop}%
\bibitem [{\citenamefont {Halimeh}\ and\ \citenamefont
  {Hauke}(2020)}]{Halimeh:2020kyu}%
  \BibitemOpen
  \bibfield  {author} {\bibinfo {author} {\bibfnamefont {J.~C.}\ \bibnamefont
  {Halimeh}}\ and\ \bibinfo {author} {\bibfnamefont {P.}~\bibnamefont
  {Hauke}},\ }\bibfield  {title} {\bibinfo {title} {{Staircase
  Prethermalization and Constrained Dynamics in Lattice Gauge Theories}},\
  }\href@noop {} {\  (\bibinfo {year} {2020})},\ \Eprint
  {https://arxiv.org/abs/2004.07248} {arXiv:2004.07248 [cond-mat.quant-gas]}
  \BibitemShut {NoStop}%
\bibitem [{\citenamefont {Jensen}\ \emph {et~al.}(2022)\citenamefont {Jensen},
  \citenamefont {Pedersen},\ and\ \citenamefont {Zinner}}]{Jensen:2022hyu}%
  \BibitemOpen
  \bibfield  {author} {\bibinfo {author} {\bibfnamefont {R.~B.}\ \bibnamefont
  {Jensen}}, \bibinfo {author} {\bibfnamefont {S.~P.}\ \bibnamefont
  {Pedersen}},\ and\ \bibinfo {author} {\bibfnamefont {N.~T.}\ \bibnamefont
  {Zinner}},\ }\bibfield  {title} {\bibinfo {title} {{Dynamical quantum phase
  transitions in a noisy lattice gauge theory}},\ }\href
  {https://doi.org/10.1103/PhysRevB.105.224309} {\bibfield  {journal} {\bibinfo
   {journal} {Phys. Rev. B}\ }\textbf {\bibinfo {volume} {105}},\ \bibinfo
  {pages} {224309} (\bibinfo {year} {2022})},\ \Eprint
  {https://arxiv.org/abs/2203.10927} {arXiv:2203.10927 [hep-lat]} \BibitemShut
  {NoStop}%
\bibitem [{\citenamefont {Gustafson}\ and\ \citenamefont
  {Lamm}(2023)}]{Gustafson:2023swx}%
  \BibitemOpen
  \bibfield  {author} {\bibinfo {author} {\bibfnamefont {E.~J.}\ \bibnamefont
  {Gustafson}}\ and\ \bibinfo {author} {\bibfnamefont {H.}~\bibnamefont
  {Lamm}},\ }\bibfield  {title} {\bibinfo {title} {{Robustness of Gauge
  Digitization to Quantum Noise}},\ }\href@noop {} {\  (\bibinfo {year}
  {2023})},\ \Eprint {https://arxiv.org/abs/2301.10207} {arXiv:2301.10207
  [hep-lat]} \BibitemShut {NoStop}%
\bibitem [{\citenamefont {Stannigel}\ \emph {et~al.}(2014)\citenamefont
  {Stannigel}, \citenamefont {Hauke}, \citenamefont {Marcos}, \citenamefont
  {Hafezi}, \citenamefont {Diehl}, \citenamefont {Dalmonte},\ and\
  \citenamefont {Zoller}}]{Stannigel:2013zka}%
  \BibitemOpen
  \bibfield  {author} {\bibinfo {author} {\bibfnamefont {K.}~\bibnamefont
  {Stannigel}}, \bibinfo {author} {\bibfnamefont {P.}~\bibnamefont {Hauke}},
  \bibinfo {author} {\bibfnamefont {D.}~\bibnamefont {Marcos}}, \bibinfo
  {author} {\bibfnamefont {M.}~\bibnamefont {Hafezi}}, \bibinfo {author}
  {\bibfnamefont {S.}~\bibnamefont {Diehl}}, \bibinfo {author} {\bibfnamefont
  {M.}~\bibnamefont {Dalmonte}},\ and\ \bibinfo {author} {\bibfnamefont
  {P.}~\bibnamefont {Zoller}},\ }\bibfield  {title} {\bibinfo {title}
  {{Constrained dynamics via the Zeno effect in quantum simulation:
  Implementing non-Abelian lattice gauge theories with cold atoms}},\ }\href
  {https://doi.org/10.1103/PhysRevLett.112.120406} {\bibfield  {journal}
  {\bibinfo  {journal} {Phys. Rev. Lett.}\ }\textbf {\bibinfo {volume} {112}},\
  \bibinfo {pages} {120406} (\bibinfo {year} {2014})},\ \Eprint
  {https://arxiv.org/abs/1308.0528} {arXiv:1308.0528 [quant-ph]} \BibitemShut
  {NoStop}%
\bibitem [{\citenamefont {Raimond}\ \emph {et~al.}(2010)\citenamefont
  {Raimond}, \citenamefont {Sayrin}, \citenamefont {Gleyzes}, \citenamefont
  {Dotsenko}, \citenamefont {Brune}, \citenamefont {Haroche}, \citenamefont
  {Facchi},\ and\ \citenamefont {Pascazio}}]{raimond2010phase}%
  \BibitemOpen
  \bibfield  {author} {\bibinfo {author} {\bibfnamefont {J.-M.}\ \bibnamefont
  {Raimond}}, \bibinfo {author} {\bibfnamefont {C.}~\bibnamefont {Sayrin}},
  \bibinfo {author} {\bibfnamefont {S.}~\bibnamefont {Gleyzes}}, \bibinfo
  {author} {\bibfnamefont {I.}~\bibnamefont {Dotsenko}}, \bibinfo {author}
  {\bibfnamefont {M.}~\bibnamefont {Brune}}, \bibinfo {author} {\bibfnamefont
  {S.}~\bibnamefont {Haroche}}, \bibinfo {author} {\bibfnamefont
  {P.}~\bibnamefont {Facchi}},\ and\ \bibinfo {author} {\bibfnamefont
  {S.}~\bibnamefont {Pascazio}},\ }\bibfield  {title} {\bibinfo {title} {Phase
  space tweezers for tailoring cavity fields by quantum zeno dynamics},\
  }\href@noop {} {\bibfield  {journal} {\bibinfo  {journal} {Physical review
  letters}\ }\textbf {\bibinfo {volume} {105}},\ \bibinfo {pages} {213601}
  (\bibinfo {year} {2010})},\ \Eprint {https://arxiv.org/abs/1007.4942}
  {arXiv:1007.4942 [quant-ph]} \BibitemShut {NoStop}%
\bibitem [{\citenamefont {Raimond}\ \emph {et~al.}(2012)\citenamefont
  {Raimond}, \citenamefont {Facchi}, \citenamefont {Peaudecerf}, \citenamefont
  {Pascazio}, \citenamefont {Sayrin}, \citenamefont {Dotsenko}, \citenamefont
  {Gleyzes}, \citenamefont {Brune},\ and\ \citenamefont
  {Haroche}}]{raimond2012quantum}%
  \BibitemOpen
  \bibfield  {author} {\bibinfo {author} {\bibfnamefont {J.-M.}\ \bibnamefont
  {Raimond}}, \bibinfo {author} {\bibfnamefont {P.}~\bibnamefont {Facchi}},
  \bibinfo {author} {\bibfnamefont {B.}~\bibnamefont {Peaudecerf}}, \bibinfo
  {author} {\bibfnamefont {S.}~\bibnamefont {Pascazio}}, \bibinfo {author}
  {\bibfnamefont {C.}~\bibnamefont {Sayrin}}, \bibinfo {author} {\bibfnamefont
  {I.}~\bibnamefont {Dotsenko}}, \bibinfo {author} {\bibfnamefont
  {S.}~\bibnamefont {Gleyzes}}, \bibinfo {author} {\bibfnamefont
  {M.}~\bibnamefont {Brune}},\ and\ \bibinfo {author} {\bibfnamefont
  {S.}~\bibnamefont {Haroche}},\ }\bibfield  {title} {\bibinfo {title} {Quantum
  zeno dynamics of a field in a cavity},\ }\href@noop {} {\bibfield  {journal}
  {\bibinfo  {journal} {Physical Review A—Atomic, Molecular, and Optical
  Physics}\ }\textbf {\bibinfo {volume} {86}},\ \bibinfo {pages} {032120}
  (\bibinfo {year} {2012})},\ \Eprint {https://arxiv.org/abs/1207.6499}
  {arXiv:1207.6499 [quant-ph]} \BibitemShut {NoStop}%
\bibitem [{\citenamefont {Signoles}\ \emph {et~al.}(2014)\citenamefont
  {Signoles}, \citenamefont {Facon}, \citenamefont {Grosso}, \citenamefont
  {Dotsenko}, \citenamefont {Haroche}, \citenamefont {Raimond}, \citenamefont
  {Brune},\ and\ \citenamefont {Gleyzes}}]{signoles2014confined}%
  \BibitemOpen
  \bibfield  {author} {\bibinfo {author} {\bibfnamefont {A.}~\bibnamefont
  {Signoles}}, \bibinfo {author} {\bibfnamefont {A.}~\bibnamefont {Facon}},
  \bibinfo {author} {\bibfnamefont {D.}~\bibnamefont {Grosso}}, \bibinfo
  {author} {\bibfnamefont {I.}~\bibnamefont {Dotsenko}}, \bibinfo {author}
  {\bibfnamefont {S.}~\bibnamefont {Haroche}}, \bibinfo {author} {\bibfnamefont
  {J.-M.}\ \bibnamefont {Raimond}}, \bibinfo {author} {\bibfnamefont
  {M.}~\bibnamefont {Brune}},\ and\ \bibinfo {author} {\bibfnamefont
  {S.}~\bibnamefont {Gleyzes}},\ }\bibfield  {title} {\bibinfo {title}
  {Confined quantum zeno dynamics of a watched atomic arrow},\ }\href@noop {}
  {\bibfield  {journal} {\bibinfo  {journal} {Nature Physics}\ }\textbf
  {\bibinfo {volume} {10}},\ \bibinfo {pages} {715} (\bibinfo {year} {2014})},\
  \Eprint {https://arxiv.org/abs/1402.0111} {arXiv:1402.0111 [quant-ph]}
  \BibitemShut {NoStop}%
\bibitem [{\citenamefont {Halimeh}\ \emph {et~al.}(2021)\citenamefont
  {Halimeh}, \citenamefont {Lang}, \citenamefont {Mildenberger}, \citenamefont
  {Jiang},\ and\ \citenamefont {Hauke}}]{Halimeh:2020ecg}%
  \BibitemOpen
  \bibfield  {author} {\bibinfo {author} {\bibfnamefont {J.~C.}\ \bibnamefont
  {Halimeh}}, \bibinfo {author} {\bibfnamefont {H.}~\bibnamefont {Lang}},
  \bibinfo {author} {\bibfnamefont {J.}~\bibnamefont {Mildenberger}}, \bibinfo
  {author} {\bibfnamefont {Z.}~\bibnamefont {Jiang}},\ and\ \bibinfo {author}
  {\bibfnamefont {P.}~\bibnamefont {Hauke}},\ }\bibfield  {title} {\bibinfo
  {title} {{Gauge-Symmetry Protection Using Single-Body Terms}},\ }\href
  {https://doi.org/10.1103/PRXQuantum.2.040311} {\bibfield  {journal} {\bibinfo
   {journal} {PRX Quantum}\ }\textbf {\bibinfo {volume} {2}},\ \bibinfo {pages}
  {040311} (\bibinfo {year} {2021})},\ \Eprint
  {https://arxiv.org/abs/2007.00668} {arXiv:2007.00668 [quant-ph]} \BibitemShut
  {NoStop}%
\bibitem [{\citenamefont {Halimeh}\ \emph {et~al.}(2022)\citenamefont
  {Halimeh}, \citenamefont {Homeier}, \citenamefont {Schweizer}, \citenamefont
  {Aidelsburger}, \citenamefont {Hauke},\ and\ \citenamefont
  {Grusdt}}]{Halimeh:2021lnv}%
  \BibitemOpen
  \bibfield  {author} {\bibinfo {author} {\bibfnamefont {J.~C.}\ \bibnamefont
  {Halimeh}}, \bibinfo {author} {\bibfnamefont {L.}~\bibnamefont {Homeier}},
  \bibinfo {author} {\bibfnamefont {C.}~\bibnamefont {Schweizer}}, \bibinfo
  {author} {\bibfnamefont {M.}~\bibnamefont {Aidelsburger}}, \bibinfo {author}
  {\bibfnamefont {P.}~\bibnamefont {Hauke}},\ and\ \bibinfo {author}
  {\bibfnamefont {F.}~\bibnamefont {Grusdt}},\ }\bibfield  {title} {\bibinfo
  {title} {{Stabilizing lattice gauge theories through simplified local
  pseudogenerators}},\ }\href
  {https://doi.org/10.1103/PhysRevResearch.4.033120} {\bibfield  {journal}
  {\bibinfo  {journal} {Phys. Rev. Res.}\ }\textbf {\bibinfo {volume} {4}},\
  \bibinfo {pages} {033120} (\bibinfo {year} {2022})},\ \Eprint
  {https://arxiv.org/abs/2108.02203} {arXiv:2108.02203 [cond-mat.quant-gas]}
  \BibitemShut {NoStop}%
\bibitem [{\citenamefont {Van~Damme}\ \emph {et~al.}(2025)\citenamefont
  {Van~Damme}, \citenamefont {Mildenberger}, \citenamefont {Grusdt},
  \citenamefont {Hauke},\ and\ \citenamefont {Halimeh}}]{VanDamme:2021njp}%
  \BibitemOpen
  \bibfield  {author} {\bibinfo {author} {\bibfnamefont {M.}~\bibnamefont
  {Van~Damme}}, \bibinfo {author} {\bibfnamefont {J.}~\bibnamefont
  {Mildenberger}}, \bibinfo {author} {\bibfnamefont {F.}~\bibnamefont
  {Grusdt}}, \bibinfo {author} {\bibfnamefont {P.}~\bibnamefont {Hauke}},\ and\
  \bibinfo {author} {\bibfnamefont {J.~C.}\ \bibnamefont {Halimeh}},\
  }\bibfield  {title} {\bibinfo {title} {{Suppressing nonperturbative gauge
  errors in the thermodynamic limit using local pseudogenerators}},\ }\href
  {https://doi.org/10.1038/s42005-025-02035-y} {\bibfield  {journal} {\bibinfo
  {journal} {Commun. Phys.}\ }\textbf {\bibinfo {volume} {8}},\ \bibinfo
  {pages} {106} (\bibinfo {year} {2025})},\ \Eprint
  {https://arxiv.org/abs/2110.08041} {arXiv:2110.08041 [quant-ph]} \BibitemShut
  {NoStop}%
\bibitem [{\citenamefont {Halimeh}\ and\ \citenamefont
  {Hauke}(2022)}]{Halimeh:2022mct}%
  \BibitemOpen
  \bibfield  {author} {\bibinfo {author} {\bibfnamefont {J.~C.}\ \bibnamefont
  {Halimeh}}\ and\ \bibinfo {author} {\bibfnamefont {P.}~\bibnamefont
  {Hauke}},\ }\bibfield  {title} {\bibinfo {title} {{Stabilizing Gauge Theories
  in Quantum Simulators: A Brief Review}}\ }(\bibinfo {year} {2022})\ \Eprint
  {https://arxiv.org/abs/2204.13709} {arXiv:2204.13709 [cond-mat.quant-gas]}
  \BibitemShut {NoStop}%
\bibitem [{\citenamefont {Ball}\ and\ \citenamefont
  {Cohen}(2024)}]{Ball:2024xmw}%
  \BibitemOpen
  \bibfield  {author} {\bibinfo {author} {\bibfnamefont {C.}~\bibnamefont
  {Ball}}\ and\ \bibinfo {author} {\bibfnamefont {T.~D.}\ \bibnamefont
  {Cohen}},\ }\bibfield  {title} {\bibinfo {title} {{Zeno effect suppression of
  gauge drift in quantum simulations}},\ }\href
  {https://doi.org/10.1103/PhysRevA.110.022417} {\bibfield  {journal} {\bibinfo
   {journal} {Phys. Rev. A}\ }\textbf {\bibinfo {volume} {110}},\ \bibinfo
  {pages} {022417} (\bibinfo {year} {2024})},\ \Eprint
  {https://arxiv.org/abs/2405.09462} {arXiv:2405.09462 [hep-lat]} \BibitemShut
  {NoStop}%
\bibitem [{\citenamefont {Wauters}\ \emph {et~al.}(2025)\citenamefont
  {Wauters}, \citenamefont {Ballini}, \citenamefont {Biella},\ and\
  \citenamefont {Hauke}}]{Wauters:2024shc}%
  \BibitemOpen
  \bibfield  {author} {\bibinfo {author} {\bibfnamefont {M.~M.}\ \bibnamefont
  {Wauters}}, \bibinfo {author} {\bibfnamefont {E.}~\bibnamefont {Ballini}},
  \bibinfo {author} {\bibfnamefont {A.}~\bibnamefont {Biella}},\ and\ \bibinfo
  {author} {\bibfnamefont {P.}~\bibnamefont {Hauke}},\ }\bibfield  {title}
  {\bibinfo {title} {{Symmetry-protection Zeno phase transition in monitored
  lattice gauge theories}},\ }\href
  {https://doi.org/10.1103/PhysRevB.111.094315} {\bibfield  {journal} {\bibinfo
   {journal} {Phys. Rev. B}\ }\textbf {\bibinfo {volume} {111}},\ \bibinfo
  {pages} {094315} (\bibinfo {year} {2025})},\ \Eprint
  {https://arxiv.org/abs/2405.18504} {arXiv:2405.18504 [quant-ph]} \BibitemShut
  {NoStop}%
\bibitem [{\citenamefont {Ball}(2025)}]{Ball:2024uqu}%
  \BibitemOpen
  \bibfield  {author} {\bibinfo {author} {\bibfnamefont {C.}~\bibnamefont
  {Ball}},\ }\bibfield  {title} {\bibinfo {title} {{Suppressing gauge drift in
  quantum simulations with gauge transformations}},\ }\href
  {https://doi.org/10.1103/PhysRevA.111.042428} {\bibfield  {journal} {\bibinfo
   {journal} {Phys. Rev. A}\ }\textbf {\bibinfo {volume} {111}},\ \bibinfo
  {pages} {042428} (\bibinfo {year} {2025})},\ \Eprint
  {https://arxiv.org/abs/2409.04395} {arXiv:2409.04395 [hep-lat]} \BibitemShut
  {NoStop}%
\bibitem [{\citenamefont {Surace}\ \emph {et~al.}(2023)\citenamefont {Surace},
  \citenamefont {Fromholz}, \citenamefont {Darkwah~Oppong}, \citenamefont
  {Dalmonte},\ and\ \citenamefont {Aidelsburger}}]{Surace:2023ycp}%
  \BibitemOpen
  \bibfield  {author} {\bibinfo {author} {\bibfnamefont {F.~M.}\ \bibnamefont
  {Surace}}, \bibinfo {author} {\bibfnamefont {P.}~\bibnamefont {Fromholz}},
  \bibinfo {author} {\bibfnamefont {N.}~\bibnamefont {Darkwah~Oppong}},
  \bibinfo {author} {\bibfnamefont {M.}~\bibnamefont {Dalmonte}},\ and\
  \bibinfo {author} {\bibfnamefont {M.}~\bibnamefont {Aidelsburger}},\
  }\bibfield  {title} {\bibinfo {title} {{Ab Initio Derivation of
  Lattice-Gauge-Theory Dynamics for Cold Gases in Optical Lattices}},\ }\href
  {https://doi.org/10.1103/PRXQuantum.4.020330} {\bibfield  {journal} {\bibinfo
   {journal} {PRX Quantum}\ }\textbf {\bibinfo {volume} {4}},\ \bibinfo {pages}
  {020330} (\bibinfo {year} {2023})},\ \Eprint
  {https://arxiv.org/abs/2301.03474} {arXiv:2301.03474 [cond-mat.quant-gas]}
  \BibitemShut {NoStop}%
\bibitem [{\citenamefont {Cheng}\ and\ \citenamefont
  {Zhang}(2025)}]{Cheng:2024pdu}%
  \BibitemOpen
  \bibfield  {author} {\bibinfo {author} {\bibfnamefont {L.-X.}\ \bibnamefont
  {Cheng}}\ and\ \bibinfo {author} {\bibfnamefont {D.-B.}\ \bibnamefont
  {Zhang}},\ }\bibfield  {title} {\bibinfo {title} {{Variational quantum
  simulation of ground states and thermal states for lattice gauge theory with
  multi-objective optimization}},\ }\href
  {https://doi.org/10.1016/j.physleta.2025.130516} {\bibfield  {journal}
  {\bibinfo  {journal} {Phys. Lett. A}\ }\textbf {\bibinfo {volume} {546}},\
  \bibinfo {pages} {130516} (\bibinfo {year} {2025})},\ \Eprint
  {https://arxiv.org/abs/2408.17300} {arXiv:2408.17300 [quant-ph]} \BibitemShut
  {NoStop}%
\bibitem [{\citenamefont {De~Paciani}\ \emph {et~al.}(2025)\citenamefont
  {De~Paciani}, \citenamefont {Homeier}, \citenamefont {Halimeh}, \citenamefont
  {Aidelsburger},\ and\ \citenamefont {Grusdt}}]{DePaciani:2025uzj}%
  \BibitemOpen
  \bibfield  {author} {\bibinfo {author} {\bibfnamefont {G.}~\bibnamefont
  {De~Paciani}}, \bibinfo {author} {\bibfnamefont {L.}~\bibnamefont {Homeier}},
  \bibinfo {author} {\bibfnamefont {J.~C.}\ \bibnamefont {Halimeh}}, \bibinfo
  {author} {\bibfnamefont {M.}~\bibnamefont {Aidelsburger}},\ and\ \bibinfo
  {author} {\bibfnamefont {F.}~\bibnamefont {Grusdt}},\ }\bibfield  {title}
  {\bibinfo {title} {{Quantum simulation of fermionic non-Abelian lattice gauge
  theories in $(2+1)$D with built-in gauge protection}},\ }\href@noop {} {\
  (\bibinfo {year} {2025})},\ \Eprint {https://arxiv.org/abs/2506.14747}
  {arXiv:2506.14747 [cond-mat.quant-gas]} \BibitemShut {NoStop}%
\bibitem [{\citenamefont {Lamm}\ \emph
  {et~al.}(2020{\natexlab{b}})\citenamefont {Lamm}, \citenamefont {Lawrence},\
  and\ \citenamefont {Yamauchi}}]{Lamm:2020jwv}%
  \BibitemOpen
  \bibfield  {author} {\bibinfo {author} {\bibfnamefont {H.}~\bibnamefont
  {Lamm}}, \bibinfo {author} {\bibfnamefont {S.}~\bibnamefont {Lawrence}},\
  and\ \bibinfo {author} {\bibfnamefont {Y.}~\bibnamefont {Yamauchi}} (\bibinfo
  {collaboration} {NuQS}),\ }\bibfield  {title} {\bibinfo {title} {{Suppressing
  Coherent Gauge Drift in Quantum Simulations}},\ }\href@noop {} {\  (\bibinfo
  {year} {2020}{\natexlab{b}})},\ \Eprint {https://arxiv.org/abs/2005.12688}
  {arXiv:2005.12688 [quant-ph]} \BibitemShut {NoStop}%
\bibitem [{\citenamefont {Kasper}\ \emph {et~al.}(2023)\citenamefont {Kasper},
  \citenamefont {Zache}, \citenamefont {Jendrzejewski}, \citenamefont
  {Lewenstein},\ and\ \citenamefont {Zohar}}]{Kasper:2020owz}%
  \BibitemOpen
  \bibfield  {author} {\bibinfo {author} {\bibfnamefont {V.}~\bibnamefont
  {Kasper}}, \bibinfo {author} {\bibfnamefont {T.~V.}\ \bibnamefont {Zache}},
  \bibinfo {author} {\bibfnamefont {F.}~\bibnamefont {Jendrzejewski}}, \bibinfo
  {author} {\bibfnamefont {M.}~\bibnamefont {Lewenstein}},\ and\ \bibinfo
  {author} {\bibfnamefont {E.}~\bibnamefont {Zohar}},\ }\bibfield  {title}
  {\bibinfo {title} {{Non-Abelian gauge invariance from dynamical
  decoupling}},\ }\href {https://doi.org/10.1103/PhysRevD.107.014506}
  {\bibfield  {journal} {\bibinfo  {journal} {Phys. Rev. D}\ }\textbf {\bibinfo
  {volume} {107}},\ \bibinfo {pages} {014506} (\bibinfo {year} {2023})},\
  \Eprint {https://arxiv.org/abs/2012.08620} {arXiv:2012.08620 [quant-ph]}
  \BibitemShut {NoStop}%
\bibitem [{\citenamefont {Tran}\ \emph {et~al.}(2021)\citenamefont {Tran},
  \citenamefont {Su}, \citenamefont {Carney},\ and\ \citenamefont
  {Taylor}}]{Tran:2020azk}%
  \BibitemOpen
  \bibfield  {author} {\bibinfo {author} {\bibfnamefont {M.~C.}\ \bibnamefont
  {Tran}}, \bibinfo {author} {\bibfnamefont {Y.}~\bibnamefont {Su}}, \bibinfo
  {author} {\bibfnamefont {D.}~\bibnamefont {Carney}},\ and\ \bibinfo {author}
  {\bibfnamefont {J.~M.}\ \bibnamefont {Taylor}},\ }\bibfield  {title}
  {\bibinfo {title} {{Faster Digital Quantum Simulation by Symmetry
  Protection}},\ }\href {https://doi.org/10.1103/PRXQuantum.2.010323}
  {\bibfield  {journal} {\bibinfo  {journal} {PRX Quantum}\ }\textbf {\bibinfo
  {volume} {2}},\ \bibinfo {pages} {010323} (\bibinfo {year} {2021})},\ \Eprint
  {https://arxiv.org/abs/2006.16248} {arXiv:2006.16248 [quant-ph]} \BibitemShut
  {NoStop}%
\bibitem [{\citenamefont {Zhao}\ \emph {et~al.}(2023)\citenamefont {Zhao},
  \citenamefont {Bukov}, \citenamefont {Heyl},\ and\ \citenamefont
  {Moessner}}]{Zhao:2022dma}%
  \BibitemOpen
  \bibfield  {author} {\bibinfo {author} {\bibfnamefont {H.}~\bibnamefont
  {Zhao}}, \bibinfo {author} {\bibfnamefont {M.}~\bibnamefont {Bukov}},
  \bibinfo {author} {\bibfnamefont {M.}~\bibnamefont {Heyl}},\ and\ \bibinfo
  {author} {\bibfnamefont {R.}~\bibnamefont {Moessner}},\ }\bibfield  {title}
  {\bibinfo {title} {{Making Trotterization Adaptive and Energy-Self-Correcting
  for NISQ Devices and Beyond}},\ }\href
  {https://doi.org/10.1103/PRXQuantum.4.030319} {\bibfield  {journal} {\bibinfo
   {journal} {PRX Quantum}\ }\textbf {\bibinfo {volume} {4}},\ \bibinfo {pages}
  {030319} (\bibinfo {year} {2023})},\ \Eprint
  {https://arxiv.org/abs/2209.12653} {arXiv:2209.12653 [quant-ph]} \BibitemShut
  {NoStop}%
\bibitem [{\citenamefont {Surace}\ \emph {et~al.}(2024)\citenamefont {Surace},
  \citenamefont {Fromholz}, \citenamefont {Scazza},\ and\ \citenamefont
  {Dalmonte}}]{Surace:2023qwo}%
  \BibitemOpen
  \bibfield  {author} {\bibinfo {author} {\bibfnamefont {F.~M.}\ \bibnamefont
  {Surace}}, \bibinfo {author} {\bibfnamefont {P.}~\bibnamefont {Fromholz}},
  \bibinfo {author} {\bibfnamefont {F.}~\bibnamefont {Scazza}},\ and\ \bibinfo
  {author} {\bibfnamefont {M.}~\bibnamefont {Dalmonte}},\ }\bibfield  {title}
  {\bibinfo {title} {{Scalable, ab initio protocol for quantum simulating
  SU($N$)$\times$U(1) Lattice Gauge Theories}},\ }\href
  {https://doi.org/10.22331/q-2024-05-23-1359} {\bibfield  {journal} {\bibinfo
  {journal} {Quantum}\ }\textbf {\bibinfo {volume} {8}},\ \bibinfo {pages}
  {1359} (\bibinfo {year} {2024})},\ \Eprint {https://arxiv.org/abs/2310.08643}
  {arXiv:2310.08643 [cond-mat.quant-gas]} \BibitemShut {NoStop}%
\bibitem [{\citenamefont {Domanti}\ \emph {et~al.}(2024)\citenamefont
  {Domanti}, \citenamefont {Zappala}, \citenamefont {Bermudez},\ and\
  \citenamefont {Amico}}]{Domanti:2023qht}%
  \BibitemOpen
  \bibfield  {author} {\bibinfo {author} {\bibfnamefont {E.~C.}\ \bibnamefont
  {Domanti}}, \bibinfo {author} {\bibfnamefont {D.}~\bibnamefont {Zappala}},
  \bibinfo {author} {\bibfnamefont {A.}~\bibnamefont {Bermudez}},\ and\
  \bibinfo {author} {\bibfnamefont {L.}~\bibnamefont {Amico}},\ }\bibfield
  {title} {\bibinfo {title} {{Floquet-Rydberg quantum simulator for confinement
  in Z2 gauge theories}},\ }\href
  {https://doi.org/10.1103/PhysRevResearch.6.L022059} {\bibfield  {journal}
  {\bibinfo  {journal} {Phys. Rev. Res.}\ }\textbf {\bibinfo {volume} {6}},\
  \bibinfo {pages} {L022059} (\bibinfo {year} {2024})},\ \Eprint
  {https://arxiv.org/abs/2311.16758} {arXiv:2311.16758 [cond-mat.quant-gas]}
  \BibitemShut {NoStop}%
\bibitem [{\citenamefont {Okuda}\ \emph {et~al.}(2024)\citenamefont {Okuda},
  \citenamefont {Parayil~Mana},\ and\ \citenamefont {Sukeno}}]{Okuda:2024jzh}%
  \BibitemOpen
  \bibfield  {author} {\bibinfo {author} {\bibfnamefont {T.}~\bibnamefont
  {Okuda}}, \bibinfo {author} {\bibfnamefont {A.}~\bibnamefont
  {Parayil~Mana}},\ and\ \bibinfo {author} {\bibfnamefont {H.}~\bibnamefont
  {Sukeno}},\ }\bibfield  {title} {\bibinfo {title} {{Anomaly inflow,
  dualities, and quantum simulation of Abelian lattice gauge theories induced
  by measurements}},\ }\href {https://doi.org/10.1103/PhysRevResearch.6.043018}
  {\bibfield  {journal} {\bibinfo  {journal} {Phys. Rev. Res.}\ }\textbf
  {\bibinfo {volume} {6}},\ \bibinfo {pages} {043018} (\bibinfo {year}
  {2024})},\ \Eprint {https://arxiv.org/abs/2402.08720} {arXiv:2402.08720
  [cond-mat.str-el]} \BibitemShut {NoStop}%
\bibitem [{\citenamefont {Schmale}\ and\ \citenamefont
  {Weimer}(2024)}]{Schmale:2024ebh}%
  \BibitemOpen
  \bibfield  {author} {\bibinfo {author} {\bibfnamefont {T.}~\bibnamefont
  {Schmale}}\ and\ \bibinfo {author} {\bibfnamefont {H.}~\bibnamefont
  {Weimer}},\ }\bibfield  {title} {\bibinfo {title} {{Stabilizing quantum
  simulations of lattice gauge theories by dissipation}},\ }\href
  {https://doi.org/10.1103/PhysRevResearch.6.033306} {\bibfield  {journal}
  {\bibinfo  {journal} {Phys. Rev. Res.}\ }\textbf {\bibinfo {volume} {6}},\
  \bibinfo {pages} {033306} (\bibinfo {year} {2024})},\ \Eprint
  {https://arxiv.org/abs/2404.16454} {arXiv:2404.16454 [quant-ph]} \BibitemShut
  {NoStop}%
\bibitem [{\citenamefont {Ballini}\ \emph {et~al.}(2024)\citenamefont
  {Ballini}, \citenamefont {Mildenberger}, \citenamefont {Wauters},\ and\
  \citenamefont {Hauke}}]{Ballini:2024qmr}%
  \BibitemOpen
  \bibfield  {author} {\bibinfo {author} {\bibfnamefont {E.}~\bibnamefont
  {Ballini}}, \bibinfo {author} {\bibfnamefont {J.}~\bibnamefont
  {Mildenberger}}, \bibinfo {author} {\bibfnamefont {M.~M.}\ \bibnamefont
  {Wauters}},\ and\ \bibinfo {author} {\bibfnamefont {P.}~\bibnamefont
  {Hauke}},\ }\bibfield  {title} {\bibinfo {title} {{Symmetry verification for
  noisy quantum simulations of non-Abelian lattice gauge theories}},\
  }\href@noop {} {\  (\bibinfo {year} {2024})},\ \Eprint
  {https://arxiv.org/abs/2412.07844} {arXiv:2412.07844 [quant-ph]} \BibitemShut
  {NoStop}%
\bibitem [{\citenamefont {Trotter}(1959)}]{trotterProductSemigroups59}%
  \BibitemOpen
  \bibfield  {author} {\bibinfo {author} {\bibfnamefont {H.~F.}\ \bibnamefont
  {Trotter}},\ }\bibfield  {title} {\bibinfo {title} {On the product of
  semi-groups of operators},\ }\href
  {https://doi.org/10.1090/S0002-9939-1959-0108732-6} {\bibfield  {journal}
  {\bibinfo  {journal} {Proc. Amer. Math. Soc.}\ }\textbf {\bibinfo {volume}
  {10}},\ \bibinfo {pages} {545} (\bibinfo {year} {1959})}\BibitemShut
  {NoStop}%
\bibitem [{\citenamefont {Suzuki}(1976)}]{Suzuki:1976be}%
  \BibitemOpen
  \bibfield  {author} {\bibinfo {author} {\bibfnamefont {M.}~\bibnamefont
  {Suzuki}},\ }\bibfield  {title} {\bibinfo {title} {{Generalized Trotter's
  Formula and Systematic Approximants of Exponential Operators and Inner
  Derivations with Applications to Many Body Problems}},\ }\href
  {https://doi.org/10.1007/BF01609348} {\bibfield  {journal} {\bibinfo
  {journal} {Commun. Math. Phys.}\ }\textbf {\bibinfo {volume} {51}},\ \bibinfo
  {pages} {183} (\bibinfo {year} {1976})}\BibitemShut {NoStop}%
\bibitem [{\citenamefont {Childs}\ \emph {et~al.}(2021)\citenamefont {Childs},
  \citenamefont {Su}, \citenamefont {Tran}, \citenamefont {Wiebe},\ and\
  \citenamefont {Zhu}}]{Childs:2019hts}%
  \BibitemOpen
  \bibfield  {author} {\bibinfo {author} {\bibfnamefont {A.~M.}\ \bibnamefont
  {Childs}}, \bibinfo {author} {\bibfnamefont {Y.}~\bibnamefont {Su}}, \bibinfo
  {author} {\bibfnamefont {M.~C.}\ \bibnamefont {Tran}}, \bibinfo {author}
  {\bibfnamefont {N.}~\bibnamefont {Wiebe}},\ and\ \bibinfo {author}
  {\bibfnamefont {S.}~\bibnamefont {Zhu}},\ }\bibfield  {title} {\bibinfo
  {title} {{Theory of Trotter Error with Commutator Scaling}},\ }\href
  {https://doi.org/10.1103/PhysRevX.11.011020} {\bibfield  {journal} {\bibinfo
  {journal} {Phys. Rev. X}\ }\textbf {\bibinfo {volume} {11}},\ \bibinfo
  {pages} {011020} (\bibinfo {year} {2021})},\ \Eprint
  {https://arxiv.org/abs/1912.08854} {arXiv:1912.08854 [quant-ph]} \BibitemShut
  {NoStop}%
\bibitem [{\citenamefont {Cuccaro}\ \emph {et~al.}(2004)\citenamefont
  {Cuccaro}, \citenamefont {Draper}, \citenamefont {Kutin},\ and\ \citenamefont
  {Moulton}}]{Cuccaro:2004xxx}%
  \BibitemOpen
  \bibfield  {author} {\bibinfo {author} {\bibfnamefont {S.~A.}\ \bibnamefont
  {Cuccaro}}, \bibinfo {author} {\bibfnamefont {T.~G.}\ \bibnamefont {Draper}},
  \bibinfo {author} {\bibfnamefont {S.~A.}\ \bibnamefont {Kutin}},\ and\
  \bibinfo {author} {\bibfnamefont {D.~P.}\ \bibnamefont {Moulton}},\
  }\bibfield  {title} {\bibinfo {title} {{A new quantum ripple-carry addition
  circuit}},\ }\href@noop {} {\  (\bibinfo {year} {2004})},\ \Eprint
  {https://arxiv.org/abs/quant-ph/0410184} {arXiv:quant-ph/0410184}
  \BibitemShut {NoStop}%
\bibitem [{\citenamefont {Draper}\ \emph {et~al.}(2006)\citenamefont {Draper},
  \citenamefont {Kutin}, \citenamefont {Rains},\ and\ \citenamefont
  {Svore}}]{Draper:2004ayj}%
  \BibitemOpen
  \bibfield  {author} {\bibinfo {author} {\bibfnamefont {T.~G.}\ \bibnamefont
  {Draper}}, \bibinfo {author} {\bibfnamefont {S.~A.}\ \bibnamefont {Kutin}},
  \bibinfo {author} {\bibfnamefont {E.~M.}\ \bibnamefont {Rains}},\ and\
  \bibinfo {author} {\bibfnamefont {K.~M.}\ \bibnamefont {Svore}},\ }\bibfield
  {title} {\bibinfo {title} {{A logarithmic-depth quantum carry-lookahead
  adder}},\ }\href {https://doi.org/10.26421/QIC6.4-5-4} {\bibfield  {journal}
  {\bibinfo  {journal} {Quant. Inf. Comput.}\ }\textbf {\bibinfo {volume}
  {6}},\ \bibinfo {pages} {351} (\bibinfo {year} {2006})},\ \Eprint
  {https://arxiv.org/abs/quant-ph/0406142} {arXiv:quant-ph/0406142}
  \BibitemShut {NoStop}%
\bibitem [{\citenamefont {Nielsen}\ and\ \citenamefont
  {Chuang}(2012)}]{Nielsen:2012yss}%
  \BibitemOpen
  \bibfield  {author} {\bibinfo {author} {\bibfnamefont {M.~A.}\ \bibnamefont
  {Nielsen}}\ and\ \bibinfo {author} {\bibfnamefont {I.~L.}\ \bibnamefont
  {Chuang}},\ }\href {https://doi.org/10.1017/cbo9780511976667} {\emph
  {\bibinfo {title} {{Quantum Computation and Quantum Information}}}}\
  (\bibinfo  {publisher} {Cambridge University Press},\ \bibinfo {year}
  {2012})\BibitemShut {NoStop}%
\bibitem [{\citenamefont {Kogut}\ and\ \citenamefont
  {Susskind}(1975)}]{Kogut:1974ag}%
  \BibitemOpen
  \bibfield  {author} {\bibinfo {author} {\bibfnamefont {J.~B.}\ \bibnamefont
  {Kogut}}\ and\ \bibinfo {author} {\bibfnamefont {L.}~\bibnamefont
  {Susskind}},\ }\bibfield  {title} {\bibinfo {title} {{Hamiltonian Formulation
  of Wilson's Lattice Gauge Theories}},\ }\href
  {https://doi.org/10.1103/PhysRevD.11.395} {\bibfield  {journal} {\bibinfo
  {journal} {Phys. Rev. D}\ }\textbf {\bibinfo {volume} {11}},\ \bibinfo
  {pages} {395} (\bibinfo {year} {1975})}\BibitemShut {NoStop}%
\bibitem [{\citenamefont {Davoudi}\ \emph {et~al.}(2023)\citenamefont
  {Davoudi}, \citenamefont {Shaw},\ and\ \citenamefont
  {Stryker}}]{Davoudi:2022xmb}%
  \BibitemOpen
  \bibfield  {author} {\bibinfo {author} {\bibfnamefont {Z.}~\bibnamefont
  {Davoudi}}, \bibinfo {author} {\bibfnamefont {A.~F.}\ \bibnamefont {Shaw}},\
  and\ \bibinfo {author} {\bibfnamefont {J.~R.}\ \bibnamefont {Stryker}},\
  }\bibfield  {title} {\bibinfo {title} {{General quantum algorithms for
  Hamiltonian simulation with applications to a non-Abelian lattice gauge
  theory}},\ }\href {https://doi.org/10.22331/q-2023-12-20-1213} {\bibfield
  {journal} {\bibinfo  {journal} {Quantum}\ }\textbf {\bibinfo {volume} {7}},\
  \bibinfo {pages} {1213} (\bibinfo {year} {2023})},\ \Eprint
  {https://arxiv.org/abs/2212.14030} {arXiv:2212.14030 [hep-lat]} \BibitemShut
  {NoStop}%
\bibitem [{\citenamefont {Beane}\ \emph {et~al.}(2014)\citenamefont {Beane},
  \citenamefont {Davoudi},\ and\ \citenamefont {Savage}}]{Beane:2012rz}%
  \BibitemOpen
  \bibfield  {author} {\bibinfo {author} {\bibfnamefont {S.~R.}\ \bibnamefont
  {Beane}}, \bibinfo {author} {\bibfnamefont {Z.}~\bibnamefont {Davoudi}},\
  and\ \bibinfo {author} {\bibfnamefont {M.~J.}\ \bibnamefont {Savage}},\
  }\bibfield  {title} {\bibinfo {title} {{Constraints on the Universe as a
  Numerical Simulation}},\ }\href {https://doi.org/10.1140/epja/i2014-14148-0}
  {\bibfield  {journal} {\bibinfo  {journal} {Eur. Phys. J. A}\ }\textbf
  {\bibinfo {volume} {50}},\ \bibinfo {pages} {148} (\bibinfo {year} {2014})},\
  \Eprint {https://arxiv.org/abs/1210.1847} {arXiv:1210.1847 [hep-ph]}
  \BibitemShut {NoStop}%
\end{thebibliography}%

\end{document}